%% file: main.tex
\documentclass[acmtog]{acmart}

\usepackage{multirow}
\usepackage{footnote}
\usepackage{booktabs} 
\newcommand{\nothing}[1]{}
\usepackage{subcaption}
\usepackage[ruled]{algorithm2e} 

\usepackage{wrapfig}
\usepackage{subcaption}

\usepackage{blindtext}
\usepackage{graphicx}
\usepackage{caption}

\usepackage{array}
\newcolumntype{L}[1]{>{\raggedright\let\newline\\\arraybackslash\hspace{0pt}}m{#1}}
\newcolumntype{C}[1]{>{\centering\let\newline\\\arraybackslash\hspace{0pt}}m{#1}}
\newcolumntype{R}[1]{>{\raggedleft\let\newline\\\arraybackslash\hspace{0pt}}m{#1}}

\SetAlFnt{  \small}
\SetAlCapFnt{\small}
\SetAlCapNameFnt{\small}
\SetAlCapHSkip{0pt}

\setcopyright{acmlicensed}
\acmJournal{TOG}
\acmYear{2020}
\acmVolume{39}
\acmNumber{6}
\acmArticle{215}
\acmMonth{12} 
\acmDOI{10.1145/3414685.3417817}

\def\showcomments{0}
\ifx\showcomments\undefined
\newcommand\todo[1]{}
\newcommand{\yajiecomment}[1]{}
\else
\newcommand{\todo}[1]{\textcolor{red}{\textbf{[-TODO-~}
    #1
    \textbf{~]}}}
\newcommand{\yajiecomment}[1]{\textbf{[-Yajie-~}
    \textcolor{purple}{#1}
    \textbf{~]}}

\fi

\begin{document}
\title{Dynamic Facial Asset and Rig Generation from a Single Scan}


\author{Jiaman Li}
\affiliation{%
 \institution{University of Southern California}
 }
\affiliation{%
 \institution{USC Institute for Creative Technologies}
}
\authornote{indicates equal contribution.}
\author{Zhengfei Kuang}
\affiliation{%
 \institution{University of Southern California}
}
\affiliation{%
 \institution{USC Institute for Creative Technologies}
}
\authornotemark[1]
\author{Yajie Zhao}
\affiliation{%
 \institution{USC Institute for Creative Technologies}
}
\authornote{indicates corresponding author.}
\author{Mingming He}
\affiliation{%
 \institution{USC Institute for Creative Technologies}
}
\author{Karl Bladin}
\affiliation{%
 \institution{USC Institute for Creative Technologies}
}
\author{Hao Li}
\affiliation{%
 \institution{University of Southern California}
}
\affiliation{%
 \institution{USC Institute for Creative Technologies}
}
\affiliation{%
 \institution{Pinscreen}
}



\begin{teaserfigure}
  \centering
  \footnotesize
  \vspace{-2mm}
  \includegraphics[width=\linewidth]{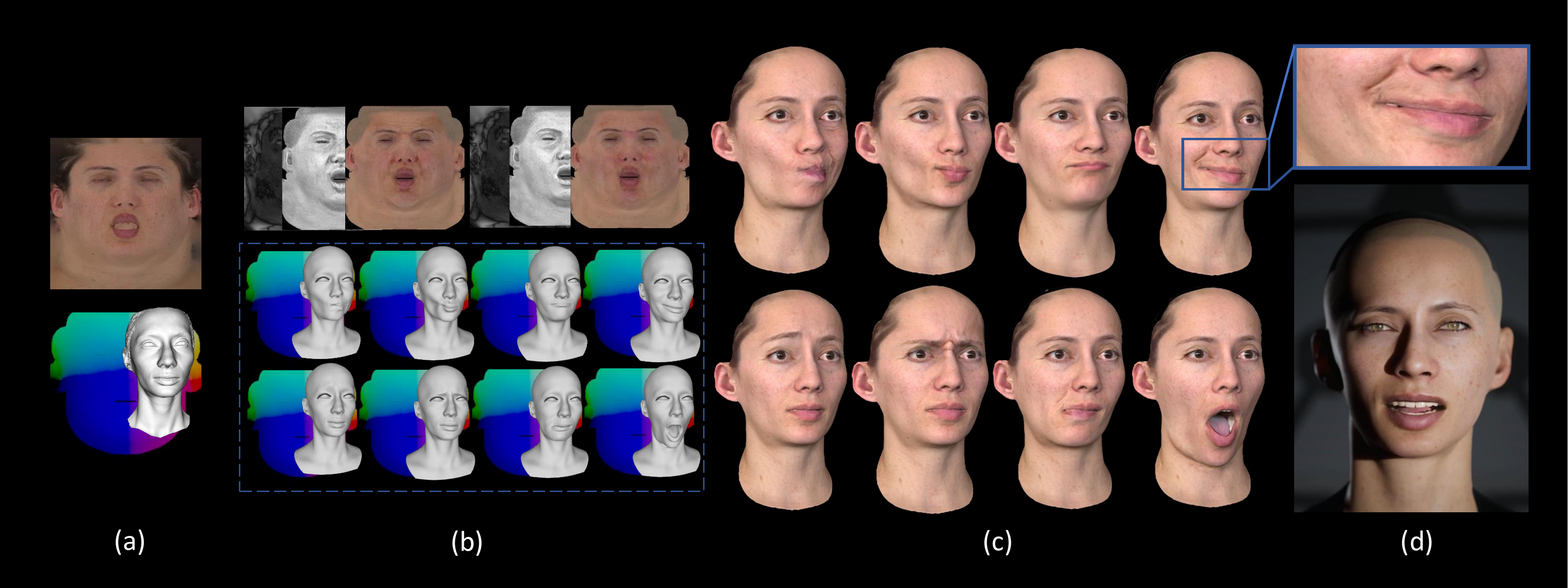}\\
  \vspace{-2mm}
  \caption{Given a single neutral scan (a), we generate a complete set of dynamic face model assets, including personalized blendshapes and physically-based dynamic facial skin textures of the input subjects (b). The results carry high-fidelity details which we render in Arnold~\cite{arnold} (c). Our generated facial assets are animation-ready as shown in (d).}
  \label{fig:teaser}
\end{teaserfigure}


\begin{abstract}
The creation of high-fidelity computer-generated (CG) characters for films and games is tied with intensive manual labor, which involves the creation of comprehensive facial assets that are often captured using complex hardware. To simplify and accelerate this digitization process, we propose a framework for the automatic generation of high-quality dynamic facial models, including rigs which can be readily deployed for artists to polish. Our framework takes a single scan as input to generate a set of personalized blendshapes, dynamic textures, as well as secondary facial components (\textit{e.g.}, teeth and eyeballs). Based on a facial database with over $4,000$ scans with pore-level details, varying expressions and identities, we adopt a self-supervised neural network to learn personalized blendshapes from a set of template expressions. We also model the joint distribution between identities and expressions, enabling the inference of a full set of personalized blendshapes with dynamic appearances from a single neutral input scan. Our generated personalized face rig assets are seamlessly compatible with professional production pipelines for facial animation and rendering. We demonstrate a highly robust and effective framework on a wide range of subjects, and showcase high-fidelity facial animations with automatically generated personalized dynamic textures. 
\end{abstract}

%
%


%
%


\maketitle

\input{Overview}

\end{document}

%% file: Overview.tex
\input{Introduction.tex}
\input{Relatedworks.tex}
\input{Method.tex}
\input{Database.tex}
\input{Results.tex}
\input{Conclusion.tex}

%
%
%
%

\bibliographystyle{ACM-Reference-Format}
\bibliography{sample-bibliography}

\appendix

%% file: Introduction.tex
\section{Introduction}
\label{sec:intro}
High-quality and personalized digital humans are relevant to a wide range of applications, such as film and game production (\textit{e.g.} Unreal Engine, Digital Doug), and virtual reality~\cite{fyffe2014driving,lombardi2018deep,wei2019vr}. To produce high-fidelity digital doubles, complex capture equipment is often needed in conventional computer graphics pipelines, and the acquired data typically undergoes intensive manual post-processing by a production team. New approaches based on deep learning-based synthesis are promising as they show how photorealistic faces can be generated from captured data directly~\cite{lombardi2018deep,wei2019vr} allowing one to overcome the notorious Uncanny Valley. In addition to their intensive GPU compute requirements and the need for large volumes of training data, these deep learning-based methods are still difficult to integrate seamlessly into virtual CG environments as they lack relighting capabilities and fine rendering controls, which prevents them from being adopted for games and film production. On the other hand, realistic digital doubles in conventional graphics pipelines require months of production and involve large teams of highly skilled digital artists as well as sophisticated scanning techniques~\cite{ghosh2011multiview}. Building facial assets of a virtual character typically requires a number of facial expression models often based on the Facial Action Coding System (FACS), as well as physically-based texture assets (\textit{e.g.}, albedo, specular maps, displacement maps) to ensure realistic facial skin reflectance in a virtual environment.

Several recent works have shown how to automate and reduce the effort for generating personalized facial rigs. The works of \citet{laine2017production,Li_2010_SIGGRAPH,pawaskar2013expression,ma2016semantically} propose to automatically build personalized blendshapes using a varying number of personalized facial scans. While effective for production pipelines, these methods either require a large number of facial scans as input and considerable post-processing, or they only focus on generating a personalized geometry for the expressions, without the textures. For consumer-accessible avatar creation techniques, the works of \citet{Hu_2017_SIGGRAPH,Nagano_2018_SIGGRAPH,Casas_2016_CASA, Thies_2016_CVPR,ichim2015dynamic} demonstrate digitization capabilities from video sequences or even a single input image. However, due to the limited input data, the resulting models often lack details or the generated assets do not contain physically-based properties for dynamic expressions. 
We propose an approach based on a 3D scan as input and our goal is to produce a fully rigged model with fixed topology, personalized blendshapes expressions along with corresponding dynamic and physically-based texture maps. 
We observe that a large amount of labeled data can enable the learning of personalized models and dynamic deformations such that wrinkle formations are specific to the shape and appearance of the subject. In particular, we extend recent deep learning approaches for high-resolution physically-based skin assets~\cite{li2020learning,yamaguchi2018high}, to generate dynamic high-resolution facial texture attributes (albedo, specular maps, and displacement maps), in order to produce effects such as plausible personalized wrinkles during animation. Existing methods transfer facial expression details from a generic database, which may lead to reasonable output for the geometry, but certainly lack dynamic texture variations. 

We present a framework to automate and simplify the generation of high-quality facial rig assets, consisting of personalized blendshapes, dynamic physically-based skin attributes (albedo, specular reflection, displacement maps), including secondary facial components (\textit{e.g.} eyes, teeth, gums, and tongue), from a single neutral geometric model and albedo map as input. Our generated assets can be directly fed into professional production pipelines. We use a high-fidelity facial scan database~\cite{li2020learning} and address both the problems of generating personalized blendshapes and inferring dynamic physically-based skin properties. We first propose an end-to-end self-supervised learning framework to overcome the lack of ground truth data for personalized blendshapes and dynamic textures. By modeling the correlation between identities and personalized expressions on the database with 178 identities, each having 19$\sim$26 different captured expressions, we eliminate the requirement of user-specific scans for personalized blendshapes generation using a trade-off between semantic meaning and personality.
Our approach uses an intermediate conversion of neutral geometry and 2D textures to a common parameterization in UV space, which enables training and inference of dynamic geometry and texture
deformation in a compact form inspired by~\citet{li2020learning}.

Learning is performed using a high-fidelity facial scan dataset with over $4000$ scans with pore-level details and different expressions. Our approach can automatically produce personalized blendshapes that reflect personalized expressions of a person from only one neutral scan.
We demonstrate the effectiveness of our framework on a wide range of subjects and showcase a number of compelling facial animations. 

In summary, our major contributions are as follows:
\begin{itemize}
\item We propose an end-to-end framework to automate the generation of high-quality facial assets and rigs. Given a single neutral face scan with albedo as input, we produce plausible personalized blendshapes, secondary facial components (\textit{e.g.} teeth, eyelashes), and most importantly, physically-based textures that are both dynamic and personalized to the appearance of the input subject.
\item We present a novel self-supervised deep learning approach to improve the personalized results using a generic facial expression template model. In particular, our approach can model the joint distribution between individual identities and their expressions in a large high-fidelity face database.
\item We also introduce a novel physically-based texture synthesis framework conditioned on neutral geometry and textures. Using a new compress and stretch map approach, we are able to synthesize dynamic expression-specific textures, including albedo, specular, and fine-scale displacement maps.
\item We will make our code, models and database with all texture assets public to facilitate further research on automating high-quality avatar generation.
\end{itemize}

%% file: Relatedworks.tex
\section{Related Work}
\label{sec:relatedworks}
\paragraph{Facial Capture}
Due to increased demands for realistic digital avatars, facial capture and performance capture have been well-studied. Based on a multi-view stereo system, fine-scale details of the captured face can be recovered in a controlled environment with multiple calibrated DSLR cameras as in the work of \citet{Beeler_2010_SIGGRAPH}. A more intricate system by \citet{ghosh2011multiview} extends the view-dependent method~\cite{Ma_2007_EG} by adopting fixed linear polarized spherical gradient illumination in front of the cameras and enables accurate acquisition of diffuse albedo, specular intensity, and pore-level normal maps. \citet{Fyffe_2016_EG} later propose a method that employs commodity hardware, while recording comparable results with off-the-shelf components and near-instant capture.
Meanwhile, works on passive facial performance capturing \cite{Bradley_2010_SIGGRAPH, Beeler_2011_SIGGRAPH,Valgaerts_2012_SIGGRAPH,fyffe2014driving} have shown impressive detailed results for highly articulated motion. Recently, \citet{Gotardo_2018_SIGGRAPH} propose a method to acquire dynamic properties of facial skin appearance, including dynamic diffuse albedo, specular intensity, and normal maps. These methods provide decent training data and set a high baseline for lightweight facial capture and modeling approaches.
\paragraph{Facial Rigging}
Creating facial animation is a well-studied problem with a plethora of methods proposed in film and video game industries.
\citet{Blanz_1999_SIGGRAPH} first introduce the Morphable Face Model to represent face shapes and textures of different identities using principal component analysis (PCA) learned from $200$ laser scan subjects. Later, the improved parametric face models are built using $10,000$ high-quality 3D face scans \cite{Booth_2016_CVPR,Booth_2017_CVPR}. A linear model generated from web images has also been demonstrated \cite{Ira_2013_ICCV}.

Modeling of variational face expressions using blendshapes is a popular approach in many applications \cite{Thies_2015_SIGGRAPH,Thies_2016_CVPR}. The approach models facial expressions as activation of shape units represented by a linear basis of facial expression vectors \cite{Sylvain_2014_EG}. \citet{Amberg_2008} combines a PCA model of a neutral face with a PCA space derived from the residual vectors of different expressions to the neutral pose. Blendshapes can either be hand-crafted by animators \cite{Alexander_2009_SIGGRAPH,Olszewski_2016_SIGGRAPH}, or be generated via statistical analysis from large facial expression datasets \cite{Vlasic_2005_SIGGRAPH,Cao_2014_TVCG,Li_2017_SIGGRAPH}. The multi-linear model~\cite{Vlasic_2005_SIGGRAPH,Cao_2014_TVCG} offers a way of capturing a joint space of expression and identity. \citet{Li_2017_SIGGRAPH} propose the FLAME model learned from thousands of scans and significantly improve the model expressiveness.
\paragraph{Personalized Blendshape Generation}
As an effort to advance and scale the production of facial animations, expression cloning \cite{Noh_2001_SIGGRAPH} has been introduced to mimic the existing deformation of a source 3D face model onto a target face. \citet{Sumner_2004_SIGGRAPH} propose deformation transfer for generic 3D triangle mesh. \citet{Onizuka_2019_ICCV} propose a landmark-guided deformation transfer method to generate expressions for any target avatar that directly maps to a generic blendshape template. These methods can generate an expression for a novel subject but might fail to capture personalized behavior due to the lack of personal information. 

To build robust face rigs, we need to reconstruct a dynamic expression model that faithfully captures the subject’s specific facial movements. A full set of personalized blendshapes for a specific subject can be built from 3D scan data of the same subject \cite{Zhang_2004_SIGGRAPH,Weise_2009_SCA,Huang_2011_SIGGRAPH,carrigan2020expression,Li_2010_SIGGRAPH}. These methods can reconstruct expressions that capture the target's personal expressions, but a large set of action units or sparse expressions are required as input. 
Some follow-up works \cite{Bouaziz_2013_SIGGRAPH,Li_2013_SIGGRAPH,Hsieh2015UnconstrainedRF} apply expression transfer on top of a generic face model and train  model correctives for the expressions during tracking with samples obtained from RGB-D video input. \citet{ichim2015dynamic} and \citet{Cao_2016_SIGGRAPH} propose a comprehensive pipeline to generate a dynamic 3D avatars based on personalized blendshapes with a monocular video of a specific expression sequence. \citet{Casas_2016_CASA} reconstruct blendshapes and each blendshape's textures with a Kinect. \citet{Garrido_2016_SIGGRAPH} introduce a video-based method, which makes blendshape generation suitable for legacy video footage.
\paragraph{Deep Face Models}
As deep learning-based methods for 3D shapes analysis have attracted increasing attention in recent years, some methods for non-linear 3D Morphable Model learning have been introduced \cite{Tewari_2017_ICCV, Bagautdinov_2018_CVPR, Tran_2018_CVPR, Tran_2019_CVPR, li2020learning}. These models are formulated as decoders using convolutional neural networks, some of these methods use fully connected layers or 2D convolutions in the image space \cite{li2020learning}, while some build decoders in the mesh domain to exploit the local geometry of 3D structures \cite{Litany_2018_CVPR, Ranjan_2018_ECCV, Zhou_2019_CVPR, Cheng_2019_MeshGAN, Abrevaya_2019_CVPR}. 
\begin{figure*}[ht!]
 \includegraphics[width=7.1in]{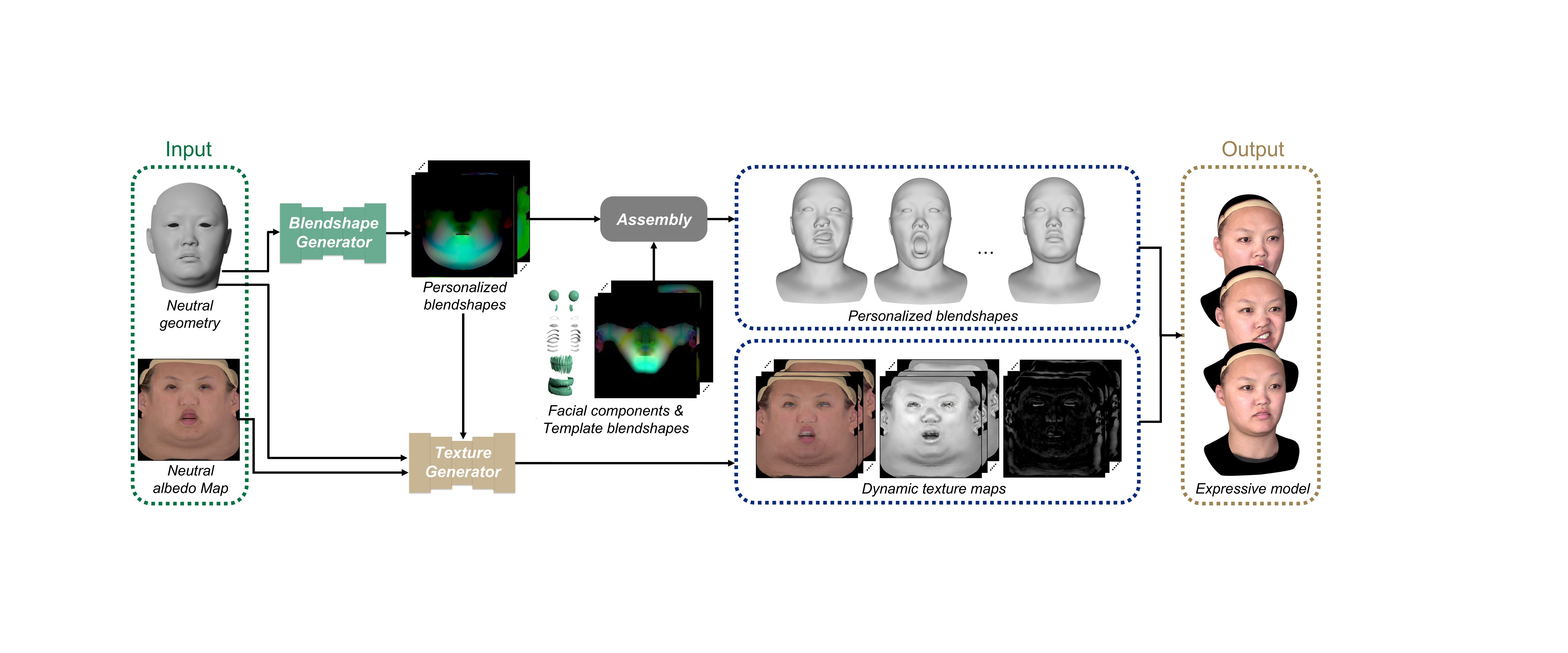}
 \caption{System Overview. Given the model from a single scan in a neutral expression, the blendshape generation module first generates its personalized blendshapes. Then, using the personalized blendshapes, along with the input neutral model and its albedo map, the texture generation module produces high-resolution dynamic texture maps including albedo, specular intensity and displacement maps. With these assets ready, we then assemble personalized blendshapes and the input neutral model into 3D models, combining other facial components (eyes, teeth, gums, and tongue) from the template models. The final output is complete face models rendered using the blendshape models and textures.}
 \label{fig:overview_system}
\end{figure*}
\paragraph{Image-to-Image Translation}
\citet{Isola_2016_CVPR} present Pix2Pix, a method to translate images from one domain to another. It consists of a generator and a discriminator, where the objective of the generator is to translate images from domain A to B, while the discriminator aims to distinguish real images from the translated ones. \citet{wang_2018_CVPR} later extend this work to Pix2PixHD to synthesize high-resolution photo-realistic images from semantic label maps. Some works \cite{wang_2018_NeurIPS, wang_2019_NeurIPS, lee2019metapix} on the learning of ``translation'' functions for videos also incorporate a spatio-temporal adversarial objective.
Image-to-image translation has also been adopted to generate 3D faces or detailed face textures. \citet{Sela_2017_ICCV} propose a Pix2Vertex framework using image-to-image translation that jointly maps the input image to a depth image and a facial correspondence map. \citet{Huynh_2018_CVPR} applies this image-to-image translation framework to infer mesoscopic facial geometry with high-quality training data captured using the Light Stage. \citet{yamaguchi2018high} presents a comprehensive method to infer facial reflectance maps from unconstrained image input. \citet{Nagano_2018_SIGGRAPH} introduces a framework to synthesize arbitrary expressions in image space and textures in UV space from a single input image. \citet{Chen_2019_ICCV} adopts a conditional GAN to synthesize geometric details (\textit{wrinkles}) by estimating a displacement map over a proxy mesh. Similarly, \citet{yang2020facescape} infers a displacement map on a base mesh generated from a single image based on a large high-quality face dataset.

%% file: Method.tex
\section{System Overview}
\label{sec:method}
Our system takes a single scanned neutral geometry with an albedo map as input and generates a set of face rig assets and texture attributes for physically based production-level rendering. As shown in Fig.~\ref{fig:overview_system}, we developed a cascaded framework, in which we first estimate a set of personalized blendshape geometries of the input subject using a Blendshape Generation network, followed by a Texture Generation network to infer a set of dynamic maps including albedo maps, specular intensity maps, and displacement maps. In the final step, we combine the obtained secondary facial components (\textit{i.e.} teeth, gums, and eye assets) from a set of template shapes, to assemble the final face model.
\section{Blendshape Generation}
\begin{figure*}[ht!]
 \includegraphics[width=7.1in]{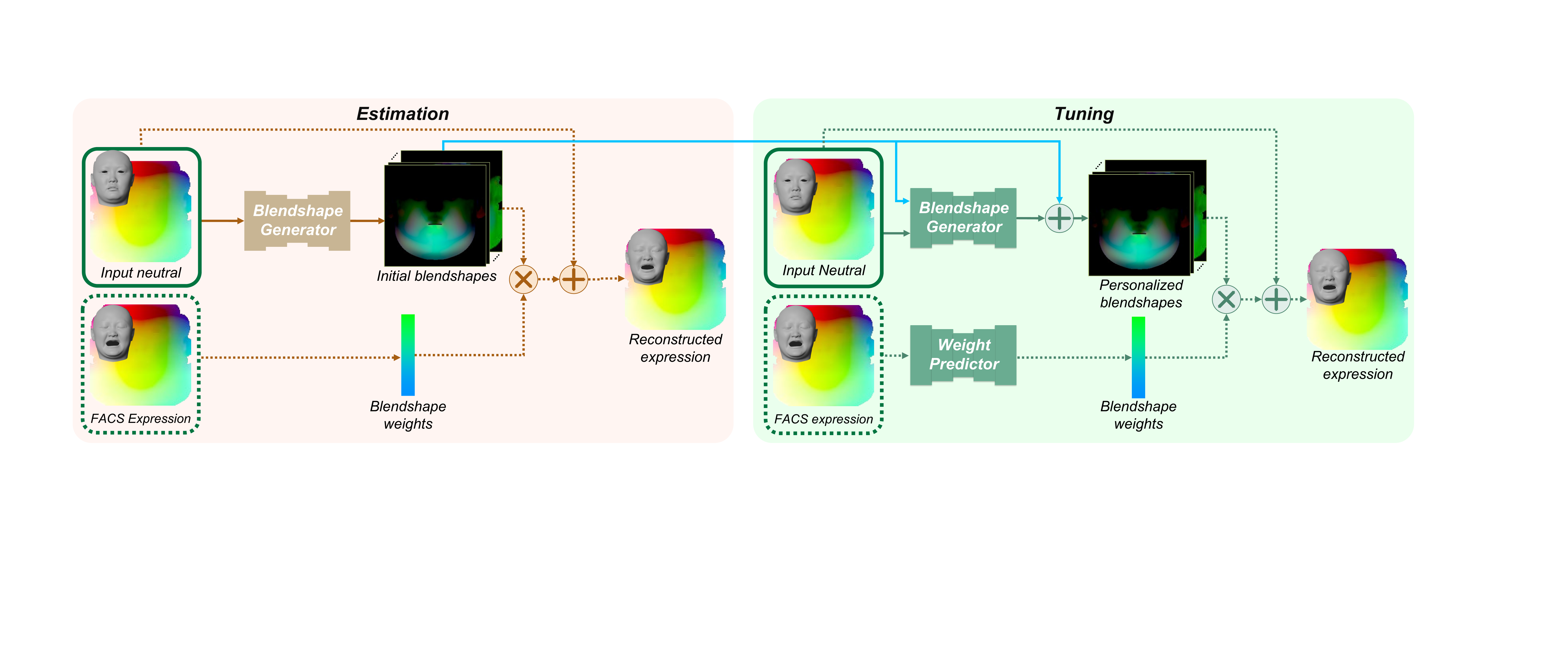}
 \caption{Two-stage self-supervised learning framework. Given a model in a neutral expression, the Estimation Stage first predicts the initial blendshapes which will work as input for the Tuning Stage to generate the final personalized blendshapes. The inference pipeline is connected by solid lines. The training architecture also involves the parts in dashed lines for computing reconstruction loss. In the Estimation Stage, the \textit{Blendshape Generator} learns to generate the initial blendshapes from the input neutral expression, which combines with the known blending weights to reconstruct the non-neutral expressions. In the Tuning Stage, the \textit{Blending Weight Predictor} is added to predict blending weights for the personalized blendshapes which will be used to reconstruct the input expression.}
 \label{fig:learning_pipeline}
\end{figure*}
Our goal is to automatically generate a full set of personalized blendshapes from a neutral 3D face of a novel subject. This is a challenging problem since generating subject-specific blendshapes usually requires different expressions of the subject. Thanks to our large-scale dataset which consists of various expressions as described in Sec. \ref{sec:data}, we introduce a self-supervised pipeline that learns to generate personalized blendshapes based on expressions. Our first task is to imitate the process followed by artists isolating scanned expressions to unit blendshapes using deep neural networks. Given a set of pre-defined generic template blendshapes as a semantic reference and multiple well-defined scan expressions of the same subject, our first goal is to automatically generate the personalized blendshapes of the input subject. 

The generic template blendshape model is defined as a generic model $S_0$ in neutral expression and a set of $N$ (in our case $N=55$) additive vector displacements $\textbf{S}=\{S_1,...,S_N\}$. Expressions can be generated as $P_k=S_0+\sum_{i=1}^N \alpha_{ik} S_i$, where $\alpha_{ik}$ are the blending weights for the expression $k$. For a new subject $j$, given his/her neutral expression model $S^{j}_{0}$ and other expressions $P^{j}_{k}$, their personalized blendshapes $S^{j}_{i}$ can be optimized by minimizing the reconstruction loss of $P^{j}_{k}{}'$ and ground truth expression $P^{j}_{k}$ if blending weights $\alpha^{j}_{ik}, i=1,...,N$ for $P^{j}_{k}$ are known:
\begin{equation}
    P^j_k{}'=S^j_0+ \sum_{i=1}^N \alpha^j_{ik} S^j_i.
    \label{eq:recon}
\end{equation}
This is the foundation of our self-supervised learning scheme.
 
Based on our template blendshape set, we also pre-defined $k=26$ FACS expressions for building the dataset (excluding neutral expression). The FACS expressions refer to a set of standardized facial poses that can be performed by a person and generally correspond to a combination of blendshapes (blending weights will be either 0 or 1) with minimum motion overlap and maximum blendshape coverage. 
We assume that our captured FACS covers all the blendshapes and they can be isolated to unit blendshapes losslessly (more details in Sec.~\ref{sec:data}). So far, for each of the training subjects, we have a set of captured FACS expressions with corresponding combinations (0 or 1 blending weights). However, it would be irresponsible to say that the blending weights of FACS can be regarded as ground truth for real scans. One can easily perform unwanted motions when trying to express a predefined FACS expression (\textit{e.g.} FACS \textit{smile} consists only 
${\textit{Left\_Lip\_Corner\_Puller}}$ and ${\textit{Right\_Lip\_Corner\_Puller}}$, ended with unexpected eye motion captured). To address this issue, we propose a two-stage learning framework as shown in Fig.~\ref{fig:learning_pipeline}. The Estimation Stage, as the first one, fixes the initial blending weights to generate a set of blendshapes that optimally preserves identity and semantics, while its counterpart, the Tuning Stage, finetunes the initial blendshapes by jointly learning blending weights to better fit captured FACS expressions.

\subsection{Estimation Stage}
\begin{figure}[t]
\centering
 \setlength{\tabcolsep}{0.25mm}{
 \begin{tabular}{cccc}
 
  \includegraphics[width=0.2\linewidth]{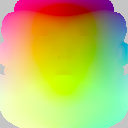} &
  \includegraphics[width=0.2\linewidth]{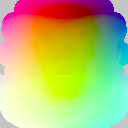} &
  \includegraphics[width=0.2\linewidth]{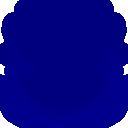} &
  \multirow{3}{*}[30pt]{\includegraphics[width=0.12\linewidth]{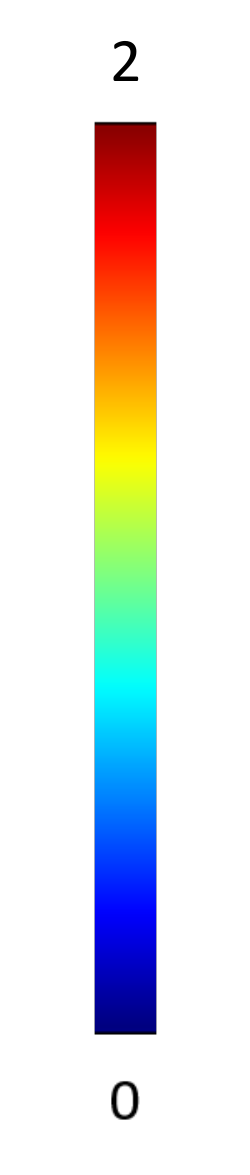}} \\
  \includegraphics[width=0.2\linewidth]{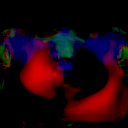} &
  \includegraphics[width=0.2\linewidth]{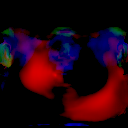} &
  \includegraphics[width=0.2\linewidth]{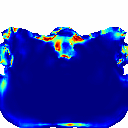} & \\
  \includegraphics[width=0.2\linewidth]{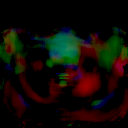} &
  \includegraphics[width=0.2\linewidth]{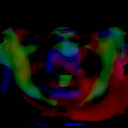} &
  \includegraphics[width=0.2\linewidth]{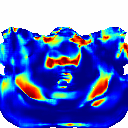} & \\
  (a) & (b) & Distance &\\
  \end{tabular}}
 \caption{Visualization of cosine distance maps between expressions, blendshapes and blendshape offsets. (a) and (b) show the same expression of different subjects represented by absolute positions in expression geometry $P^j_i$ (Row 1), blendshape offsets from neutral expression $S^j_i$ (Row 2) and offsets from the template blendshape $\Delta S^j_i$ (Row 3). Note that the distance map in Row 1 is almost filled with zeroes. This is because the average difference of the same expression between different individuals is much less than the scale of the human head.}
 \label{fig:offset}
\end{figure}

As shown in Fig.~\ref{fig:learning_pipeline}, the Estimation Stage takes a model with neutral expression $S^j_0$ and pre-defined blending weights for FACS expression $P^j_k$ as its input. It contains of a \textit{Blendshape Generator}, which learns to generate personalized blendshapes that are used to reconstruct the expression $P^j_k$ using Eq.~\ref{eq:recon}. We define a reconstruction loss in Eq.~\ref{eq:L_recon} between the reconstructed expression and the input expression.
\begin{equation}
    L_{rec} = \sum_{x\in P^j_k}{\left\lVert P^j_k{}'(x) - P^j_k(x) \right\rVert}_1.
    \label{eq:L_recon}
\end{equation}
Inspired by the idea in~\citet{Li_2010_SIGGRAPH} which emphasizes the importance of relative change between the template and the target models, we propose to learn blendshape offsets instead of blendshapes themselves because: (1) blendshape offsets are distributed in a nearly standard normal distribution which is easy for the network to learn; (2) blendshape offsets can better demonstrate the identity difference. For the example in Fig.~\ref{fig:offset}, the same expression of two different subjects are presented, where their difference is most obviously shown by the blendshape offsets. Thus, the output of the \textit{Blendshape Generator}, $\{\Delta S^j_{1},...,\Delta S^j_{n}\}$, are the offsets from the template blendshape to the target, which can be used to reconstruct the target personalized blendshapes by adding the template blendshapes as:
\begin{equation}
    S^j_{i} = \Delta S^j_{i} + S_{i}, \forall i \geq 1.
\label{eq:grad}
\end{equation}

To make the target blendshapes semantically consistent with the template blendshapes, we define a regularization term on blendshape offsets to minimize their relative difference. 
\begin{equation}
    L_{reg} = \sum_{i=1}^N \sum_{x \in S_i} g_i m_{i}(x) \left\lvert \left\lvert\Delta S^j_{i}(x) \right\rvert \right\rvert_1, \forall i \geq 1.
    \label{eq:L_reg}
\end{equation}
where $g_i$ are global weights for different kinds of blendshapes and $m_{i}(x)$ are local weights for each vertex $x$ in the blendshape $S_i$, defined as Eq.~\ref{eq:global_reg} and Eq.~\ref{eq:map_reg}.

The global weights are defined as:
\begin{equation}
    g_i = \frac{\lambda_g}{\sum_{x \in S_i}\left\lVert S_i(x)\right\rVert_{2}}, \forall i \geq 1.
    \label{eq:global_reg}
\end{equation}
where $\lambda_g$ is a scale factor restricting the maximum $g_i$ to 1. Considering the scale difference in different blendshapes, we introduce global weights to balance the influence of each blendshape for regularization loss. For example, the shape $\textit{Jaw\_Open}$ involves more moving vertices than $\textit{Left\_Eye\_Open}$. If the same weight is assigned to both, the regularization loss will be dominated by ${\textit{Jaw\_Open}}$, thus underestimating less pronounced shapes. Thus, we adopt a strategy that assigns a smaller regularization weight to blendshapes with larger offset scale. A similar strategy is used in \citet{ChenBLR18}, where adaptive weights for multi-objective loss are applied to balance the gradients in the training.

The local weights $m_{i}$ are defined by normalized norms of template blendshapes in which the vertex values are normalized to $(0, 1]$:
\begin{equation}
    m_{i}(x) = \frac{\lambda_l^i}{{\left\lVert S_i(x)\right\rVert}_{2}}, \forall x \in S_i.
\label{eq:map_reg}
\end{equation}

where $\lambda_l^i$ is a scale factor restricting the maximum $m_i$ to 1 (excluding fixed vertices), as for fixed vertices in blendshape $S_i$ (where $S_i(x) = 0$), we manually assign a relative large weight to constrain their movements (we used $4$ in our experiments). For each blendshape, the changes from the input neutral face are dominated by only a subset of vertices while the remaining vertices remain unchanged. The local weights are used to penalize large movements of the unchanged vertices and ensure the overall isolation of the generated blendshapes. 

Finally, we combine the reconstruction loss $L_{rec}$ and the regularization term $L_{reg}$ to yield the loss function for the \textit{Blendshape Generator}:
\begin{equation}
    L_{G} = L_{rec} + \omega_{reg} L_{reg},
\label{eq:L_G}
\end{equation}
where $\omega_{reg}$ is the regularization weight which is set to $1$ in the training. 

The \textit{Blendshape Generator} is a 2D convolutional neural network (CNN), similar to the image translator in \citet{liu2019few}, consisting of an identity encoder and a blendshape decoder. The encoder, same as the content encoder in \citet{liu2019few}, is made of a few 2D convolutional layers followed by several residual blocks. It takes a neutral expression $S^j_{0}$ as input and maps it into a content latent code that is a spatial feature map. The decoder consists of several instance normalization residual blocks followed by a couple of upscale convolutional layers. It decodes the feature vector into the blendshape offsets. To adapt 3D models to a compact representation which is friendly for the 2D CNN, we represent every 3D model as a 2D geometry image by first registering all the input 3D models with a same topology and aligning them in UV space (implementation details in Sec.~\ref{sec:data}), in which each pixel stores the $x-y-z$ coordinates of one vertex. 

\begin{figure}[t]
\centering
 \setlength{\tabcolsep}{0.5mm}{
 \begin{tabular}{ccc}
  \includegraphics[width=0.32\linewidth]{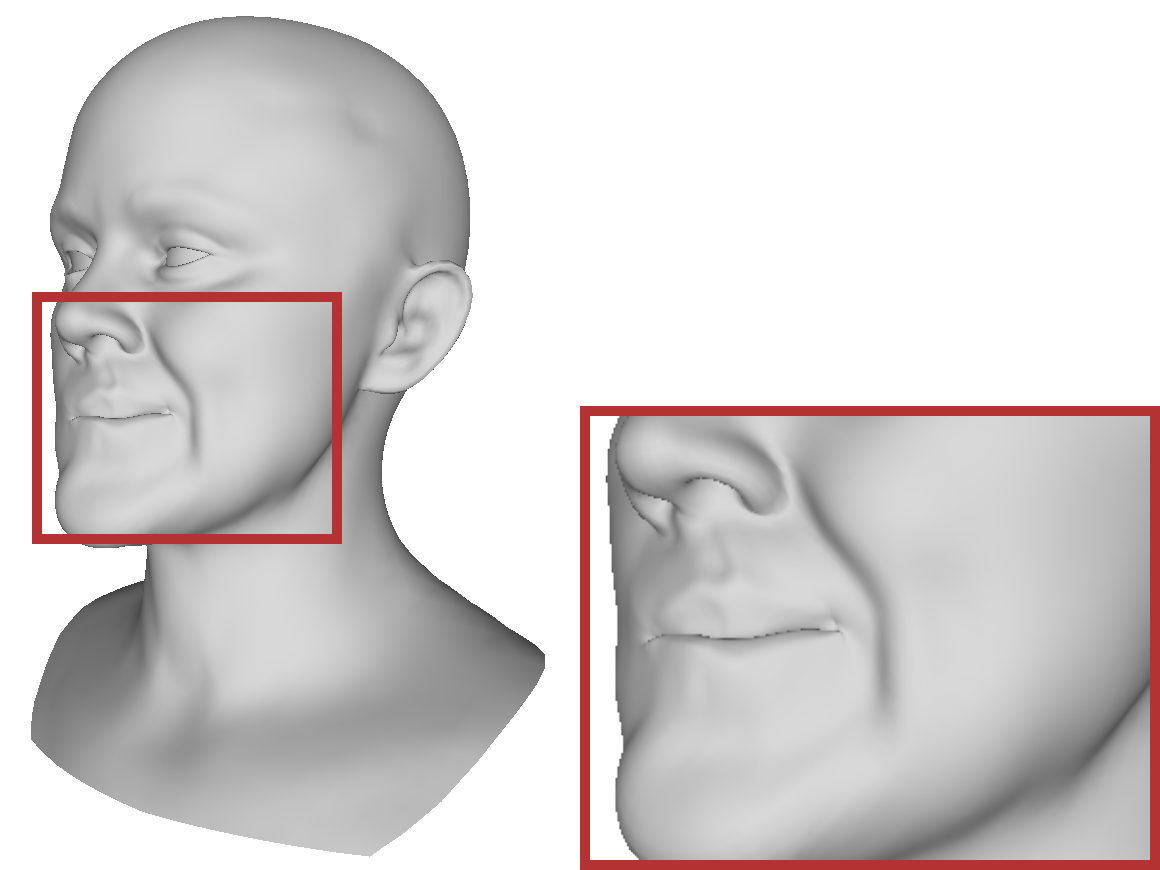} &
  \includegraphics[width=0.32\linewidth]{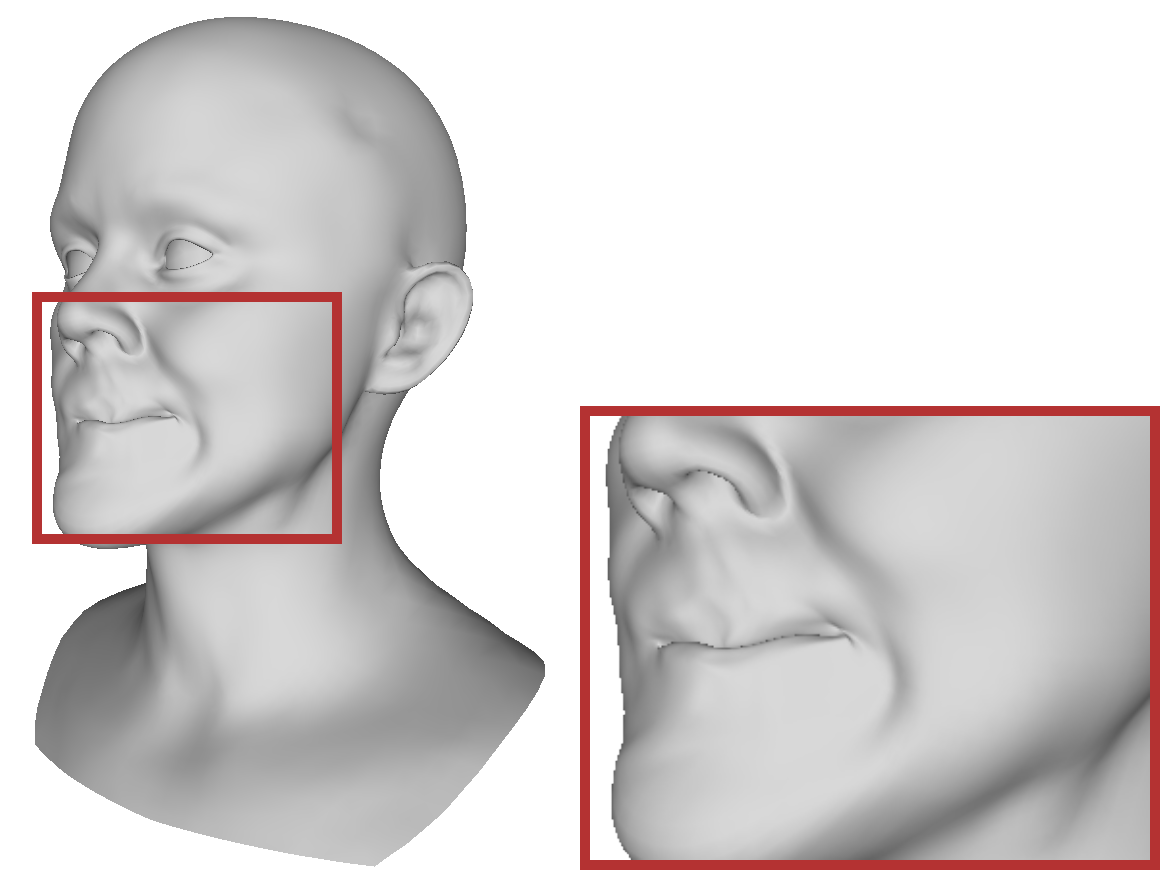} &
  \includegraphics[width=0.32\linewidth]{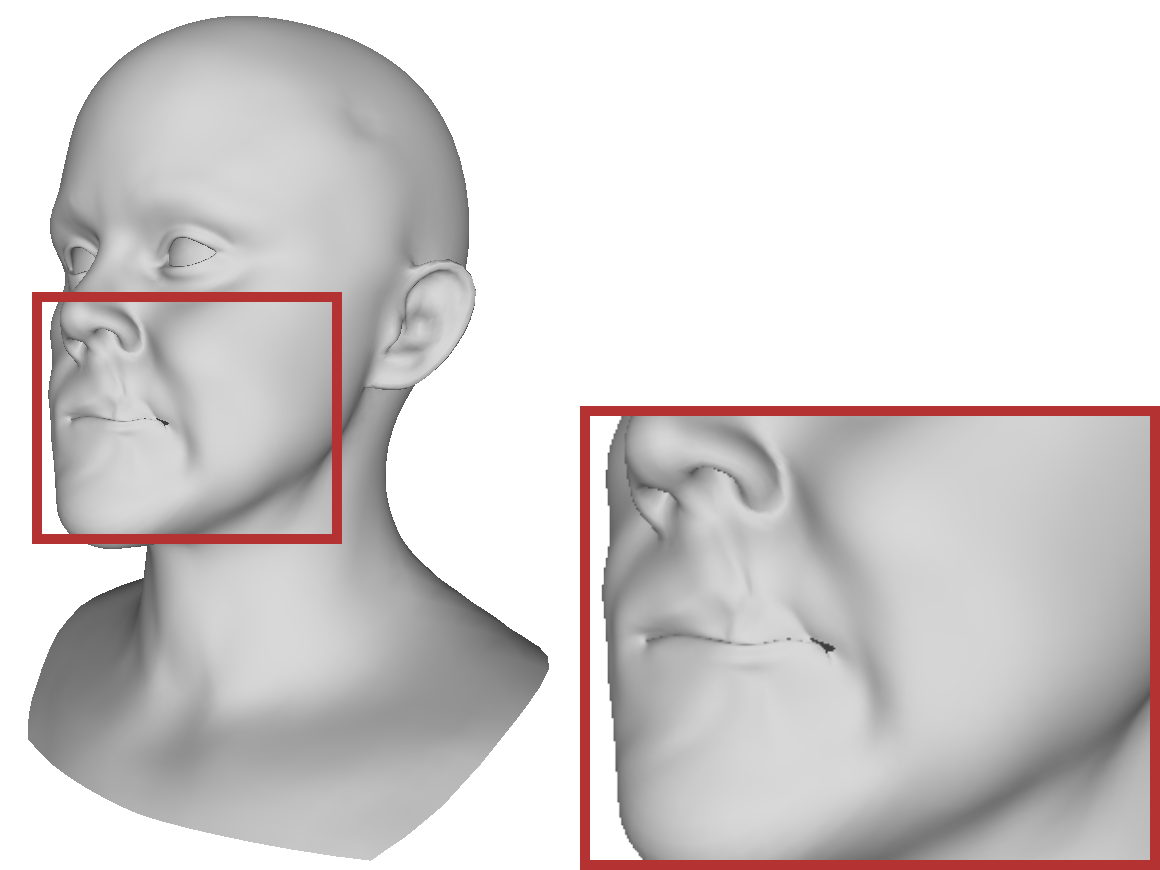} \\
  \includegraphics[width=0.32\linewidth]{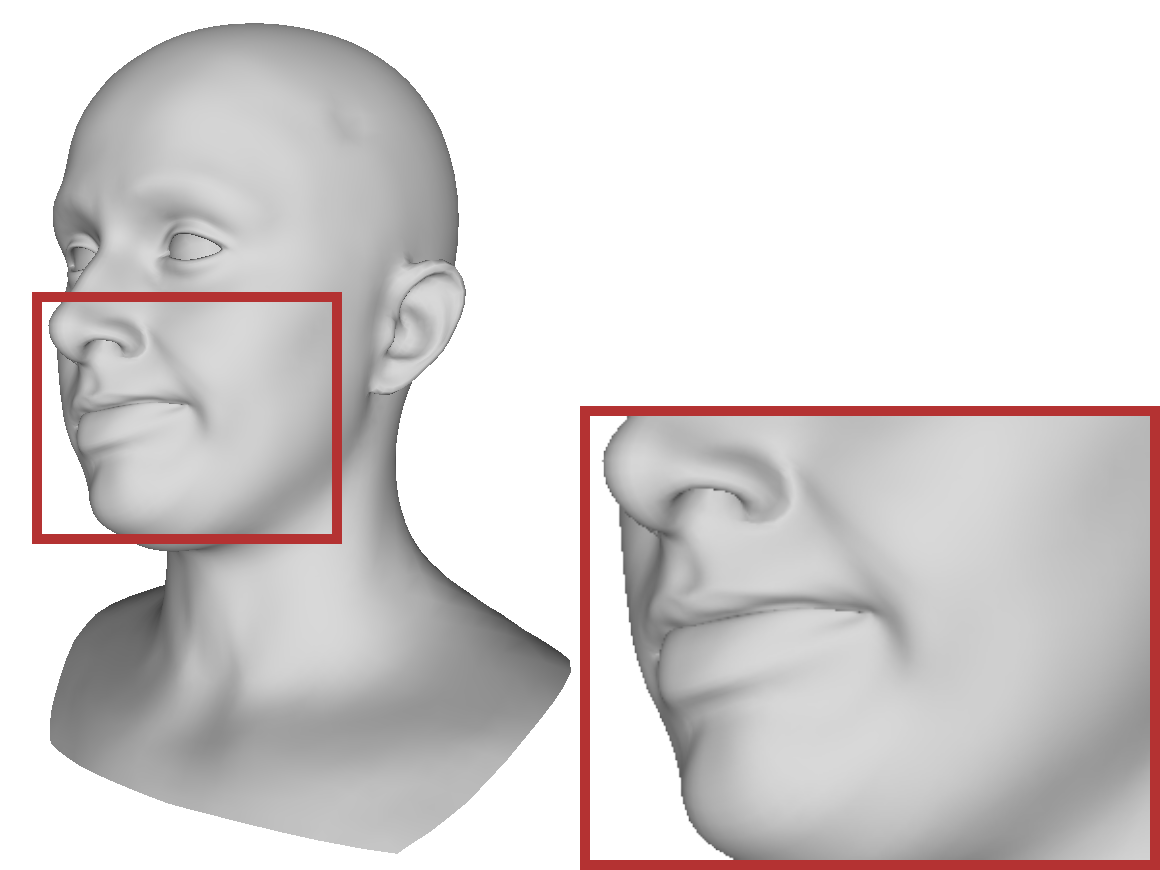} &
  \includegraphics[width=0.32\linewidth]{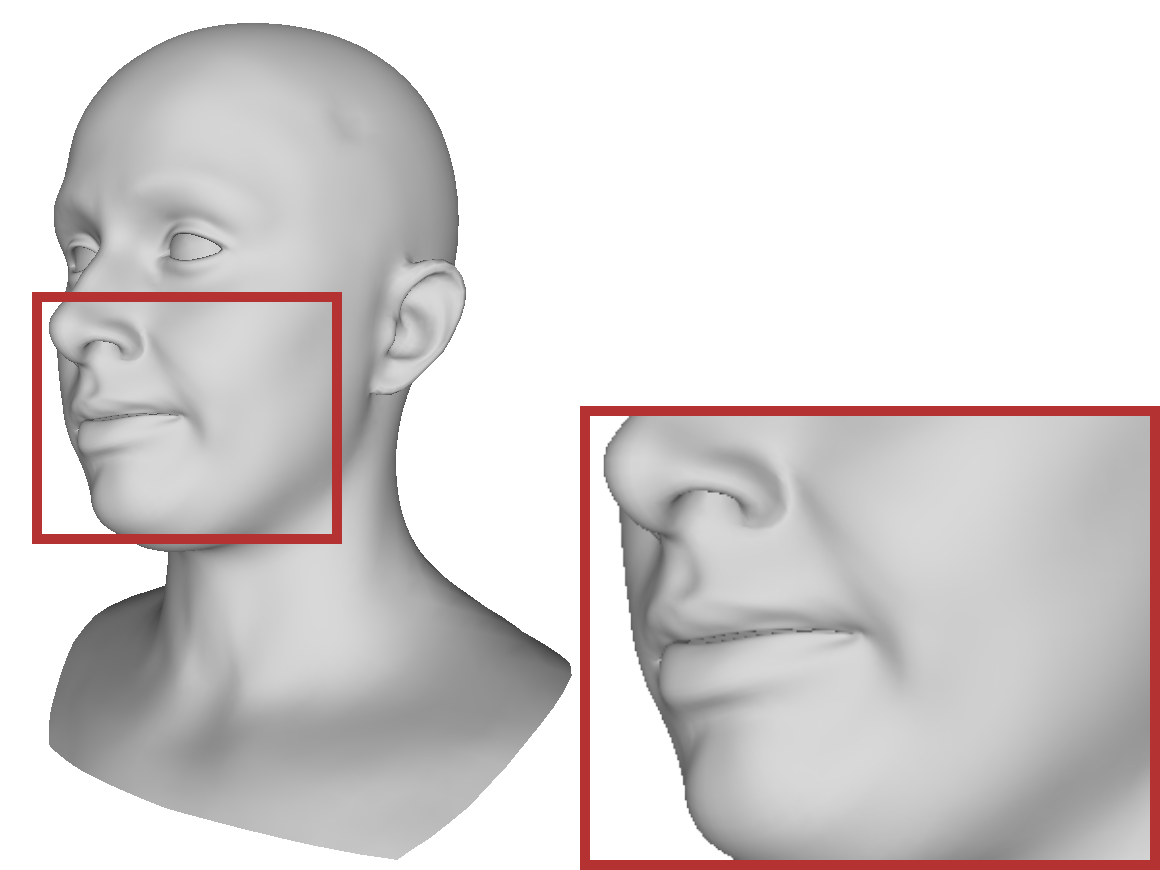} &
  \includegraphics[width=0.32\linewidth]{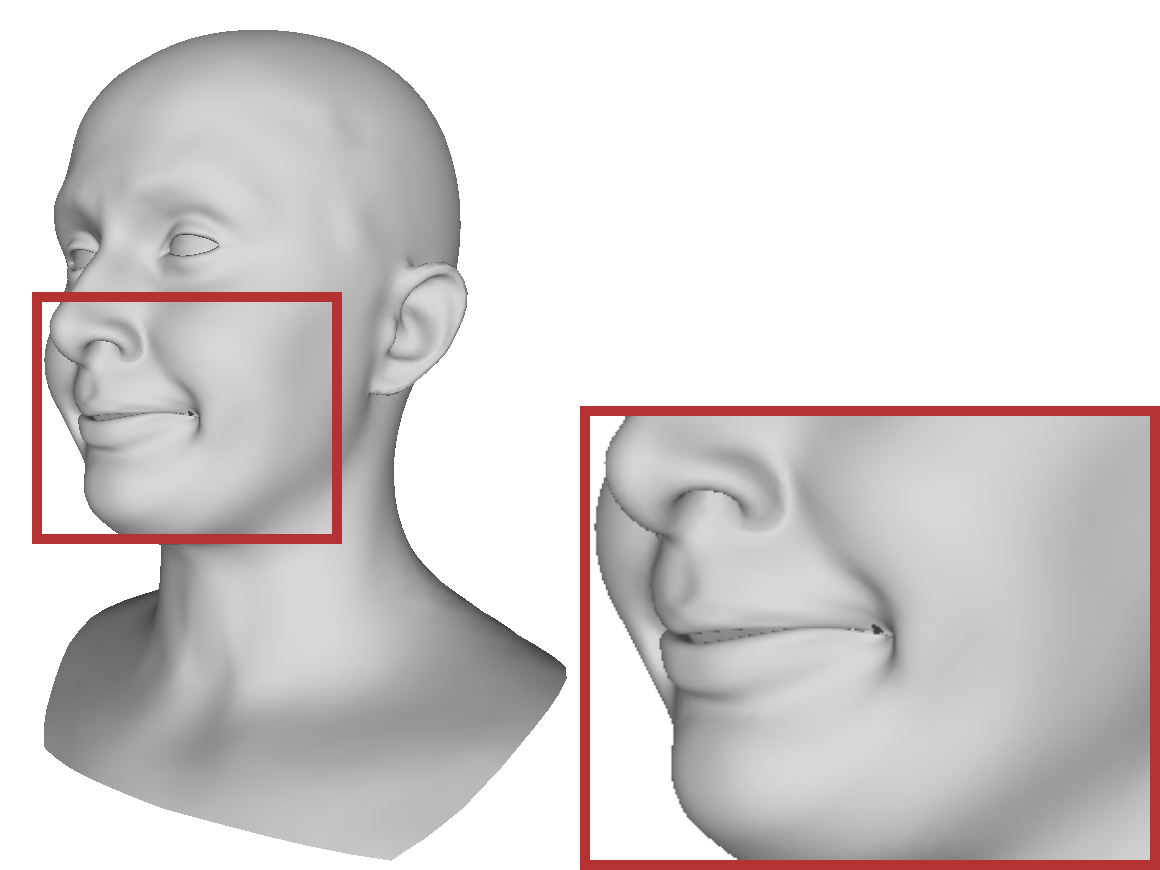} \\
  Single-branch network & Two-branch network & GT expression \\
  \end{tabular}}
 \caption{Comparison of two blendshape models generated by the \textit{Blendshape Generator} with a single-branch network and a two-branch network in the Estimation Stage. GT expression represents the reference FACS expression which is most semantically similar to the corresponding blendshape. Compared to the single-branch results, the two-branch results are more similar to the reference FACS expressions while keeping the semantic meaning of the generic blendshapes.}
 \label{fig:s1_cmp}
\end{figure}

Instead of training the generator in one network, we adopt a two-branch architecture inspired by \citet{bai2017multi} which uses a multi-branch network for face detection and tracking with different face size.

We observe that the scale of different blendshapes varies greatly. Thus we came up with a two-branch training strategy. We separate our blendshapes into two categories: 14 extreme blendshapes with relatively large motion and the rest with small motion. As shown in Fig.~\ref{fig:s1_cmp}, the two-branch network makes the generated blendshapes more personalized and closer to the reference FACS expression.

\subsection{Tuning Stage}
In the Estimation Stage, the blending weights are given, and consistent for all subjects, but practically it is hard to guarantee that different subjects can realize the same exact expressions. 
In this scenario, the fixed blending weights lead to inaccuracy when fitting such expressions for different subjects. Therefore, we relax constraints on the blending weights and instead learn them with a neural network. As shown in Fig.~\ref{fig:learning_pipeline}, compared to the Estimation Stage, the initial blendshapes work as additional input to the \textit{Blendshape Generator}, and another \textit{Blending Weight Predictor} is introduced to predict blending weights from the input expression in the Tuning Stage.

The \textit{Blending Weight Predictor} shares a similar network architecture as the \textit{Blendshape Generator} which consists of an expression encoder and a blending weight decoder. Given an input expression $P^j_k$, the encoder maps it to an expression latent code, followed by the decoder which decodes the latent code into a vector of $N$ blending weights whose values are constrained in $[0, 1]$. Combining the blending weights with the personalized blendshapes generated by the \textit{Blendshape Generator}, we reconstruct the input expression using Eq.~\ref{eq:recon}. The loss used to constrain the output of the \textit{Blending Weight Predictor} is the reconstruction loss defined in Eq.~\ref{eq:L_recon}. 

\begin{figure}[t]
\centering
 \setlength{\tabcolsep}{0.6mm}{
 \begin{tabular}{cccc}
  \includegraphics[width=0.34\linewidth]{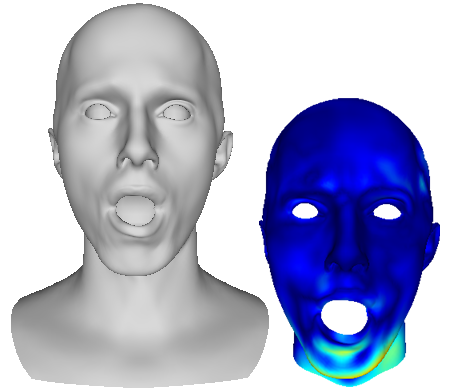} &
  \includegraphics[width=0.34\linewidth]{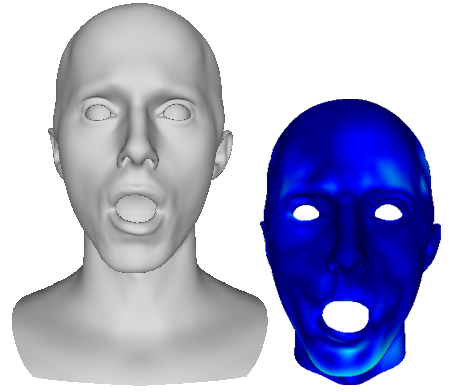} &
  \includegraphics[width=0.20\linewidth]{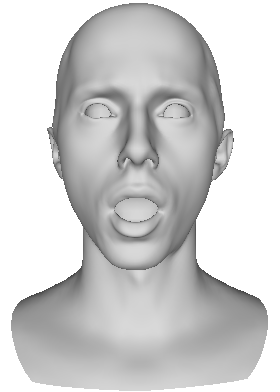} & 
  \multirow{2}{*}[40pt]{\includegraphics[height=0.38\linewidth]{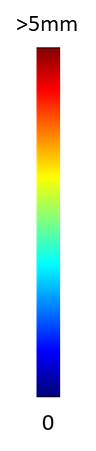}} \\
  
  \includegraphics[width=0.34\linewidth]{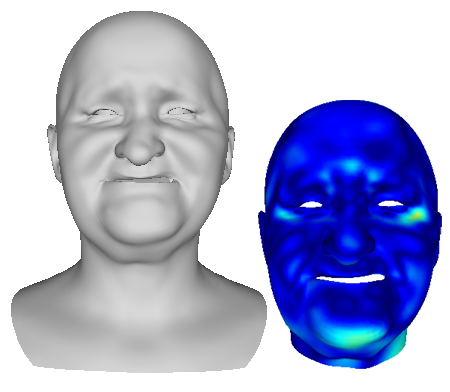} &
  \includegraphics[width=0.34\linewidth]{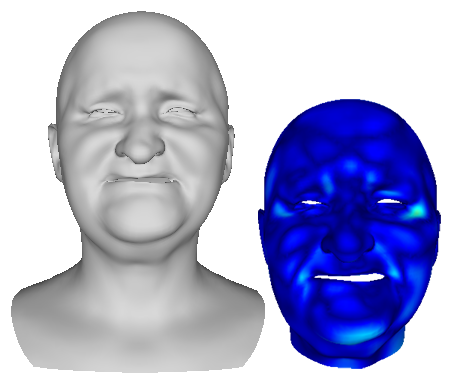} &
  \includegraphics[width=0.20\linewidth]{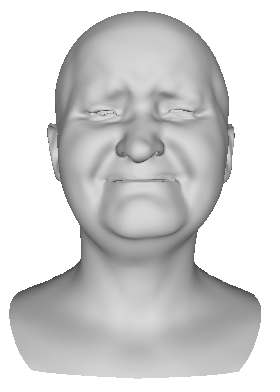} & \\
  
  Estimation Stage & Tuning Stage & GT expression &  \\
  \end{tabular}}
 \caption{Comparison of two reconstructed expressions by the Estimation Stage alone and with the addition of the Tuning Stage, along with error maps between the reconstructed expressions and the ground truth expressions. The output from the Tuning Stage results in better reconstruction with smaller fitting errors. }
 \label{fig:s1_s2_cmp}
\end{figure}

In order to preserve the semantics and personality of the initial blendshapes generated by the Estimation Stage, we define the regularization term as follows:
\begin{equation}
    L_{{reg}_{FT}} = \sum_{i=1}^N {\left\lVert \Delta S^j_{{i}_{FT}} - \Delta S^j_{i}\right\rVert}_1,
\label{eq:L_reg}
\end{equation}
where $\Delta S^j_{{i}_{FT}}$ are the target blendshape offsets and $\Delta S^j_{{i}}$ are initial blendshape offsets generated in the Estimation Stage. Thus, the loss function used in the Tuning Stage is:
\begin{equation}
    L_{G_{FT}} = L_{rec} + \omega_{{reg}_{FT}}L_{{reg}_{FT}},
\label{eq:L_tuning}
\end{equation}
where $\omega_{{reg}_{FT}}=0.1$. In our implementation, we add skip connections from the initial blendshape to the generator output (as shown in the red line in Fig.~\ref{fig:learning_pipeline}) such that the generator predicts $\Delta S^j_{{i}_{FT}} - \Delta S^j_{i}$. 
Examples of with and without tuning are shown in Fig.~\ref{fig:s1_s2_cmp}, we observe that the Tuning Stage achieves better fitting results by fine-tuning the blendshapes, and jointly optimizing blending weights while preserving the semantics and personality.

\begin{figure}[t]
 \includegraphics[width=3.2in]{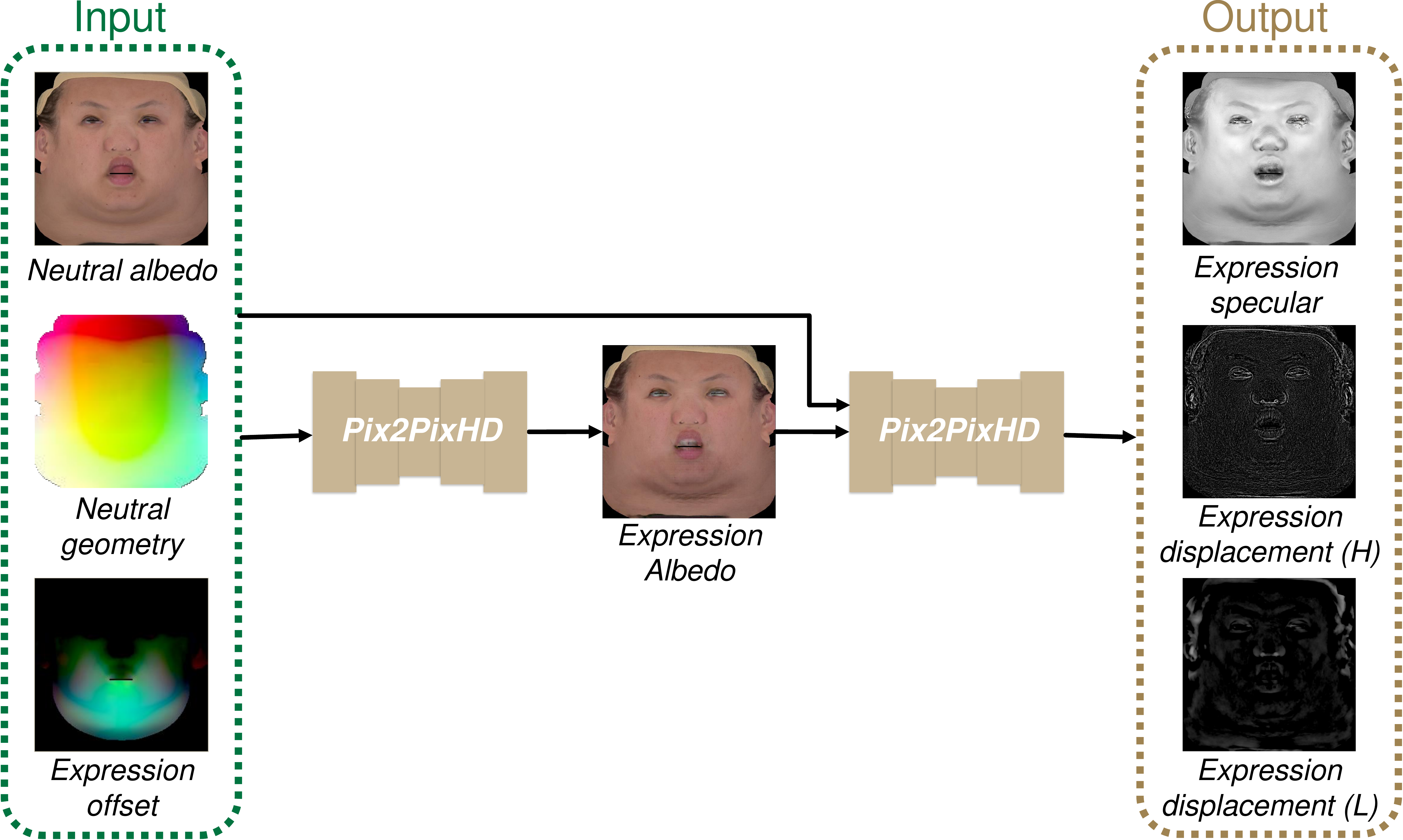}
 \caption{Texture Generative Network. Given the albedo map and the geometry image of the input model in neutral expression and the geometry image of the target expression offset, the first network generates the albedo map of the expression using pix2pixHD~\cite{wang_2018_CVPR}. Then, combining the initial input and prediected albedo map, the second network infers specular intensity, low-frequency, and high-frequency displacement maps.}
 \label{fig:texture_generator}
\end{figure}
\begin{figure}[t]
 \includegraphics[width=2.6in]{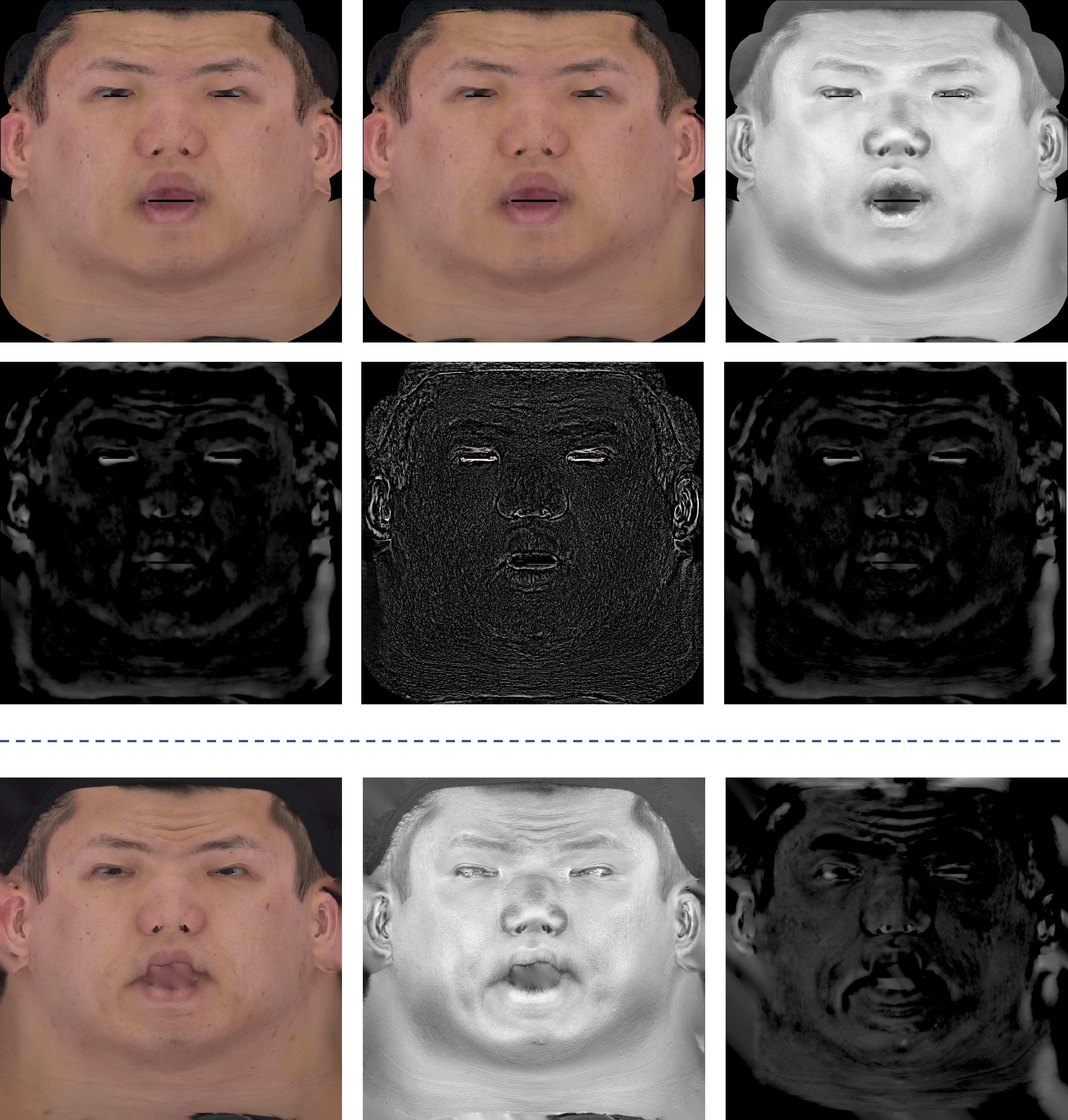}
 \caption{Generated textures and ground truth textures of an expression. Row 1 from left to right: low-resolution albedo map (1K $\times$ 1K), high-resolution albedo map (4K $\times$ 4K), and specular intensity map.
 Row 2 from left to right: low-frequency, high-frequency and combined displacement maps.
 Row 3 from left to right: ground truth of albedo, specular intensity and displacement maps.}
 \label{fig:generated_texture}
\end{figure}
 
\section{Dynamic Texture Generation}
In this section, we first introduce our compact representation of dynamic texture assets- Compress and Stretch maps, followed by a learning-based method to infer/extract them. Finally, we demonstrate the utilization of our Compress and Stretch maps for rendering at run-time.
\subsection{The Representation}

\begin{figure}[t]
 \includegraphics[width=3in]{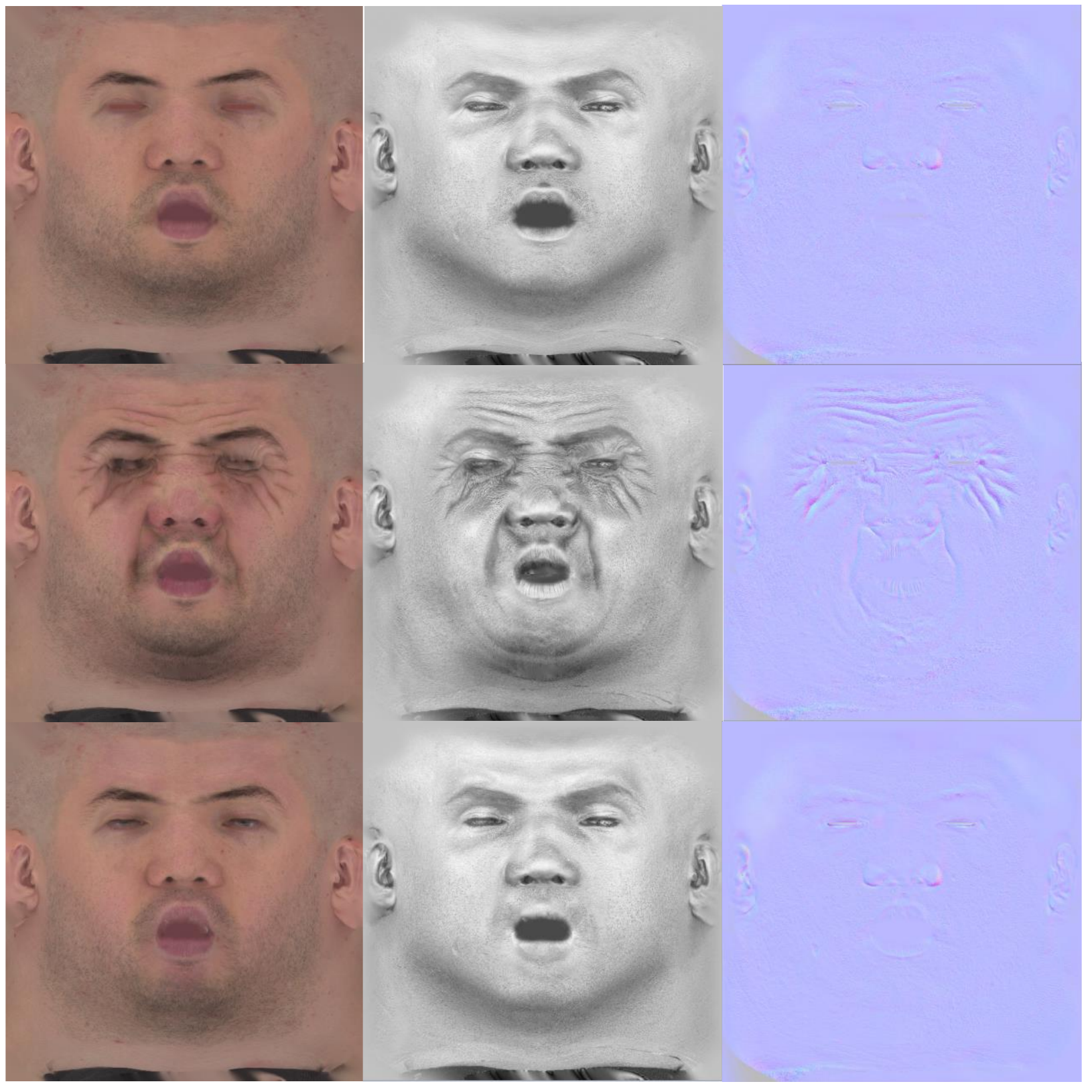}
 \caption{Illustration of Compress and Stretch Maps. From Top to Down Rows: Neutral Static maps, Compress maps, Stretch maps. From Left to Right Columns: Diffuse Albedo maps, Specular maps, Normals maps (in tangent space) computed from Displacement maps.}
 \label{fig:compressStretch2}
\end{figure}

\begin{figure*}
 \centering
 \includegraphics[width=0.95\linewidth]{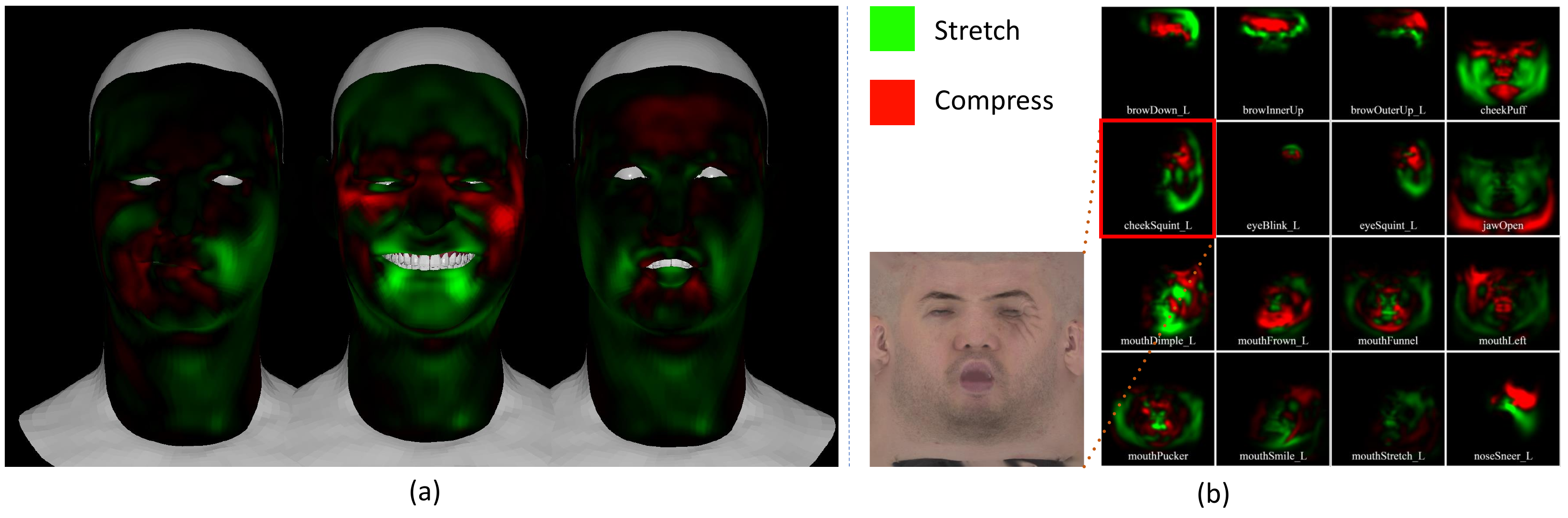}
 \vspace{-3mm}
 \caption{Illustration of Influence maps. (a). Influence value rendered in geometries with different expressions (\textit{Mouth Right}, \textit{Smile} and \textit{Lip Funnel}). (b). Selected Influence maps from a set of blendshapes and an example of dynamic albedo with its corresponding influence map in the blendshape \textit{CheckSquint\_L}. Note that we store compress and stretch influence maps as $R$ and $G$ channels and set $B$ channel to zeros. }
 \label{fig:InfluenceMap2}
\end{figure*}

\begin{figure}[t]
 \includegraphics[width=3.3in]{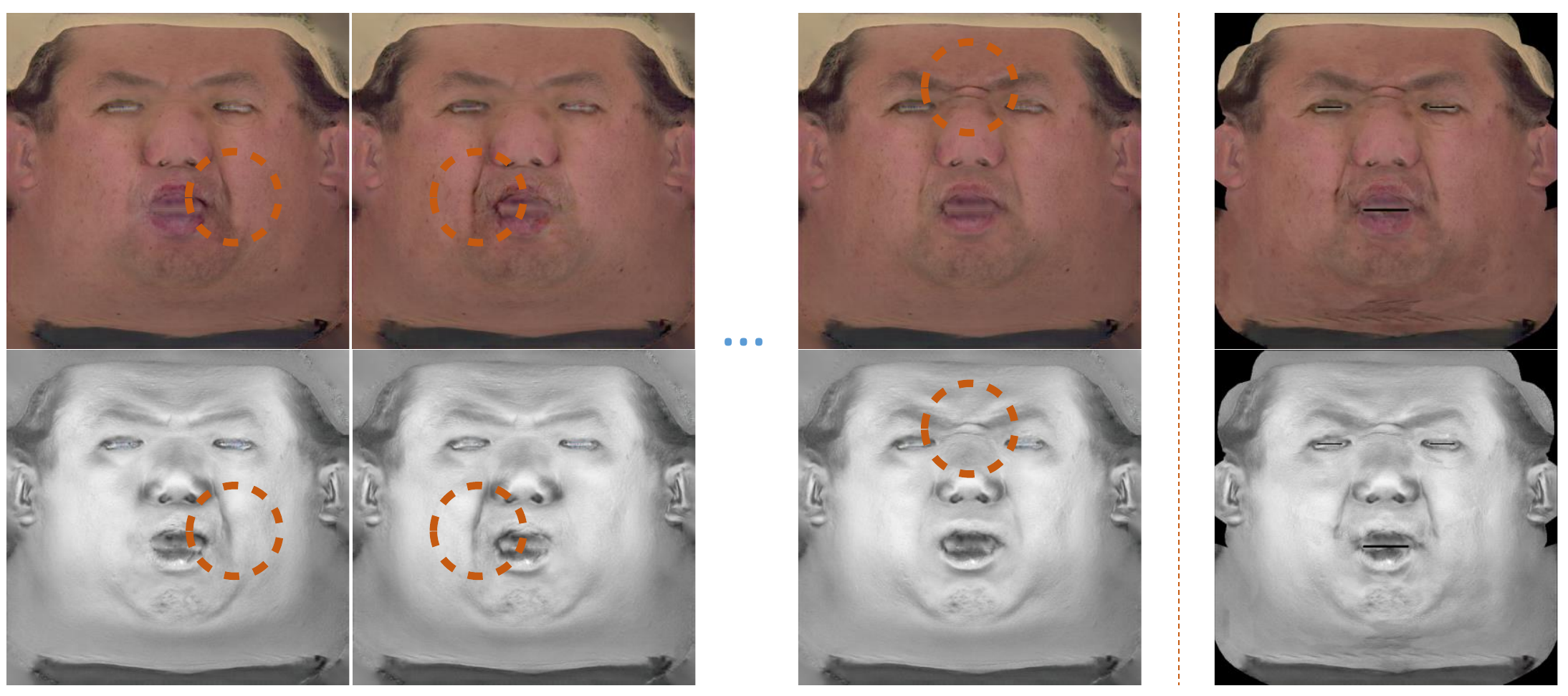}
 \caption{Illustration of Compress Maps Extraction. Left: expression textures generated from networks. Right: compress maps extracted by blending expression textures based on the influence maps. Note that the final compress maps gather all the dynamic details caused by skin local compression (in the orange circles) from all the expressions.}
 \label{fig:compressGeneration_v2}
\end{figure}

\paragraph{Compress and Stretch Maps.}
When static textures (obtained from a neutral expression) are used to render extensive expressions, the missing details (\textit{e.g.} wrinkles) caused by facial motion will significantly reduce the photo-realism of rendering results. Especially for the extreme/exaggerated expressions, high-fidelity muscle movement and micro-expressions make big differences. A natural way to solve this problem is to customize a set of dynamic textures for blendshapes. However, the number of blendshapes used in high-end industries may be of the magnitude of hundreds or thousands. The creation of such large dynamic textures is costly and requires substantial computational power. More importantly, it is 
difficult to load such a vast collection of dynamic textures into a rendering engine at once, in particular, with multiple layers (\textit{e.g.} albedo, specular intensity, displacement maps) at high resolution. A memory-efficient, compact, and easy-to-compute dynamic representation is needed.  Moreover, it should also be expressive enough to cover all the possible dynamic details of facial motion losslessly. We adopt \textit{Compress and Stretch 
Maps} as shown in Fig.~\ref{fig:compressStretch2} along with a static neutral texture to be the dynamic texture library, which is a commonly adopted format in the industry~\cite{oat2007animated}. Guided by \textit{Influence Maps}, compress and stretch maps gather the most prominent features caused by the local compression/stretching movement of all the available expressions.

\paragraph{Influence Maps.} 
Influence maps are computed based on the geometry changes between the expressions and the neutral face. For each of the vertices $x$ on the neutral mesh $N$, we define the average edge length of its one-ring neighbors as $E_{N}(x)$, and then for an arbitrary expression mesh $P$ of the same subject, the influence value of each vertex on $P$ in compress maps can be computed as: 
\begin{equation}
I_{{P}_{Compress}}(x) = \begin{cases}
\parallel  E_{N}(x) - E_{P}(x) \parallel, & E_{P}(x) < E_{N}(x)\\
	0, & E_{P}(x) \ge E_{N}(x)
		   \end{cases}.
    \label{eq:compress}
\end{equation}

Similarly, the influence value of each vertex on $P$ in stretch maps is as follows:
\begin{equation}
I_{{P}_{Stretch}}(x) = \begin{cases}
\parallel E_{P}(x) - E_{N}(x) \parallel, & E_{P}(x) > E_{N}(x)\\
	0, & E_{P}(x) \le E_{N}(x)	
		   \end{cases}.
    \label{eq:compress}
\end{equation}

Based on the per-vertex influence values, we interpolate a per-pixel compress and stretch influence map as shown in Fig.~\ref{fig:InfluenceMap2}. Note that we store compress and stretch influence maps as $R$ and $G$ channels separately. The influence maps provide the weights to blend and extract dynamic textures. 

\subsection{Compress and Stretch Map Generation}
In the standard industry pipeline, the compress and stretch maps are handcrafted by skilled artists using numerous captured expressions as reference. To automate this procedure, especially when only a single scan is provided in our scenarios, we came up with a two-step solution. Firstly, we predict the texture maps (\textit{i.e.} albedo, specular intensity, and displacement) of the input subject's pre-defined expressions using a deep neural network. Then a blending step is introduced to fuse them into compress and stretch maps. 
\paragraph{Expression Texture Generation Networks.}
Given a single neutral scan with an albedo map, in order to predict the high-fidelity albedo, specular intensity, and displacement maps of different expressions, we propose a cascade architecture, as shown in Fig.~\ref{fig:texture_generator}. We first take the neutral geometry with its albedo map and the target expression offset from the neural geometry as input to predict the albedo map offset of the target expression. The predicted offset is then added to the neutral albedo map to generate the expression albedo map as the intermediate results, further combining the input of the first network to be fed into the second network. The second network then infers the specular intensity and displacement maps. Both of the networks are the Pix2pixHD~\cite{wang_2018_CVPR} model, which contains an encoder with several CNN layers, followed by a couple of Resnet blocks, and a decoder with similar architecture. The reason of using a cascade network with an expression albedo map as intermediate results include: (1) the specular intensity and displacement maps generated using the albedo map as a prior have fewer artifacts and higher quality; (2) this architecture allows us to handle incomplete training data (some of the subjects do not have the specular intensity and displacement maps). 
In particular, we separate the displacement map into low-frequency and high-frequency during training, following~\citet{yamaguchi2018high,Huynh_2018_CVPR} to make the problem more tractable and merge them together before using. Both input and output of the two networks have $1K \times 1K$ resolution. Furthermore, with all these $1K$ result maps, we up-scale them into $4K \times 4K$ using a pre-trained super-resolution network~\cite{ledig2017photo}. In Fig.~\ref{fig:generated_texture}, we show a complete set of expression textures generated by our networks.
\paragraph{Compress and Stretch Map Extraction.}
We design an algorithm to extract compress and stretch maps based on the influence maps from the above predicted expression textures as shown in Fig.~\ref{fig:compressGeneration_v2}. Let $I_i$ be the influence map of the $i$\textit{th} expression, and the influence value of each pixel $(x,y)$ is $I_i(x,y)$. We first normalize the influence map of all the expressions with a weighted sum strategy to ensure the spatial consistency among all the expressions as follows (take the compress map as an example):
\begin{equation}
    \hat{I}_{i_{Compress}}(x,y)=\frac{e^{I_{i_{Compress}}(x,y)}}{\sum\limits_{i}e^{I_{i_{Compress}}(x,y)}},
    \label{eq:compress_influence}
\end{equation}
in which $\hat{I}_{i_{Compress}}$ is the normalized influence map of $I_{i_{Compress}} (i=1 \dots N)$ where $N$ is the number of  expressions. 

Once we get the normalized influence maps, the compress map is computed as follows:
\begin{equation}
    T_{Compress}(x,y)=\sum\limits_{i}\hat{I}_{i_{Compress}}(x,y)T_i(x,y),
    \label{eq:compress}
\end{equation}

Where $T_i$ is the texture of the $i$\textit{th} expression, and it can be one of the albedo, specular, and displacement. The stretch maps are computed similarly. Finally, we obtain compress and stretch maps for albedo, specular, and displacement maps, respectively.

\subsection{Runtime Dynamic Texture Generation}
When using dynamic assets for rendering in runtime applications, such as tracking, animation, we first solve the blending weights of each input expression using personalized blendshapes. Those blending weights combined with a set of pre-defined influence maps of blendshapes, will be used to sample the current dynamic textures from compress and stretch maps. The dynamic textures are generated as follows:
\begin{align}
    T(x,y) = &T_{N}(x,y) \nonumber \\
    + & \sum_i^N \Big (\alpha_i \hat{I}_{i_{Compress}} (x,y) (T_{Compress}(x,y)-T_{N}(x,y))  \nonumber \\
     + &\alpha_i \hat{I}_{i_{Stretch}} (x,y) (T_{Stretch}(x,y)-T_{N}(x,y)) \Big )
    \label{eq:blend}
\end{align}
where $T_{N}$ is the static texture of neutral expression, $T_{Compress}$ and $T_{Stretch}$ correspond to the compress and stretch textures, $\hat{I}_{i_{Compress}}$ and $\hat{I}_{i_{Stretch}}$ are the influence maps of the $i$\textit{th} blendshape and $\alpha_i$ indicates its blending weight.

\section{Assembly}
In addition to the primary dynamic assets (face geometry and textures) generated using networks, we also include secondary components (\textit{e.g.} eyeballs, lacrimal fluid, eyelashes, teeth, and gums) in our avatar as shown in Fig.~\ref{fig:face_parts}.  We handcrafted a set of generic blendshapes with all the primary and secondary parts. We further use this set of generic blendshapes to linearly fit each expression generated by our networks based on corresponding vertices on the facial regions. The computed coefficients based on the primary parts drive the secondary components, such that eyelashes will travel with eyelids. The linearly fitted secondary elements will be combined with the primary facial parts to get an integrated face model. Except for eyeball, other secondary parts share a set of generic textures for all the subjects. For eyeball textures, we adopt an eyeball assets database~\cite{3Dhumaneye} with 90 difference eye textures (pupil patterns) to match with input subjects. 

%% file: Database.tex
\section{Dataset}
\label{sec:data}
\begin{figure}[t]
\begin{center}
\includegraphics[width=0.48\textwidth]{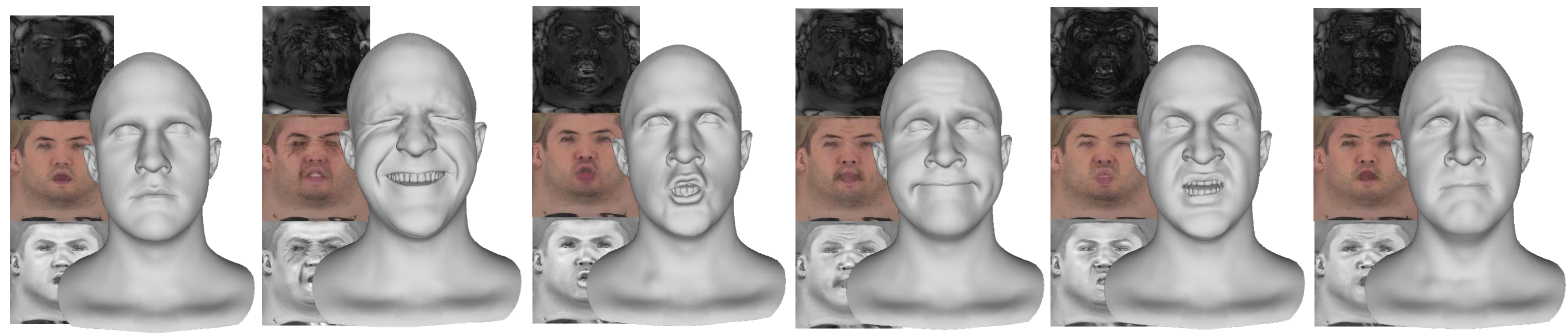}
\end{center}
  \vspace*{-8pt}
  \caption{Selected FACS units from Light Stage Dataset. From left to right: \textit{Neutral, Eye\_close\_Lip\_corner\_Puller, Eyes\_Up\_Lip\_Funneler, Inner\_Brow\_Raiser\_Dimpler, Upper\_Lip\_Raiser\_Lower\_Lip\_Depressor\_Outer\_
  Brow\_Raiser,
  Brow\_Lowerer\_Inner\_Brow\_Raiser\_Lip\_Presser.}}
    \vspace{-10pt}
\label{fig:expressionDescription}
\end{figure}

The facial scan dataset used in training comes from a combined source of aligned face models with 4k resolution textures and geometries aligned to a known topology \cite{li2020learning}. The dataset consists of 178 scan subjects divided into two sets, one of 78 (Light Stage), and one of 100 subjects (\cite{triplegangers}); performing 26 and 20 static FACS expressions respectively. The FACS expressions are fixed, which enables labeling of corresponding weights in our set of template blendshapes. This feature is particularly useful when isolating orthogonal shapes that are combined under the scanning session. One such example may be the combination of \textit{action unit 1 (Inner brow raiser)}, and \textit{action unit 14 (Dimpler)} \cite{Ekman1978FacialAC}. This makes it possible to significantly reduce the number of scans needed (Fig.~\ref{fig:expressionDescription}).

The assumptions that have to be realized under the learning of corresponded face morphologies described in section \ref{sec:method} are (1) a rigid transformation of each subject's skull shape can be found for every expression the subject performs, (2) sparse correspondence among subjects need to be established for a common parameterization to be usable, and (3) dense correspondence among expressions need to be established for each subject to track minute changes in skin deformation using texture maps. Next, we describe how these problems are solved to generate the desired dataset.
\begin{figure}[t!]
    \centering
    \begin{subfigure}[t]{0.15\textwidth}
        \centering
        \includegraphics[width=1\textwidth]{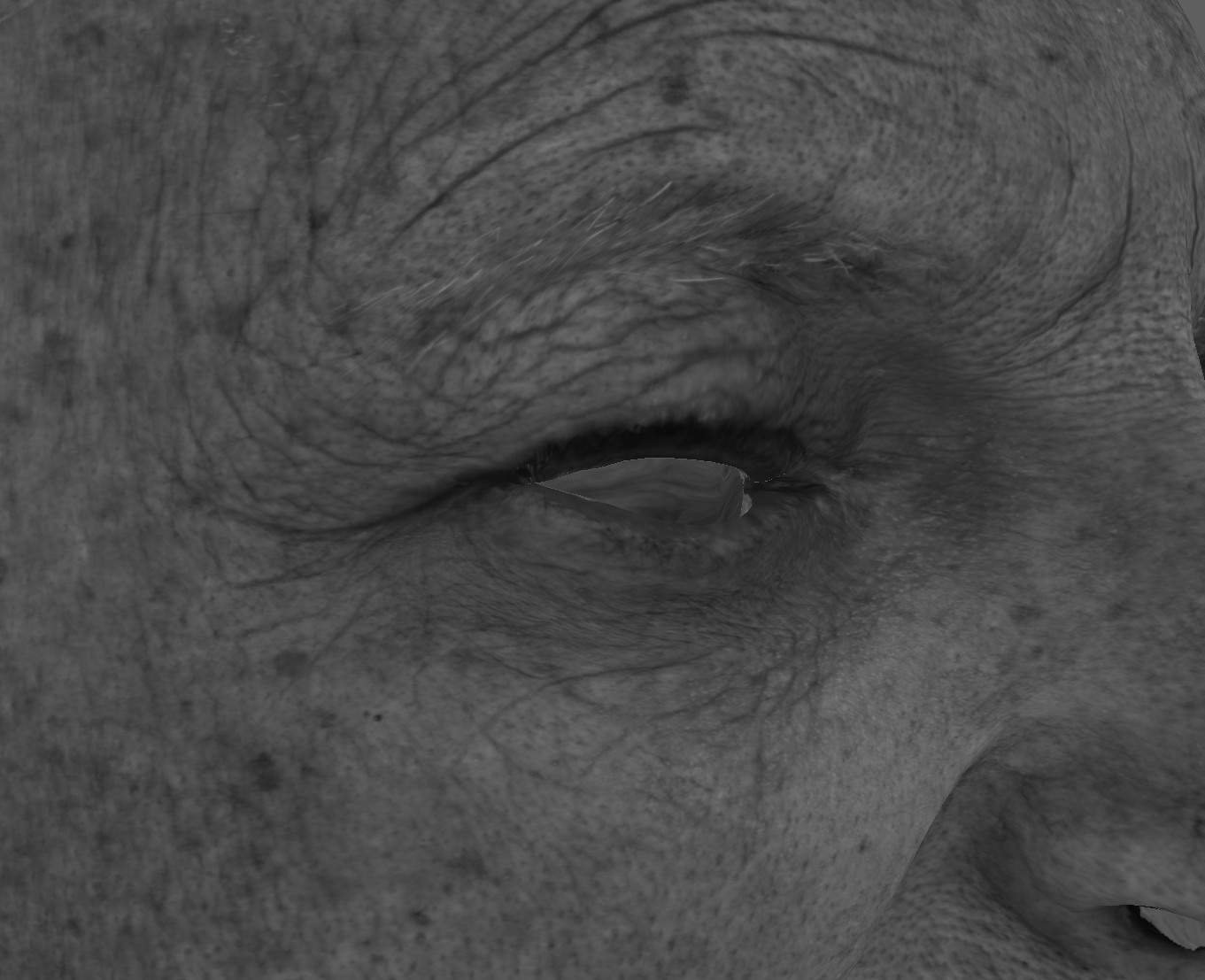}
        \caption{}
    \end{subfigure}
    \begin{subfigure}[t]{0.15\textwidth}
        \centering
        \includegraphics[width=1\textwidth]{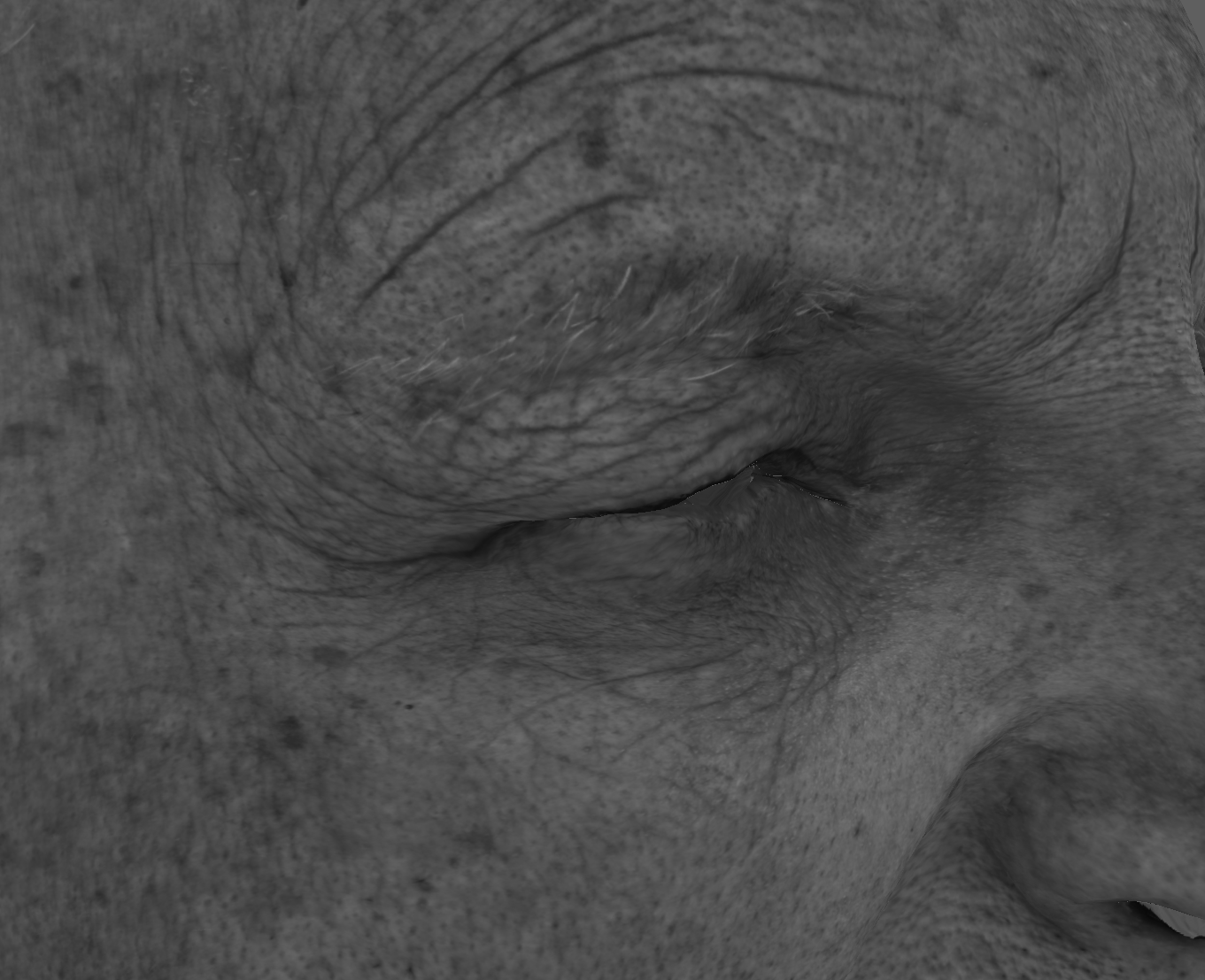}
       \caption{}
    \end{subfigure}
    \begin{subfigure}[t]{0.15\textwidth}
        \centering
        \includegraphics[width=1\textwidth]{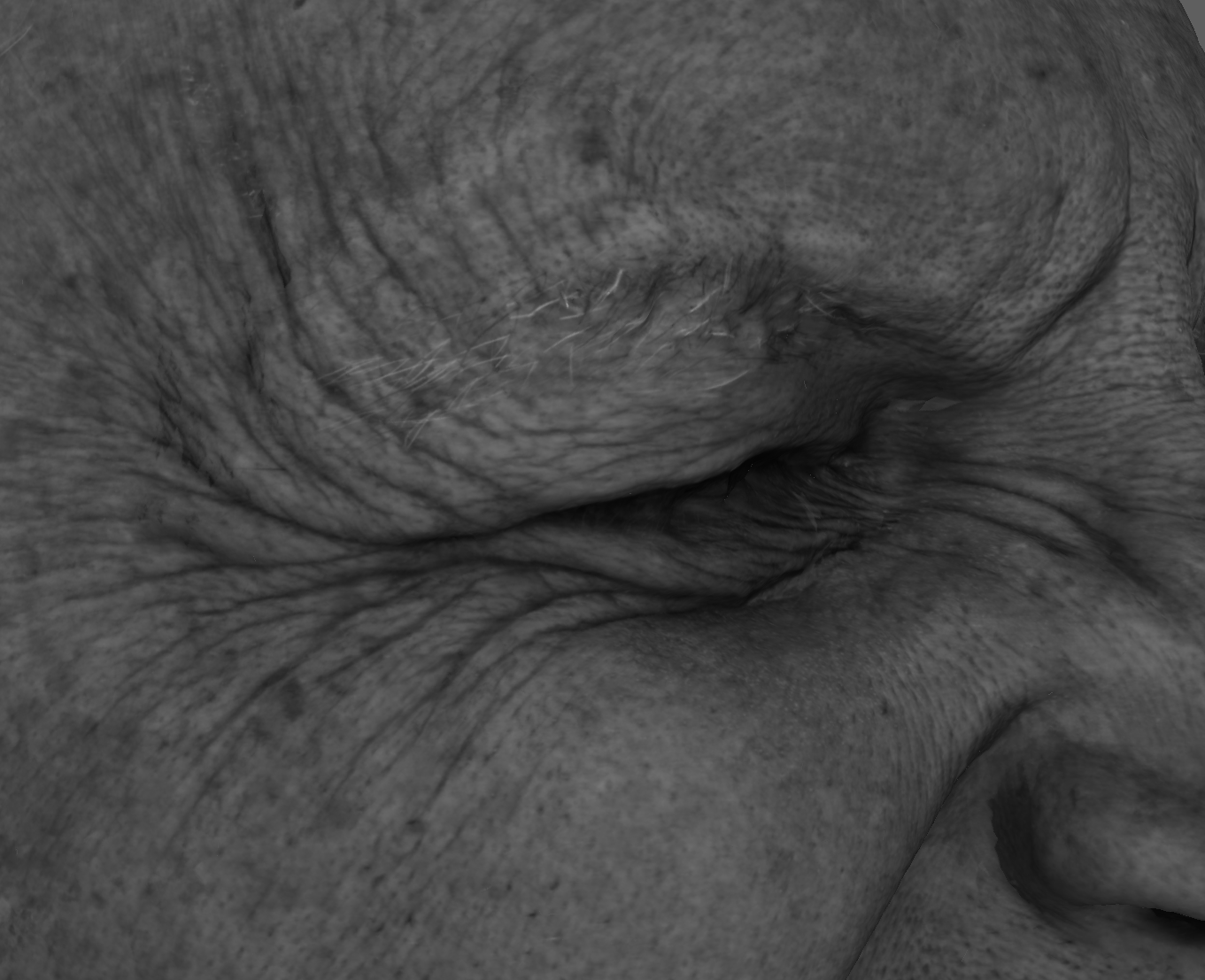}
       \caption{}
    \end{subfigure}
    \vspace{-4mm}
    \caption{Laplacian deformation results of neutral mesh to target expression model using (a) landmarks only, and (b) dense optical flow correspondence. (c) Target expression.}
    \label{fig:textureDrifting}
\end{figure}

\begin{figure}[t]
\begin{center}
\includegraphics[width=0.30\textwidth]{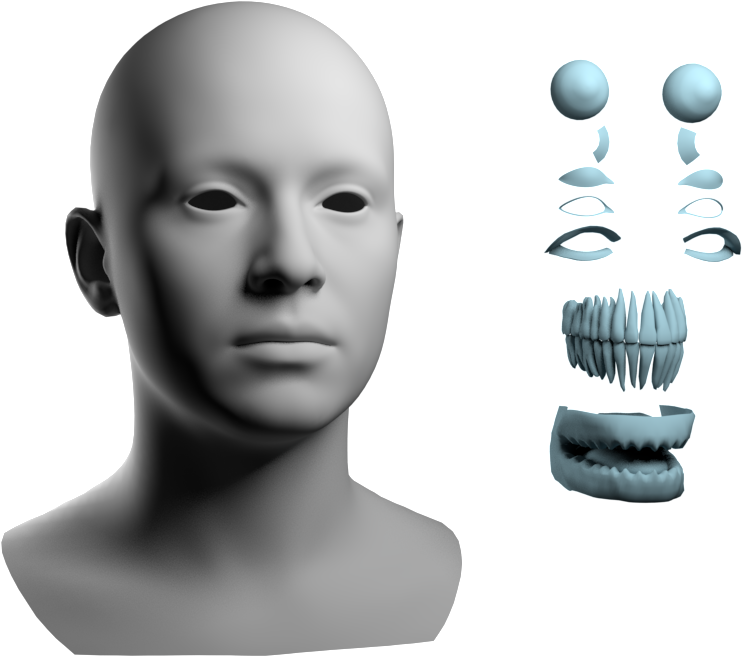}
\end{center}
  \vspace*{-8pt}
  \caption{Our face model consists of multiple parts including face, eyes, eye blend mesh, lacrimal fluid, eye occlusion, eyelashes, teeth, gums and tongue.
  }
    \vspace{-10pt}
\label{fig:face_parts}
\end{figure}

\begin{figure*}
 \centering
 \includegraphics[width=0.9\linewidth]{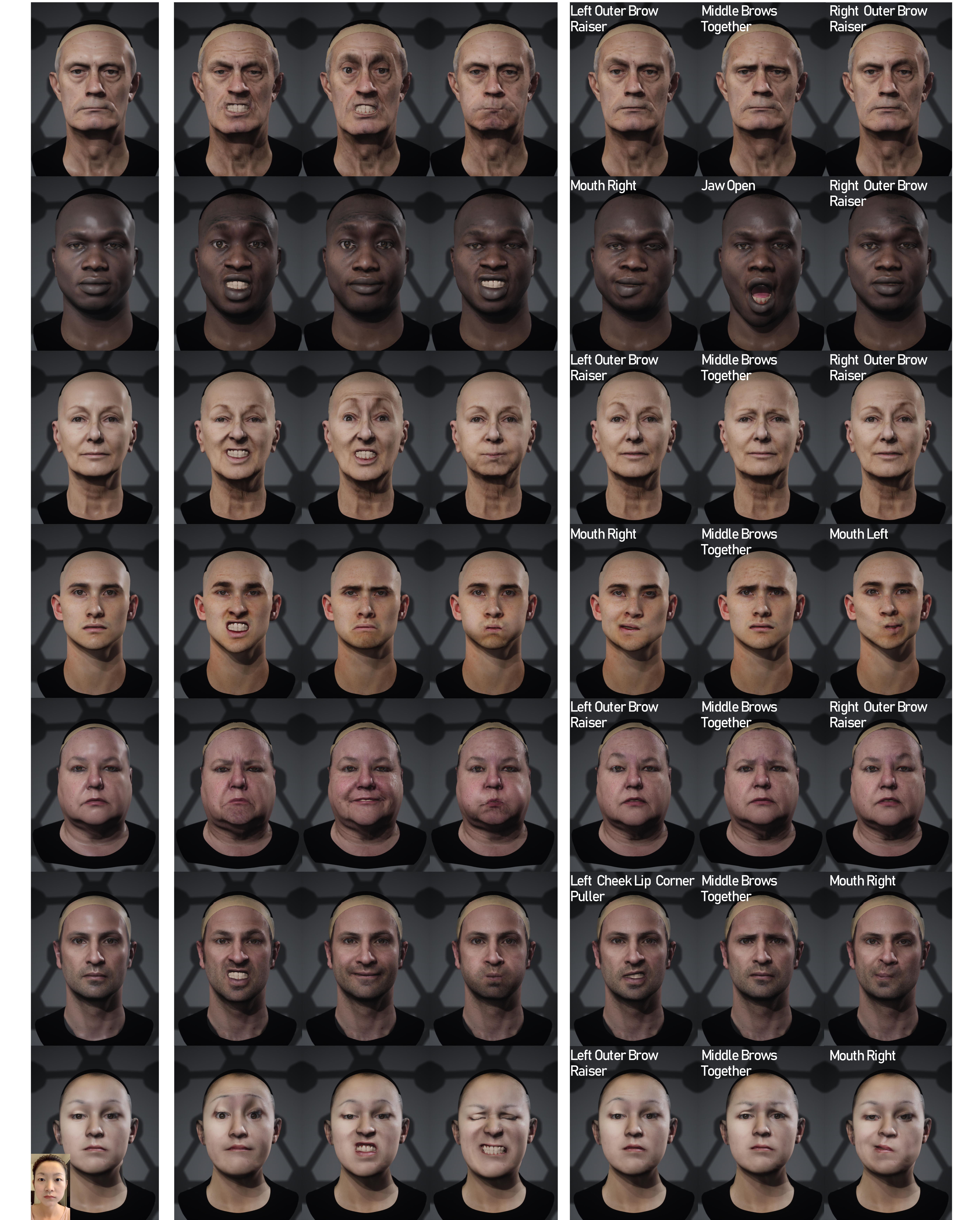}
 \vspace{-3mm}
 \caption{Expressions reconstructed by face rig assets generated by our framework with inputs from multiple sources. From left to right: Column 1: input neutral including geometry and albedo. Column 2 to Column 4: selected reconstructed expressions. Column 5 to Column 7: selected blendshape units. From top to bottom: Row 1 and Row 2: input neutral from Triplegangers~\cite{triplegangers}, Row 3 and Row 4: input neutral from online resources~\cite{3DScanstore}, Row 5 and Row 6: input neutral from Light Stage testing set. Row 7: Input neutral from iPhone X Arkit. The last example shows that our method can also be applied to data captured by a low-quality device despite that low-resolution input image may reduce the resulting quality.}{}
 \label{fig:render_res}
\end{figure*}

\begin{figure*}
 \centering
 \includegraphics[width=0.8\linewidth]{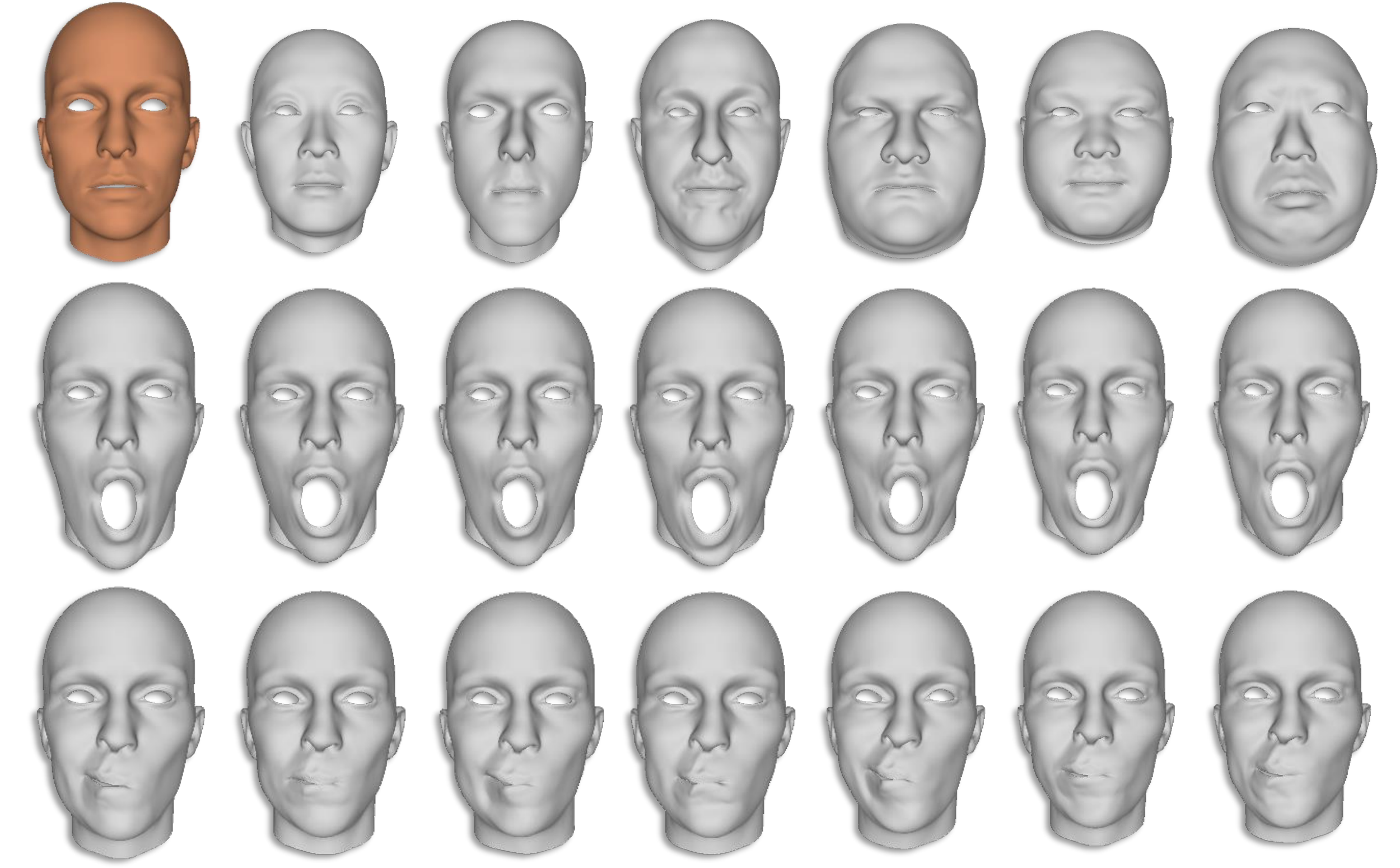}\\
 \setlength{\tabcolsep}{0\linewidth}
 \caption{Demonstration of customized identity of individuals on our generated blendshapes expressions. We combine blendshapes units from different individuals with the same template neutral (shown in orange). Row 1: source individuals. Row 2: combine personalized \textit{Jaw\_Open} of individuals in row one with template neutral. Row 3: combine personalized \textit{Mouth\_Right} of individuals in row one with template neutral.}
 \label{fig:diff_bs_combo}
\end{figure*}

\begin{figure}
 \includegraphics[width=3.0in]{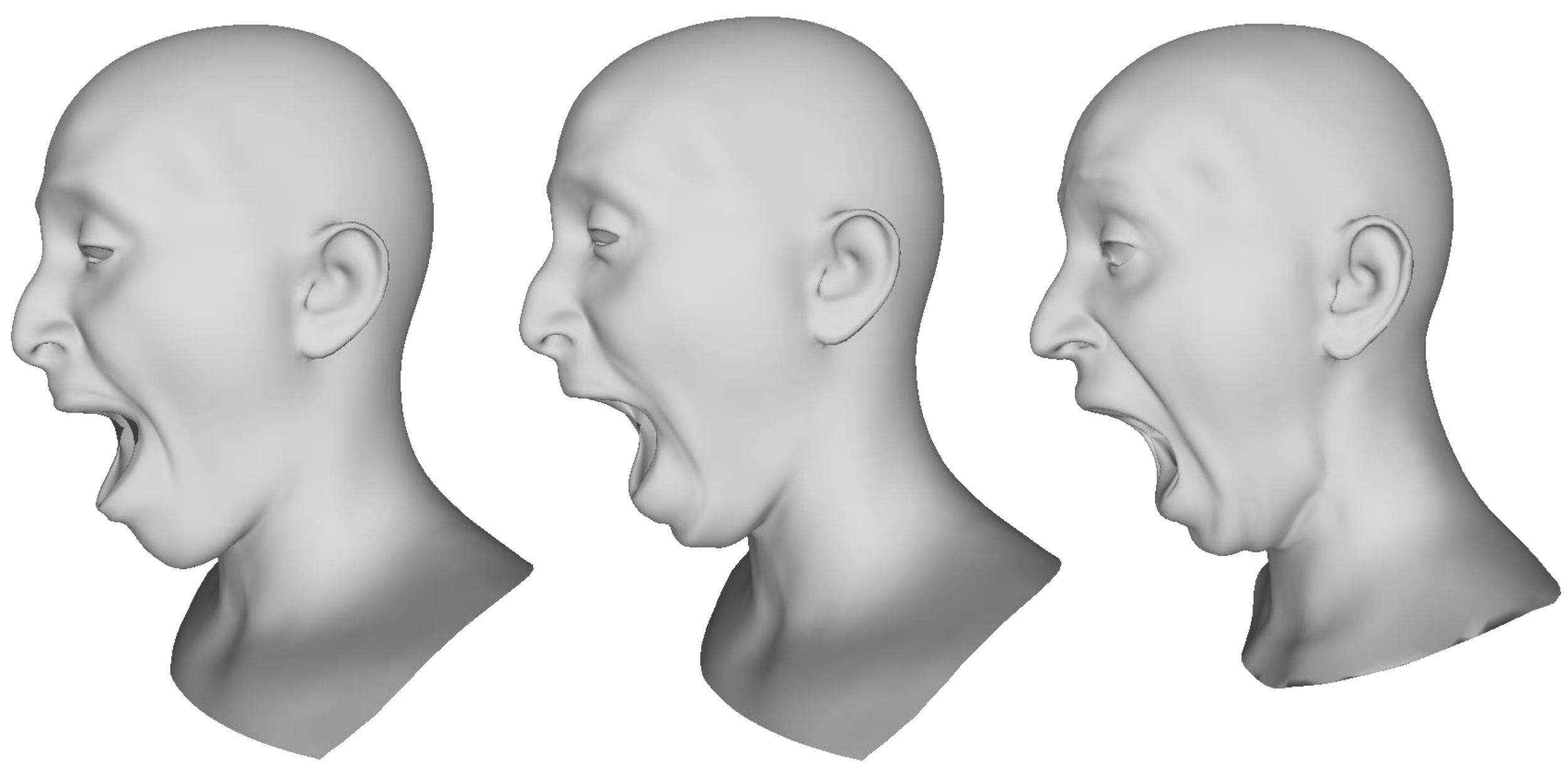}
 \caption{Comparison of extreme expression fitting using template blendshapes and our generated personalized blendshapes. Left: fitting results using template blendshapes. Middle: fitting results using our generated Personalized blendshapes. Right: ground truth expression.}
 \label{fig:extreme_exp_fit}
\end{figure}

\begin{figure}[ht]
 \centering
 \includegraphics[width=\linewidth]{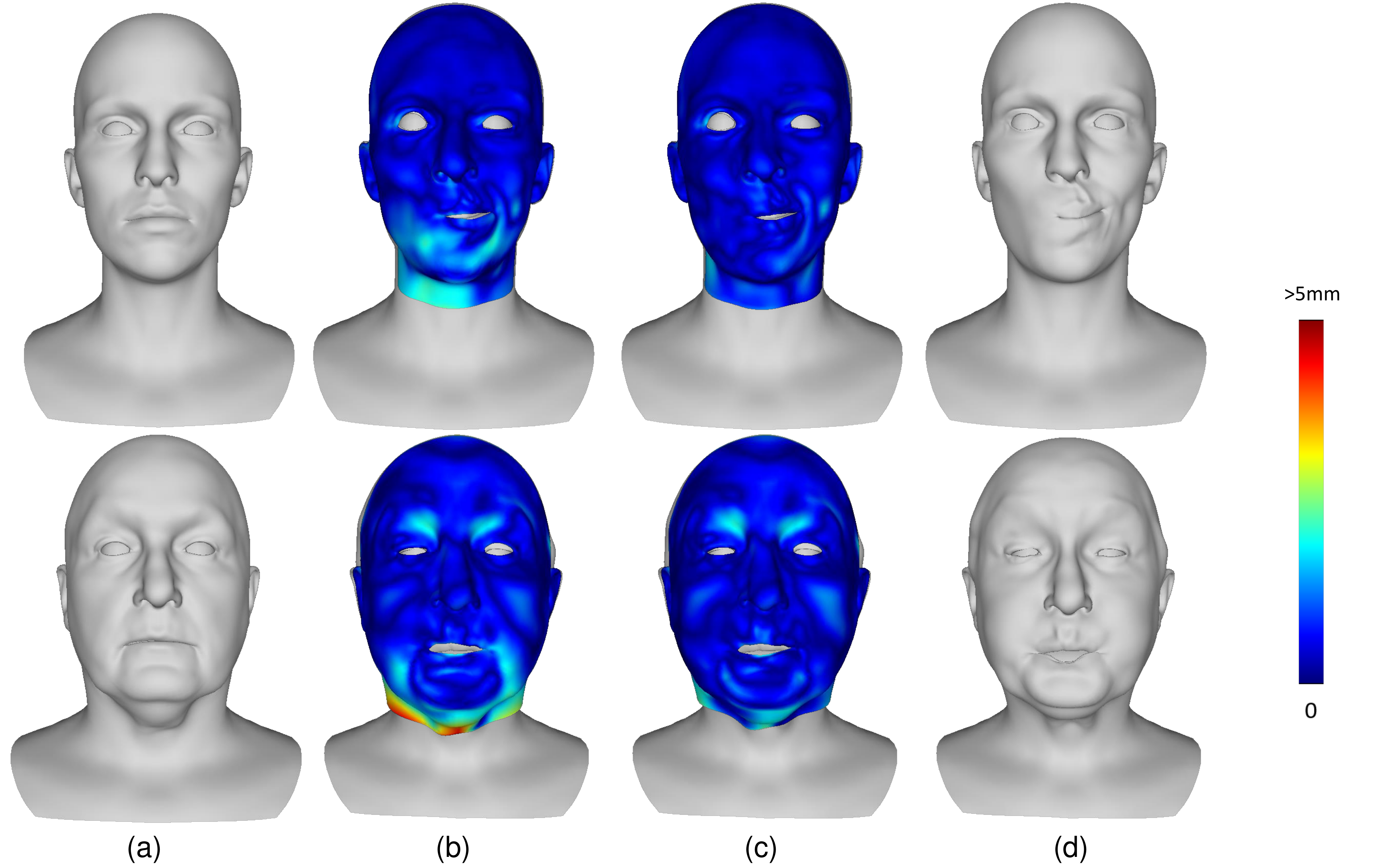}\\
 \setlength{\tabcolsep}{0\linewidth}
 \caption{Numerical analysis of the expressiveness of personalized blendshapes on expression tracking by swapping Blendshapes. (a) Neutrals of two individuals. (b) Reconstruction error using personalized Blendshapes from counterpart individuals. (c) Reconstruction error using their own personalized Blendshapes. (d) Target expressions.}
 \label{fig:diff_sub_bs_cmp}
\end{figure}

 \begin{figure*}
 \centering
 \setlength{\tabcolsep}{0.2mm}{
 \begin{tabular}{cccc|ccc}
     & \multicolumn{3}{c}{Mouth\_Open} & \multicolumn{3}{c}{Mouth\_Left}\\
  \vspace{-0.01 in} Template & \includegraphics[width=0.145\linewidth]{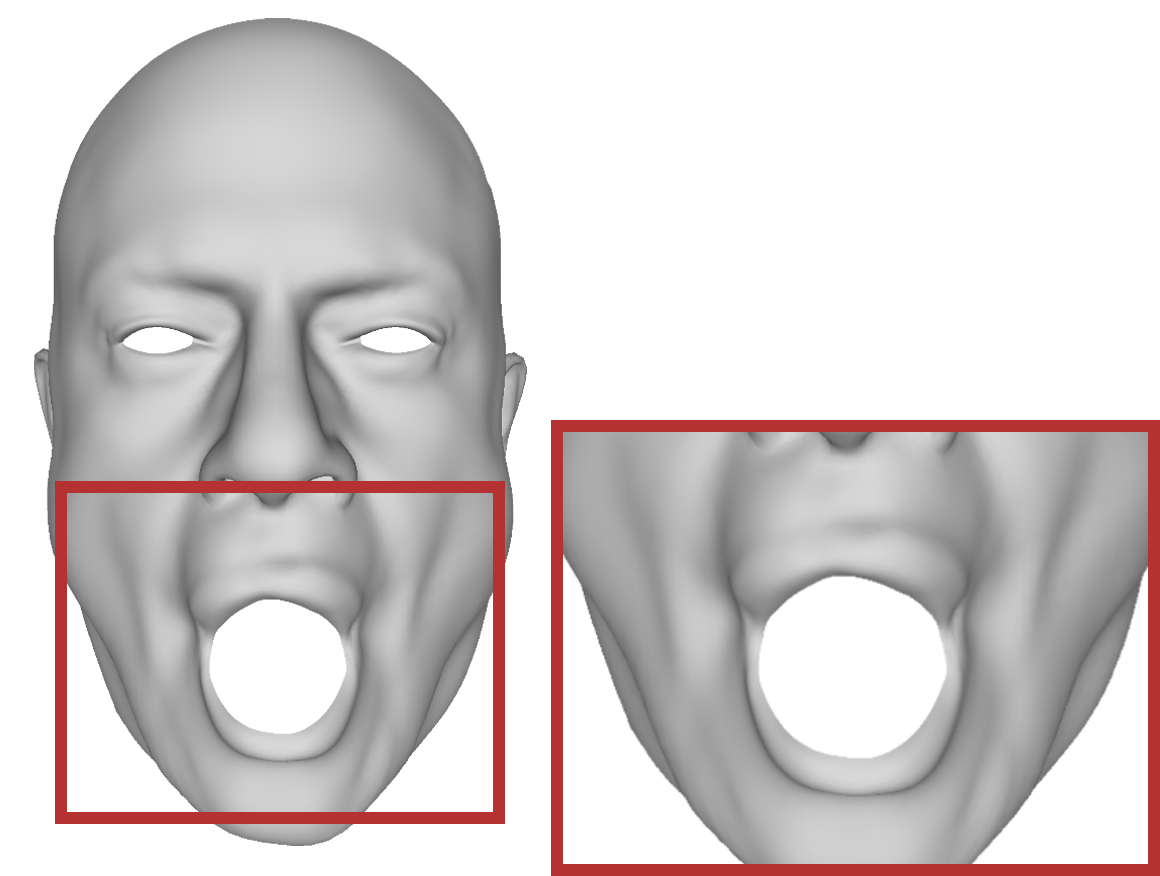} &
  \includegraphics[width=0.145\linewidth]{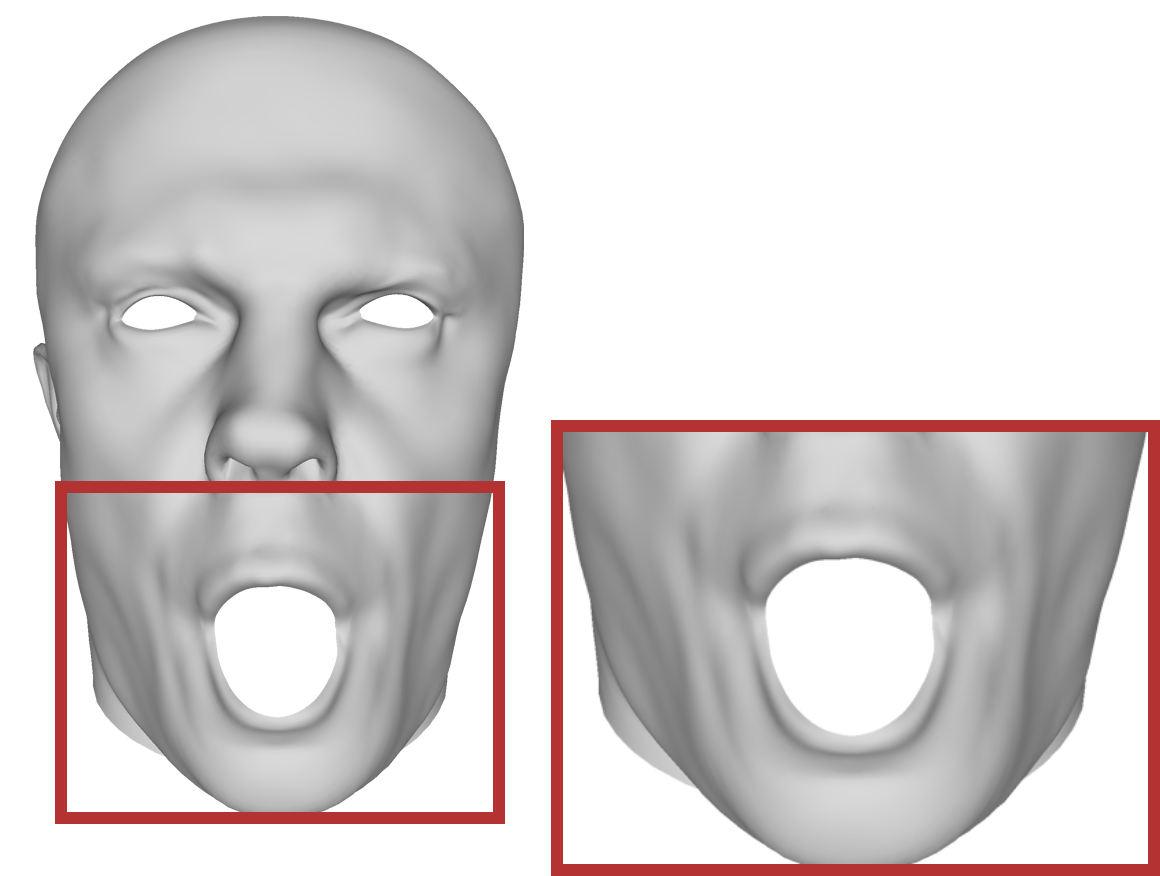} &
  \includegraphics[width=0.145\linewidth]{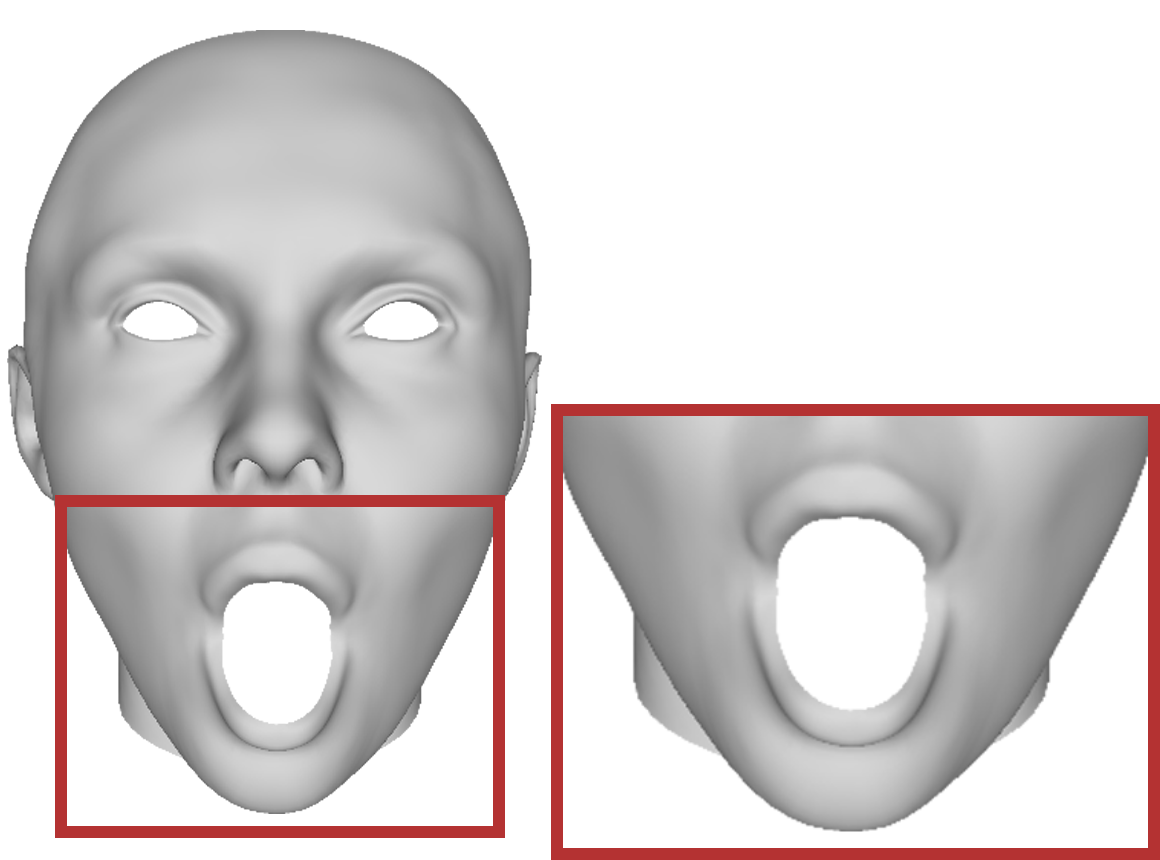}  \vspace{0.001 in} & 
  \includegraphics[width=0.145\linewidth]{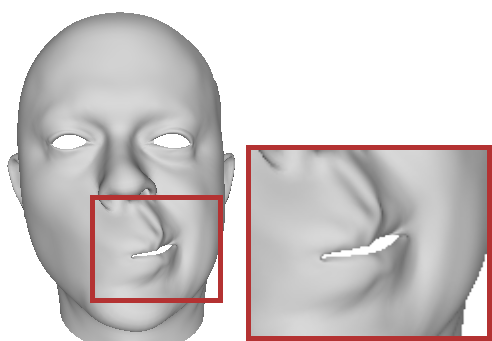} &
  \includegraphics[width=0.145\linewidth]{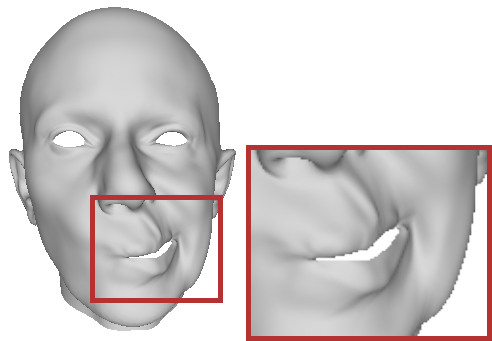} &
  \includegraphics[width=0.145\linewidth]{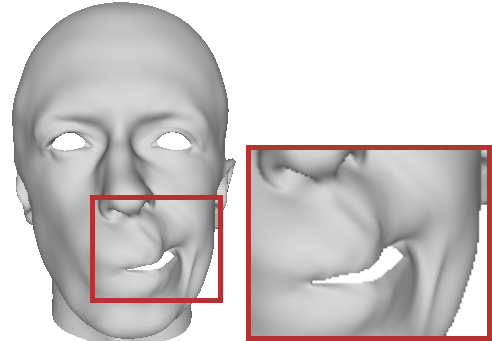} \\

   ~\citet{Li_2010_SIGGRAPH} & \includegraphics[width=0.145\linewidth]{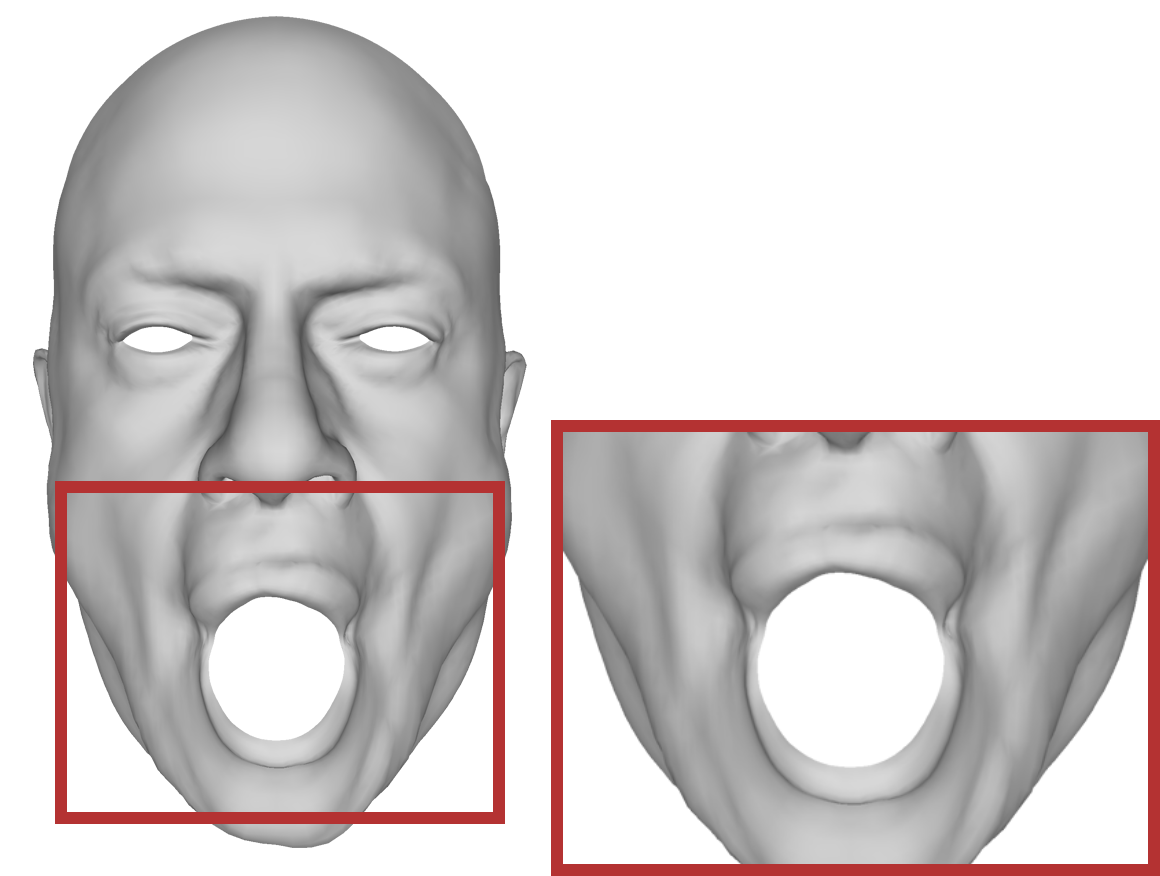} &
  \includegraphics[width=0.145\linewidth]{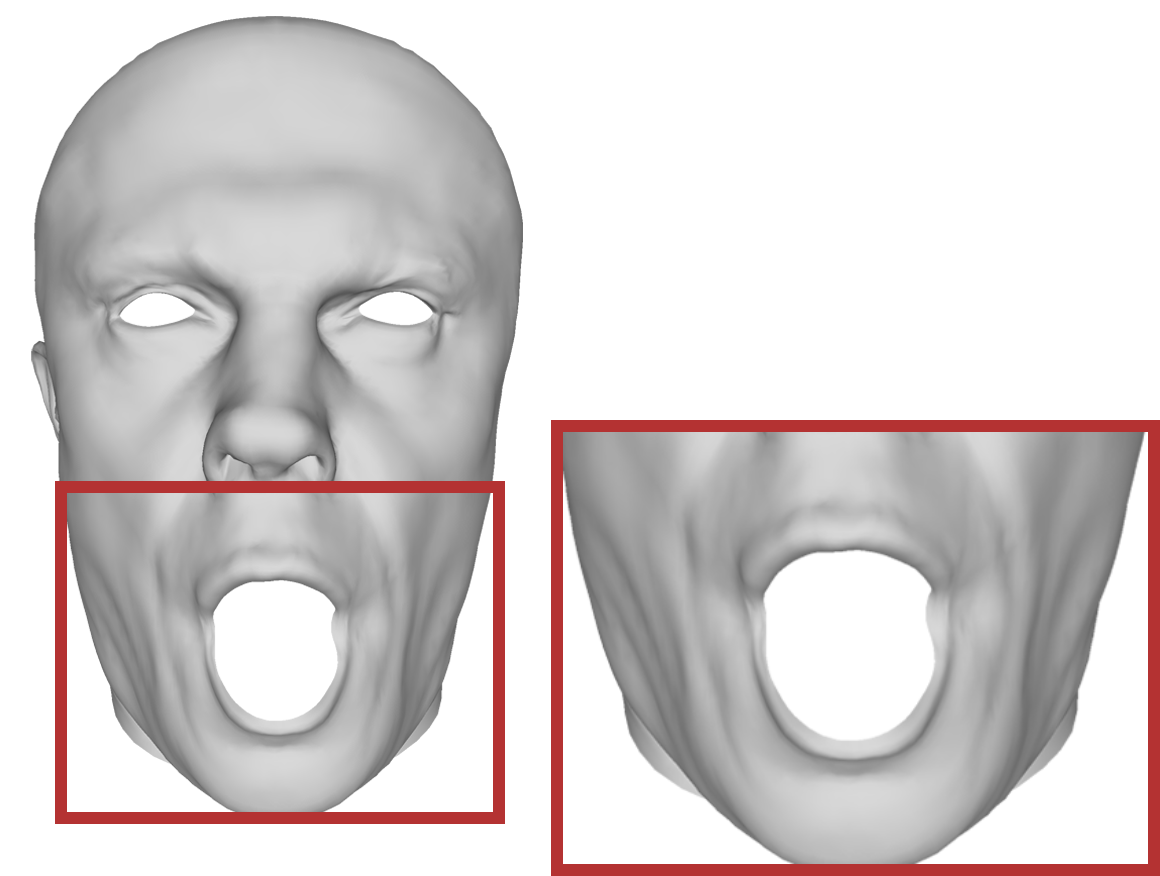} &
  \includegraphics[width=0.145\linewidth]{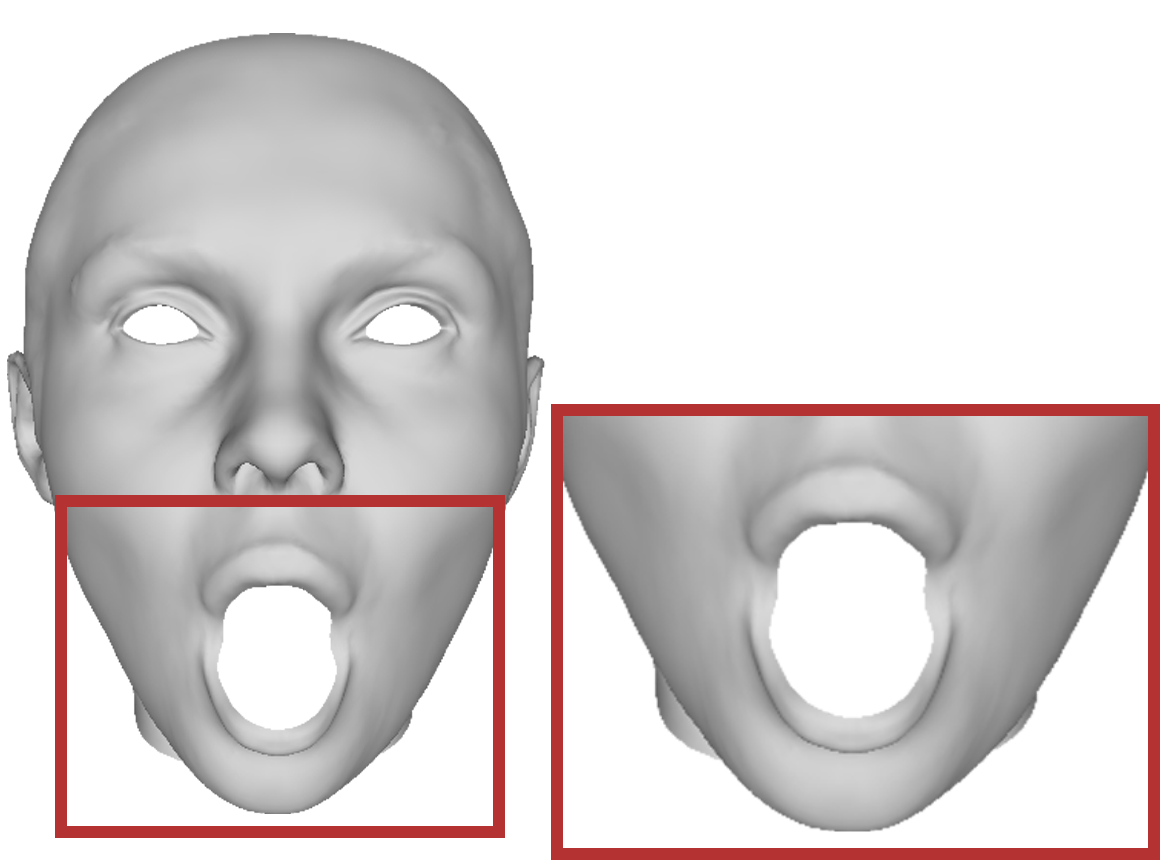} \vspace{0.001 in} & 
  \includegraphics[width=0.145\linewidth]{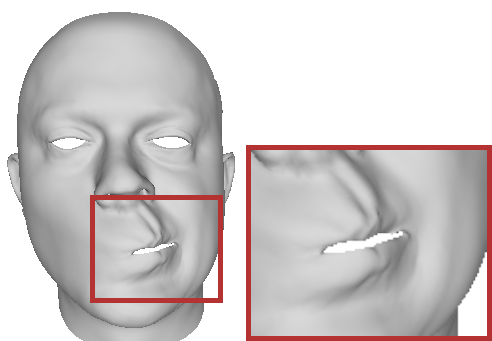} &
  \includegraphics[width=0.145\linewidth]{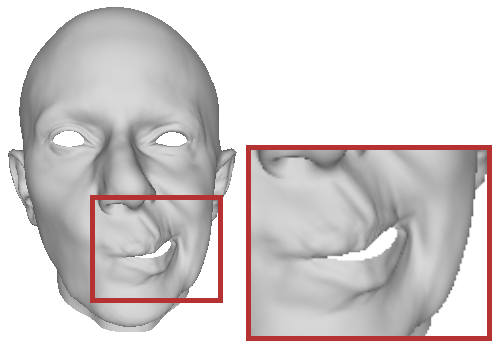} &
  \includegraphics[width=0.145\linewidth]{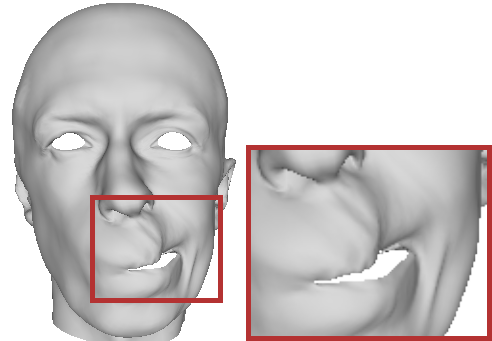} \\
  
   Ours & \includegraphics[width=0.145\linewidth]{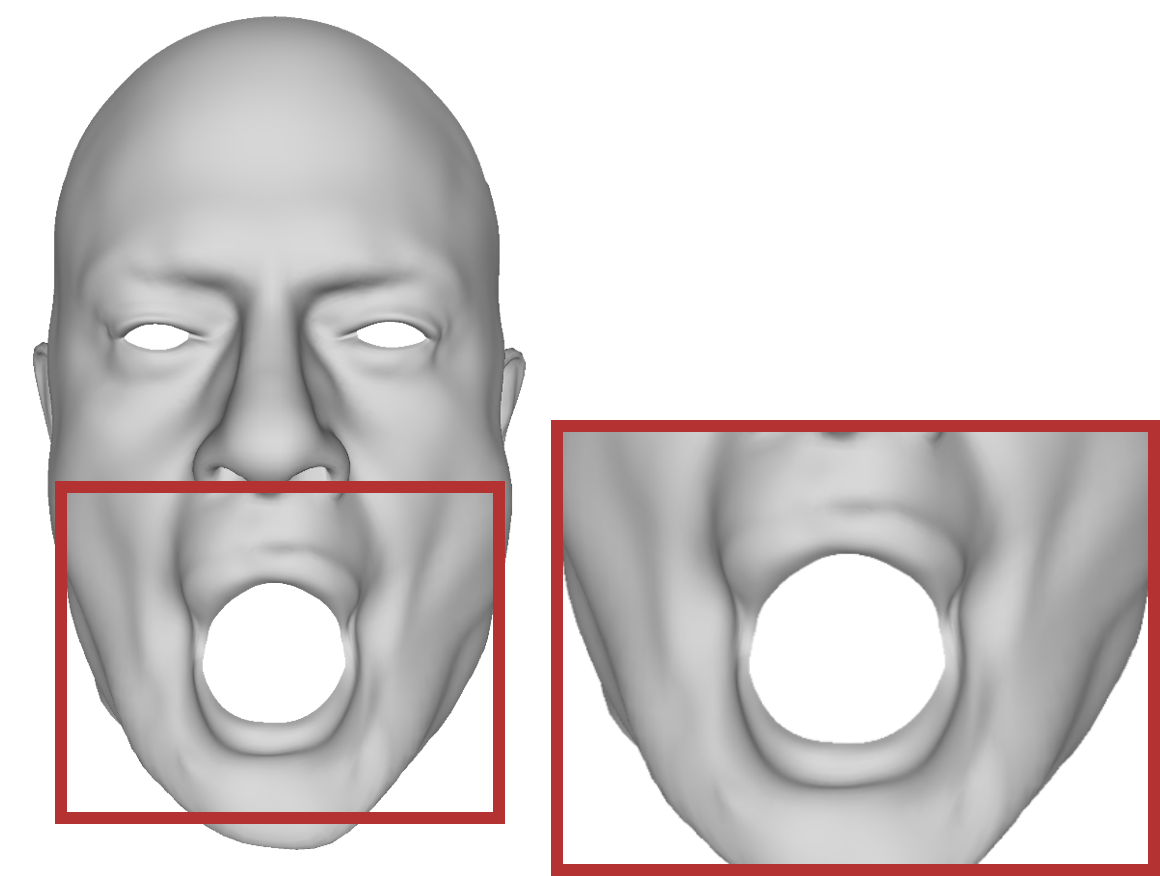} &
  \includegraphics[width=0.145\linewidth]{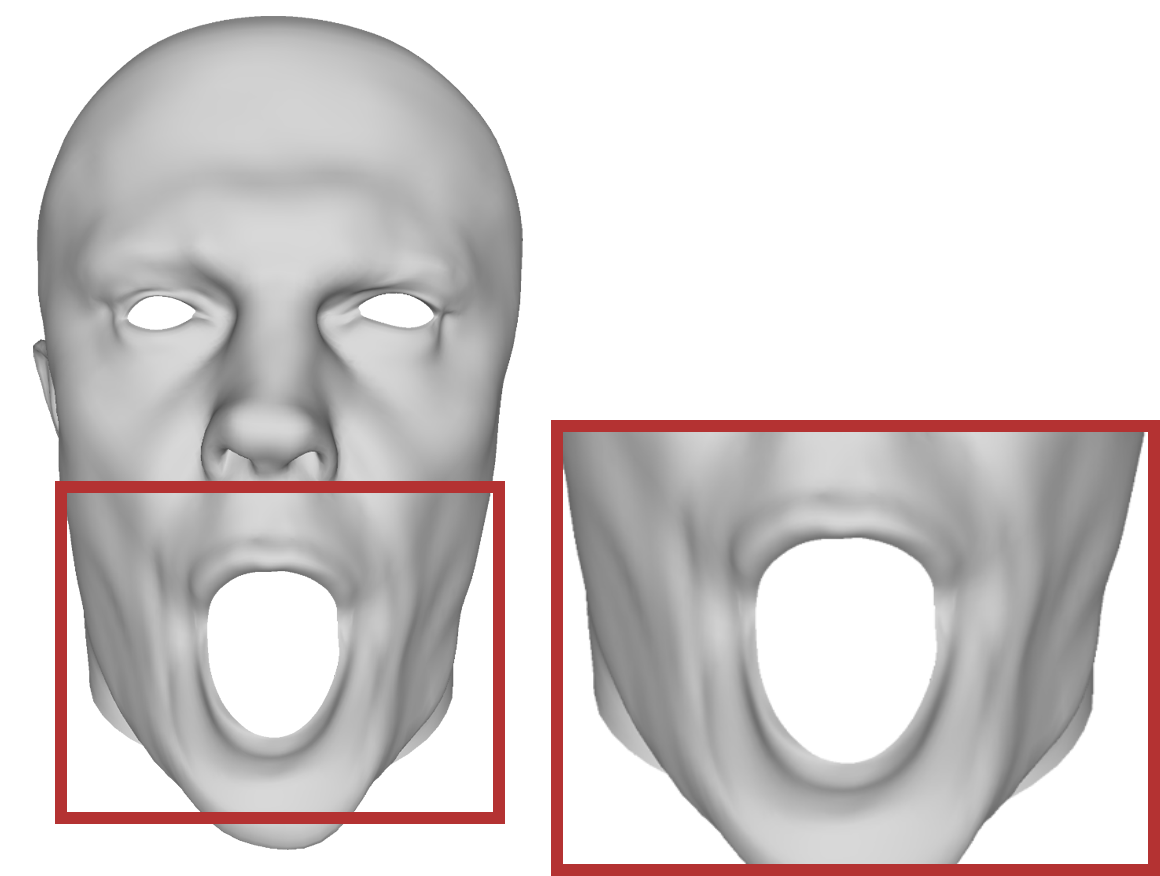} &
  \includegraphics[width=0.145\linewidth]{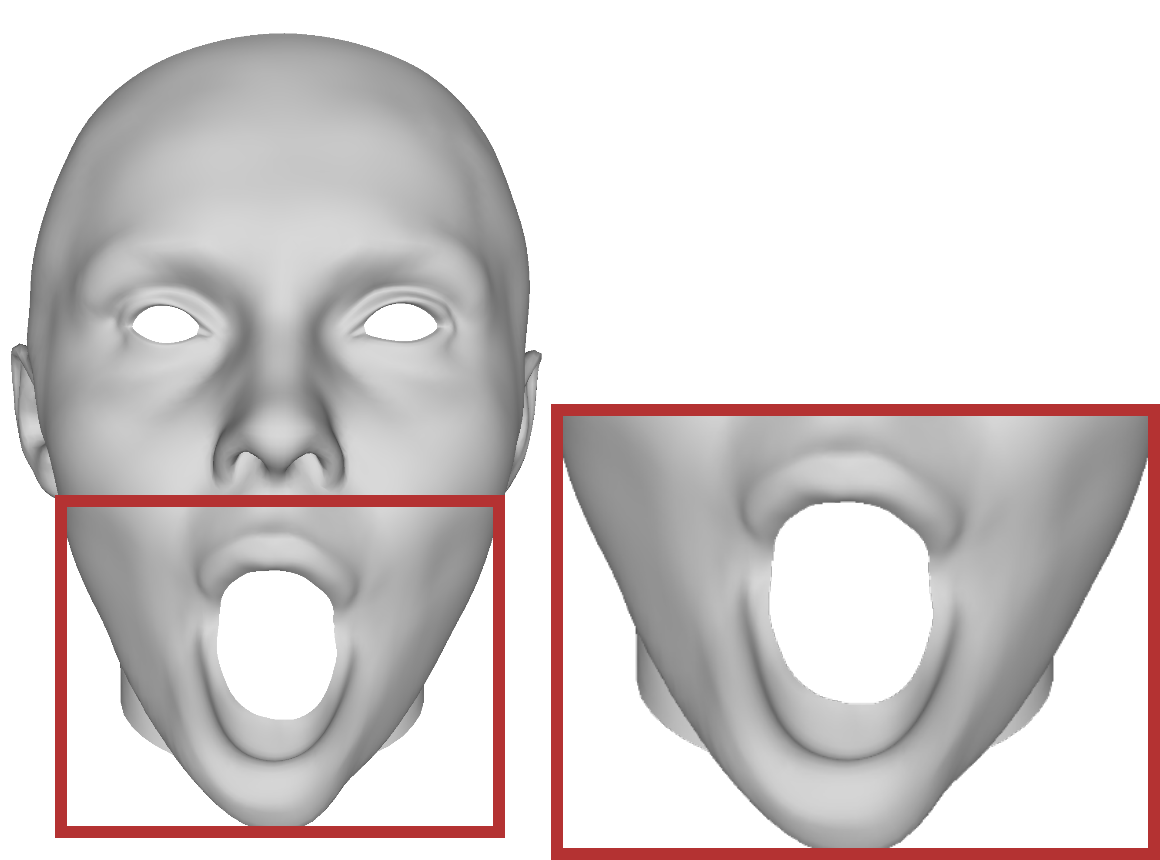} \vspace{0.001 in} & 
  \includegraphics[width=0.145\linewidth]{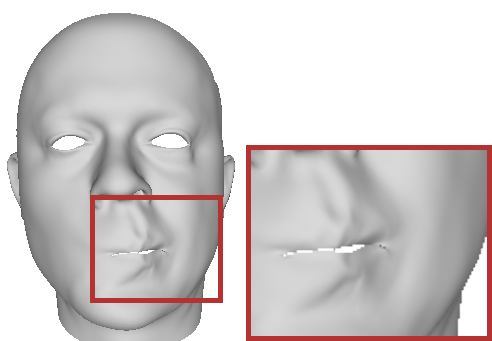} &
  \includegraphics[width=0.145\linewidth]{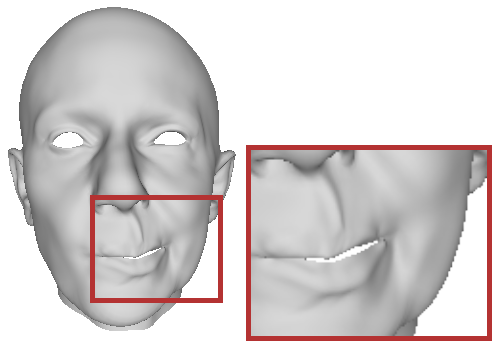} &
  \includegraphics[width=0.145\linewidth]{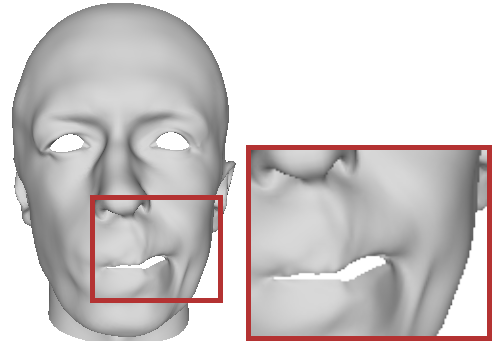} \\
  \end{tabular}}
  
   \caption{Comparison of selected generated Blendshapes units with template generic and \citet{Li_2010_SIGGRAPH}.~Row 1: template blendshapes generated by expression transferred from a set of generic blendshapes using method in \citet{Sumner_2004_SIGGRAPH}.~Row 2: blendshapes optimized with method in \citet{Li_2010_SIGGRAPH}.~Row 3: our method. Note that the results generated in \citet{Li_2010_SIGGRAPH} are from 26 scanned expressions, ours are from a single neutral input.} 
 \label{fig:bs_cmp}
\end{figure*}

\begin{figure}[t]
 \includegraphics[width=3in]{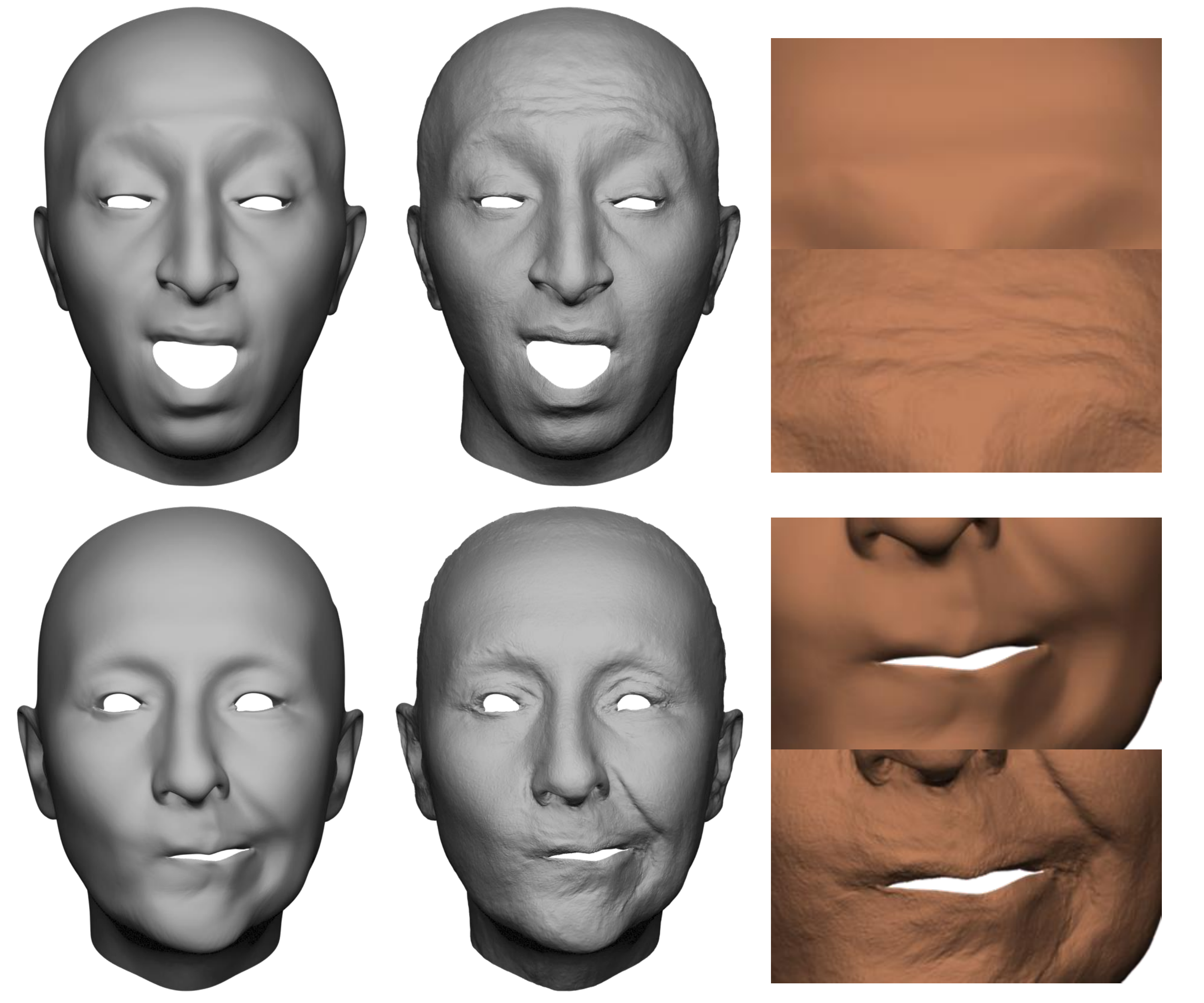}
 \caption{Dynamic displacement map predicted by our framework. Left: base geometries. Middle: results by applying generated displacement to base geometries. Right: closed-up comparison before and after applying dynamic displacement maps. 
 }
 \label{fig:displacement}
\end{figure}

\begin{figure}
  \centering
 \setlength{\tabcolsep}{-0.01mm}{
 \begin{tabular}{cccc}
     Reference & \multicolumn{3}{c}{Avatars} \\
     \includegraphics[width=0.20\linewidth]{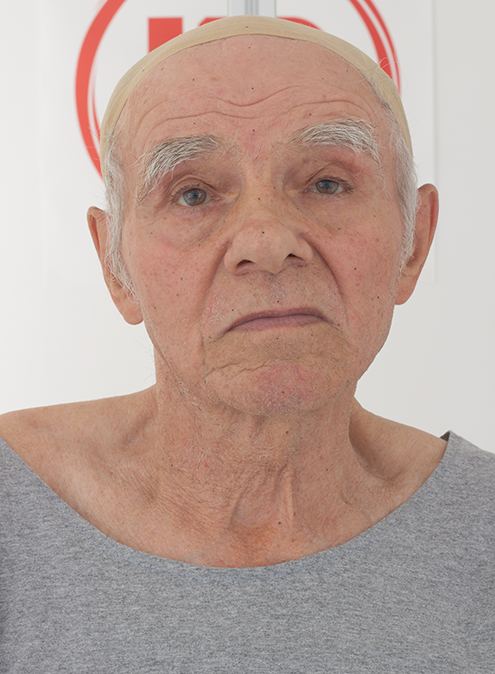} \vspace{0.001 in} &
     \includegraphics[width=0.18\linewidth]{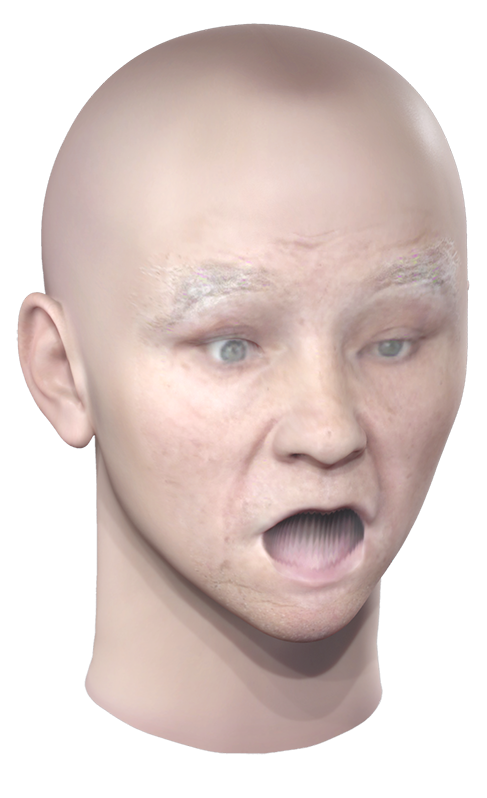} &
     \includegraphics[width=0.18\linewidth]{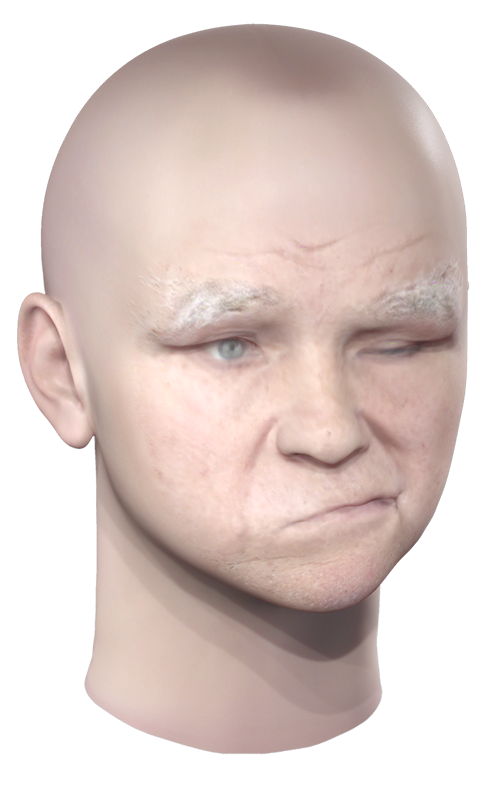} &
     \includegraphics[width=0.18\linewidth]{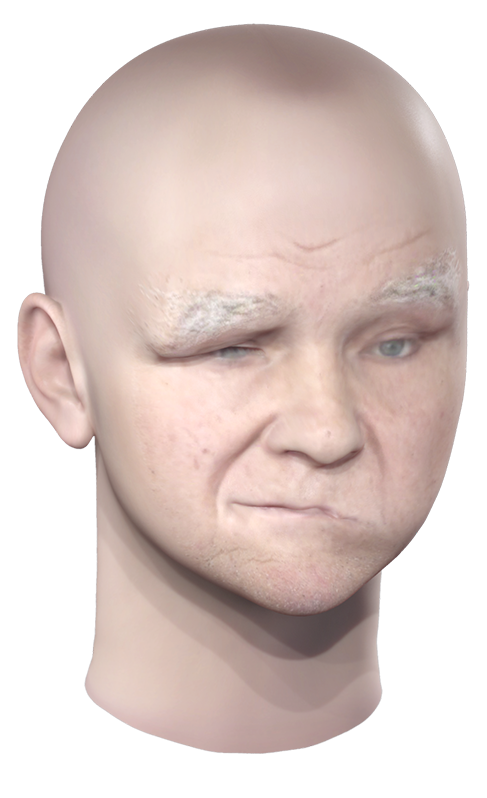} \\
     & \includegraphics[width=0.22\linewidth]{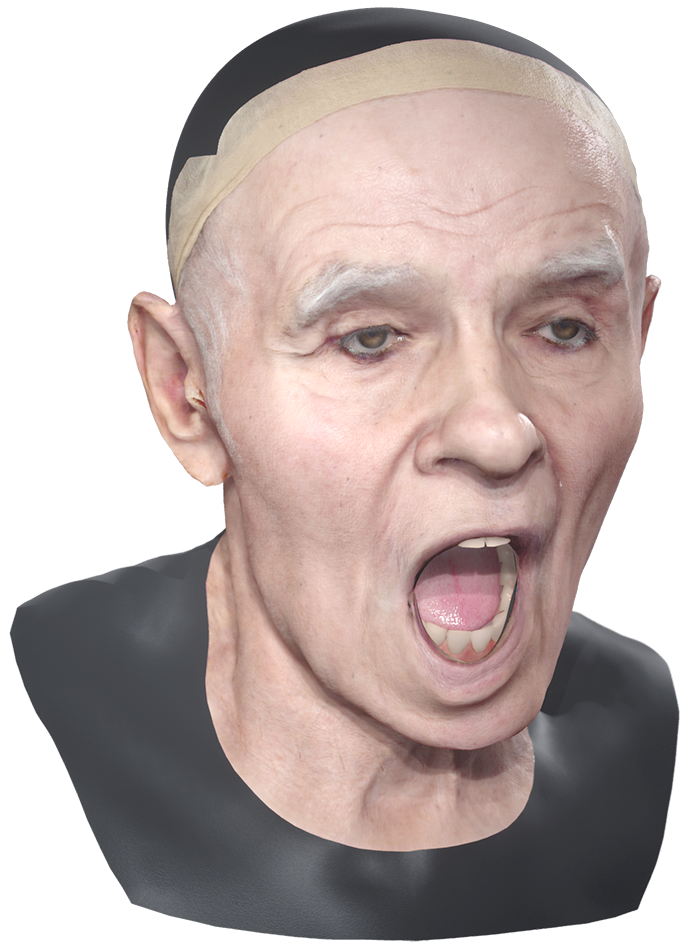} &
     \includegraphics[width=0.22\linewidth]{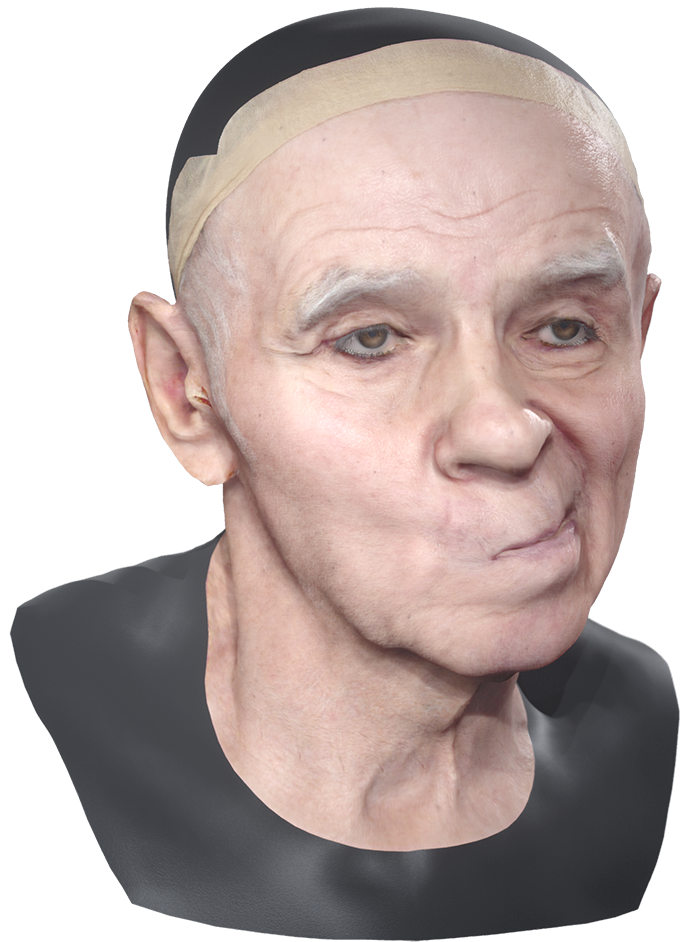} &
     \includegraphics[width=0.22\linewidth]{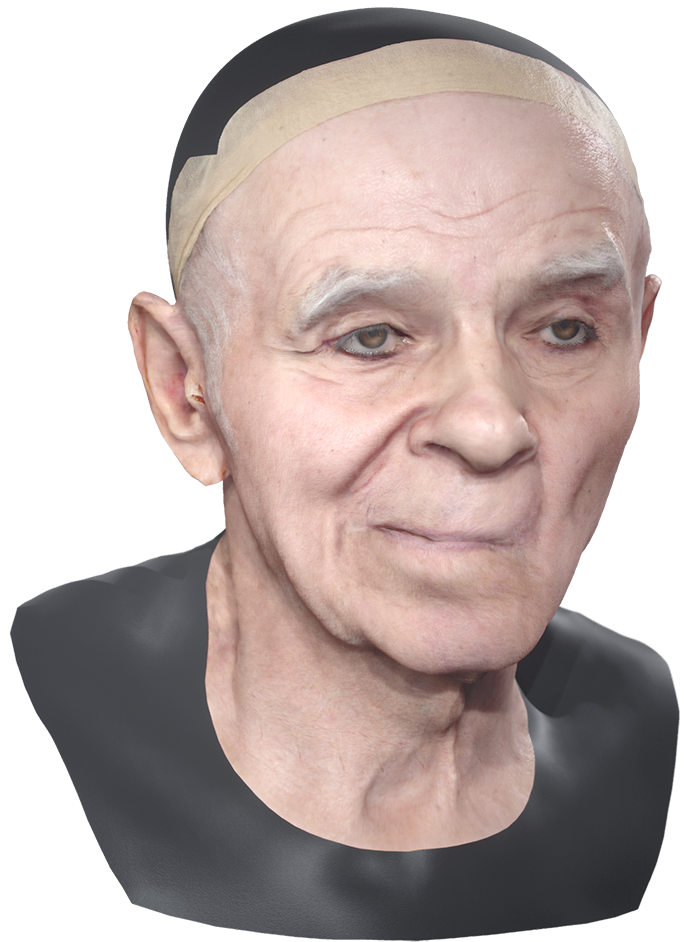}\\
     & \multicolumn{3}{c}{(a)} \\
     \includegraphics[width=0.20\linewidth]{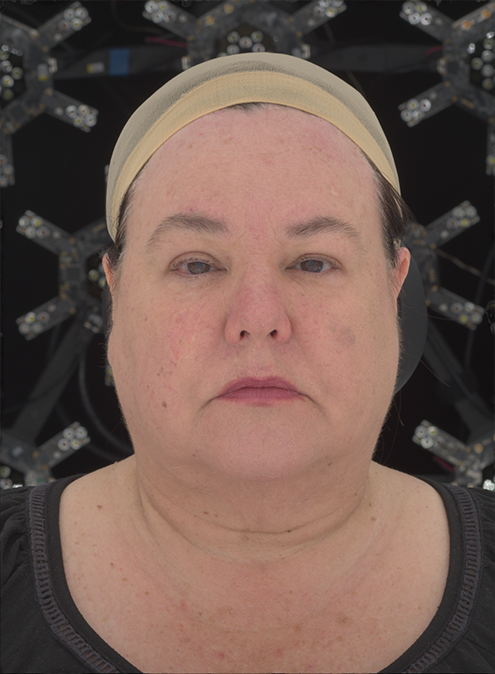} \vspace{0.001 in} & \includegraphics[width=0.18\linewidth]{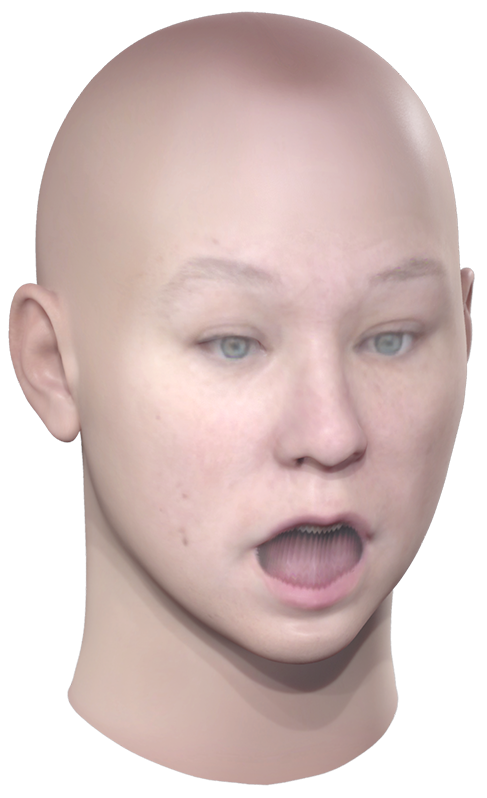} &
     \includegraphics[width=0.18\linewidth]{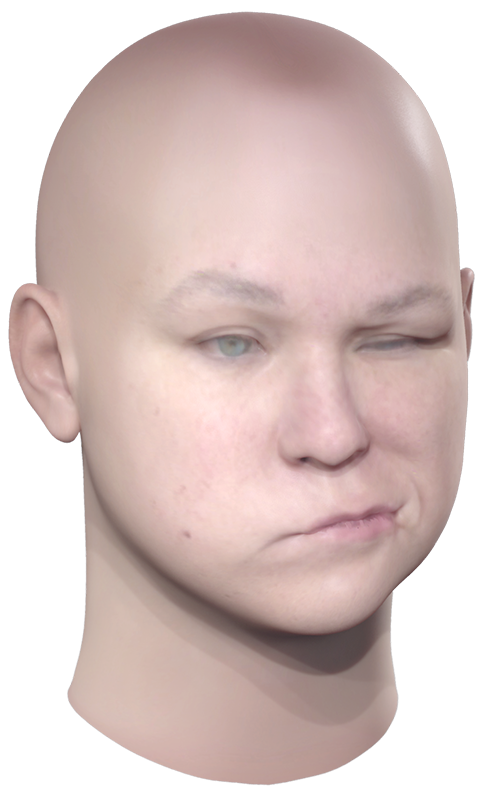} &
     \includegraphics[width=0.18\linewidth]{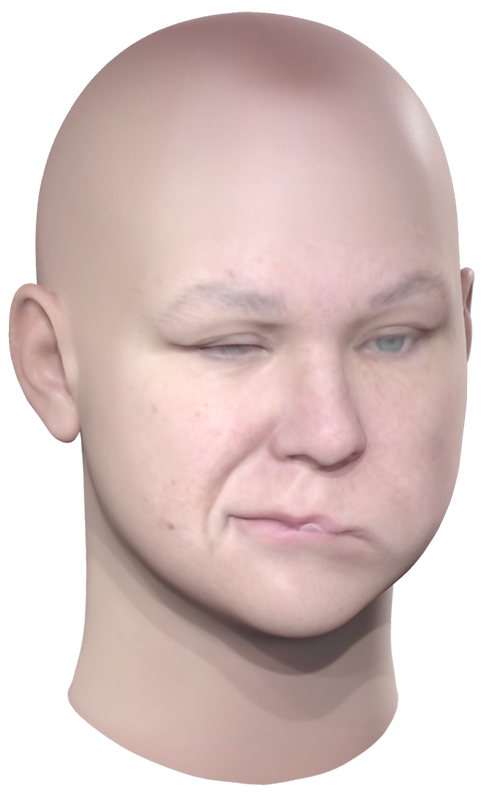} \\
     & \includegraphics[width=0.22\linewidth]{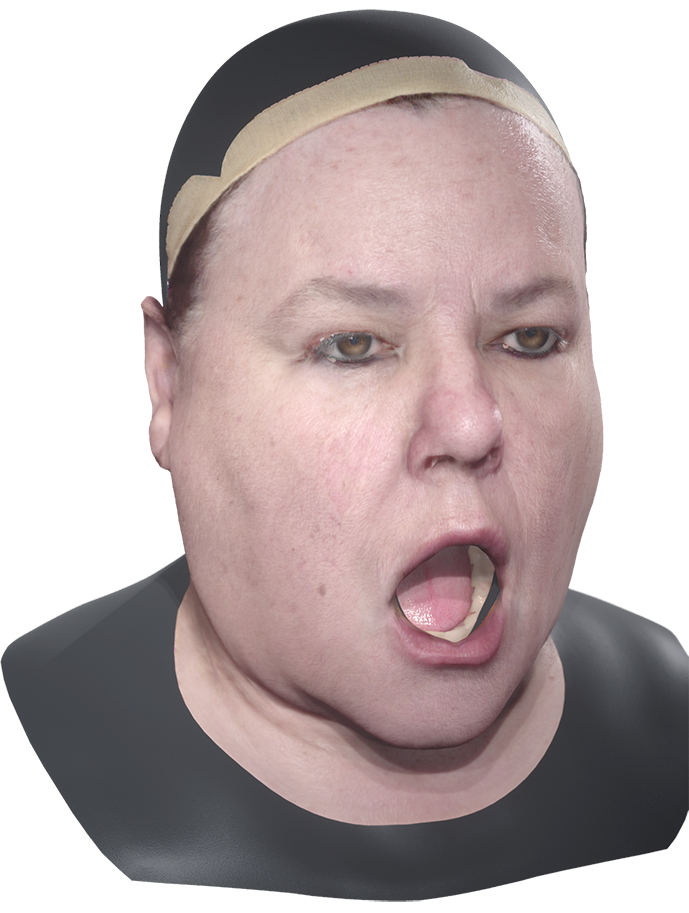} &
     \includegraphics[width=0.22\linewidth]{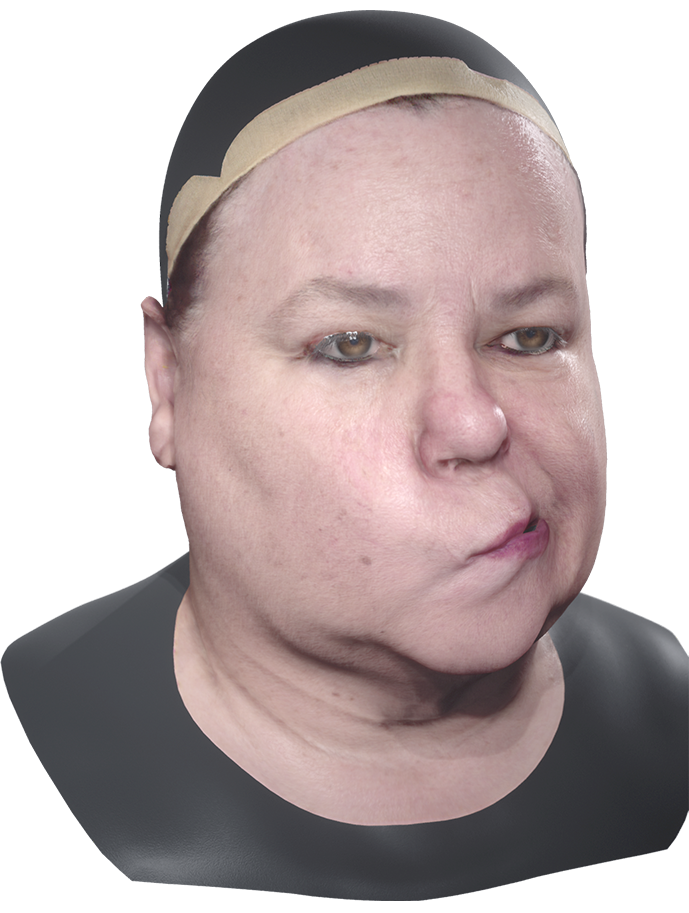} &
     \includegraphics[width=0.22\linewidth]{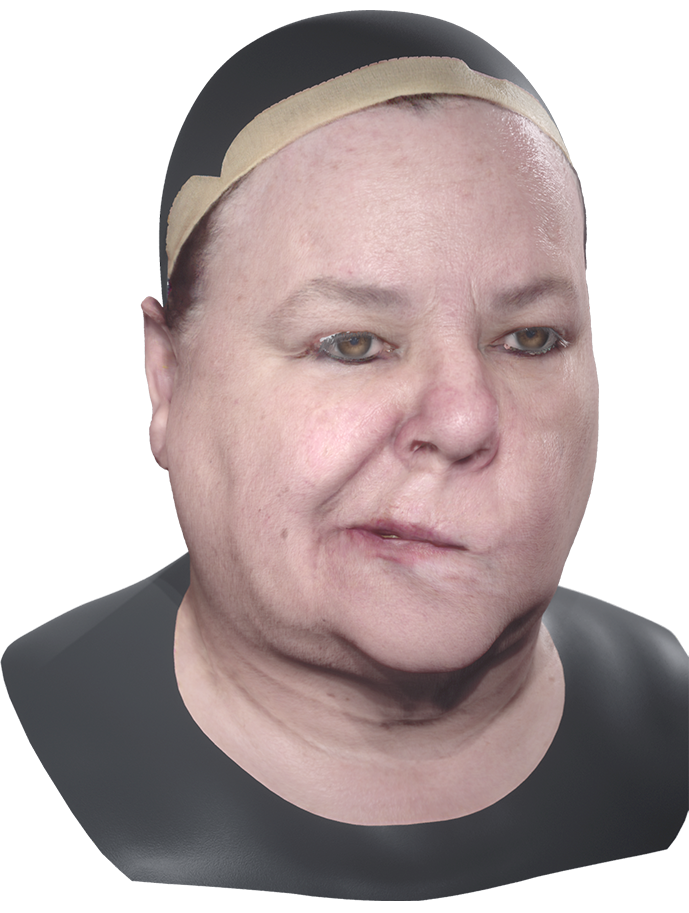}\\
     & \multicolumn{3}{c}{(b)} \\
  \end{tabular}}
 \caption{Comparison of generated face avatars between paGAN~\cite{Nagano_2018_SIGGRAPH} and our method. (a) and (b) show two cases of generated avatars from the neutral model in reference images. In each case, Row 1 shows the avatars generated by paGAN~\cite{Nagano_2018_SIGGRAPH} while Row 2 shows our results.}{}
 \label{fig:avatar_cmp}
\end{figure}

 \begin{figure*}
 \centering
  \centering
 \setlength{\tabcolsep}{0.6mm}{
 \begin{tabular}{ccccccc}
  \includegraphics[height=0.14\linewidth]{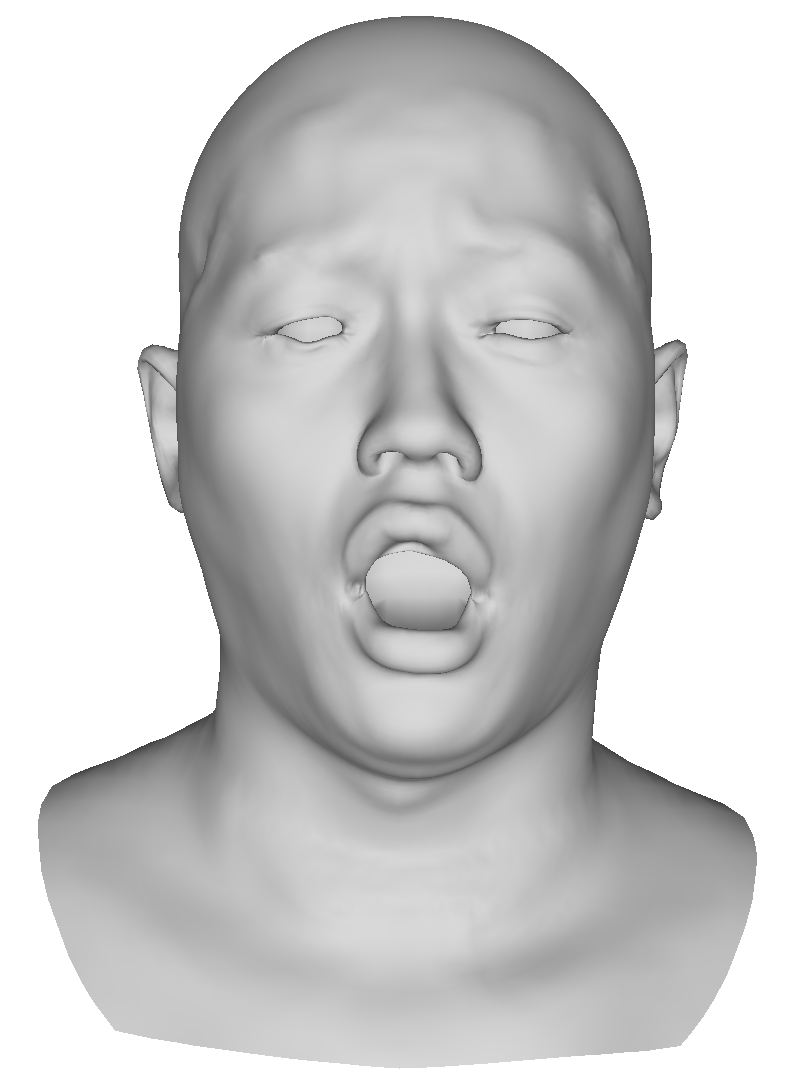} &
  \includegraphics[height=0.14\linewidth]{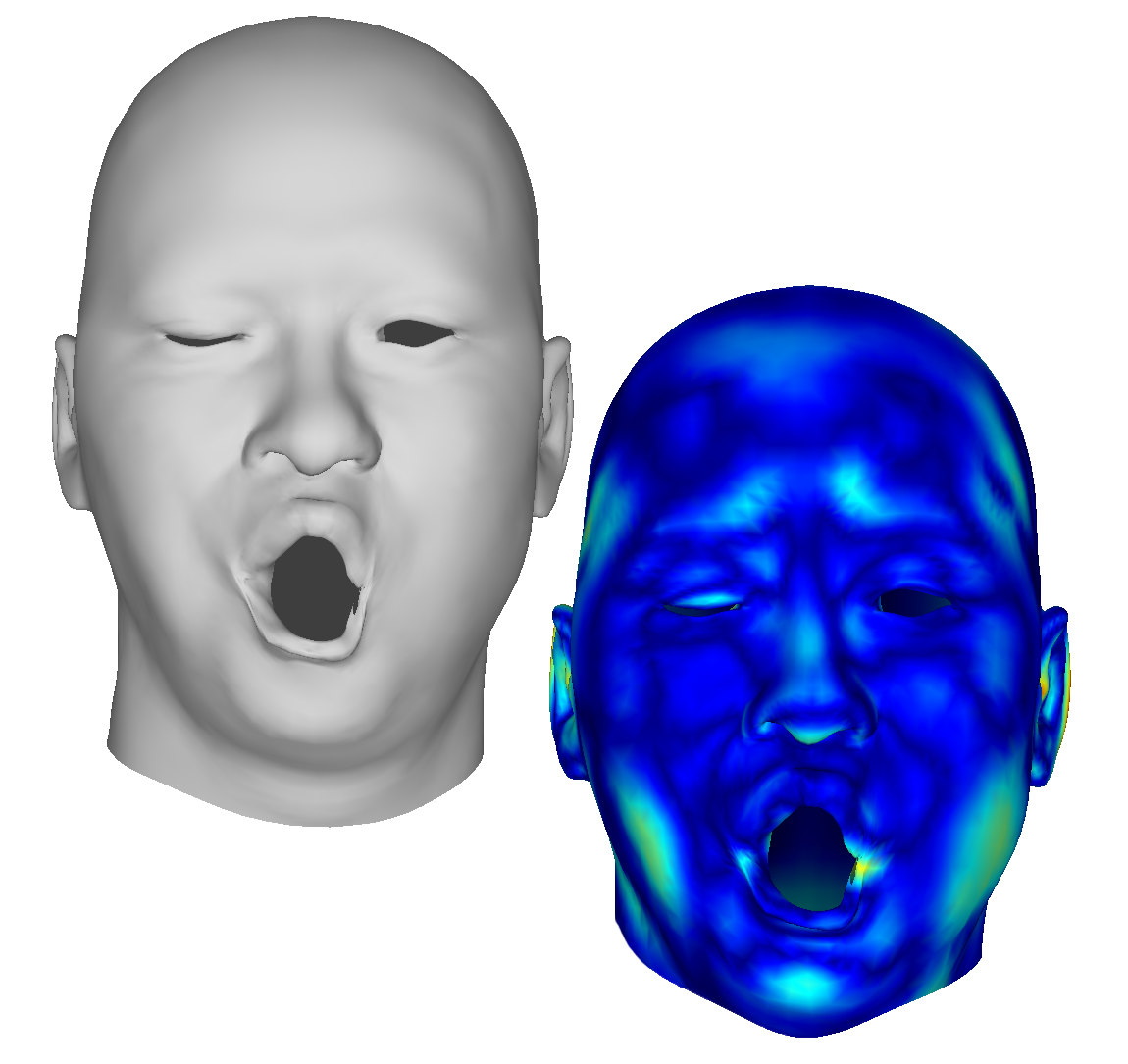} &
  \includegraphics[height=0.14\linewidth]{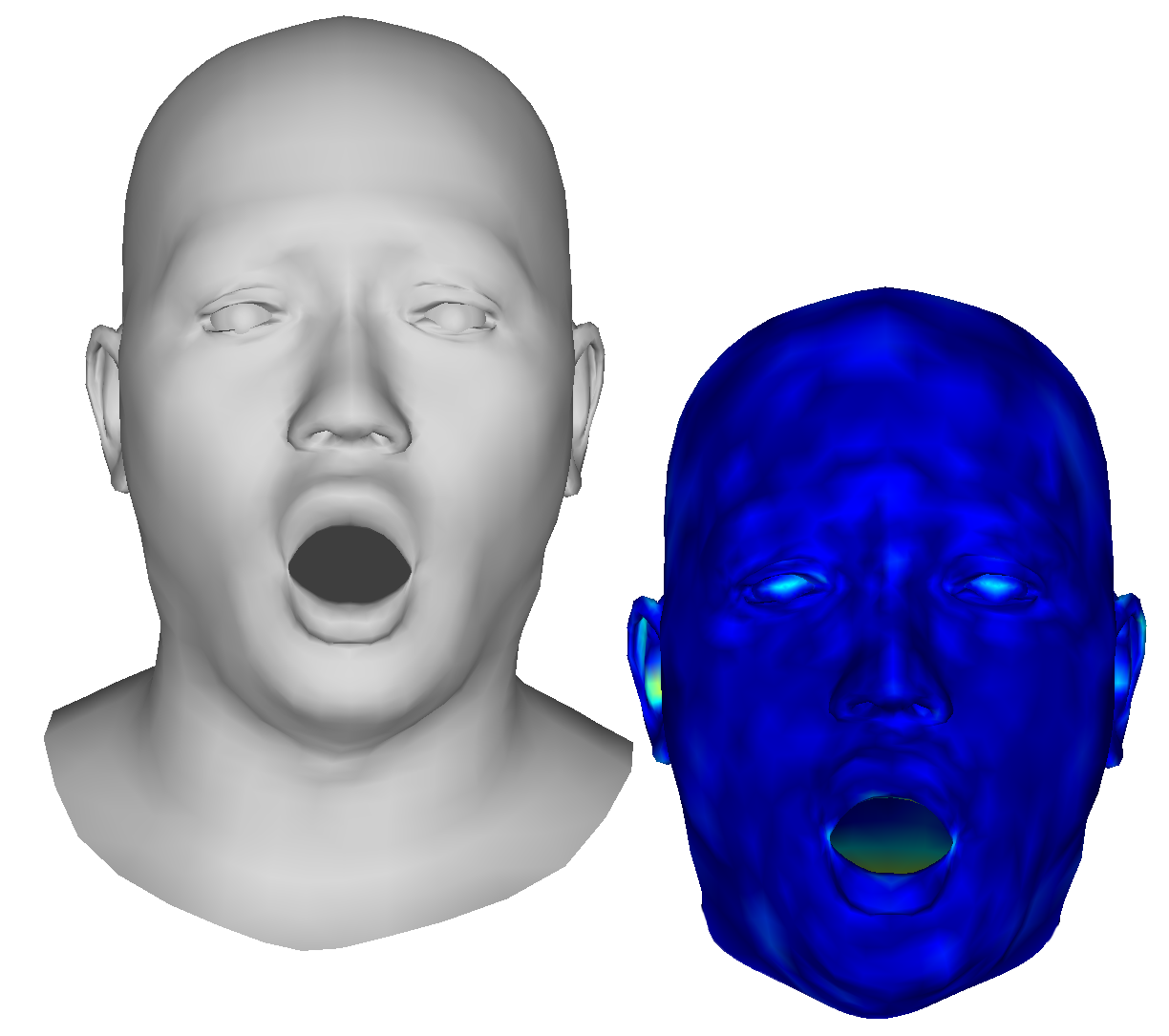} & 
  \includegraphics[height=0.14\linewidth]{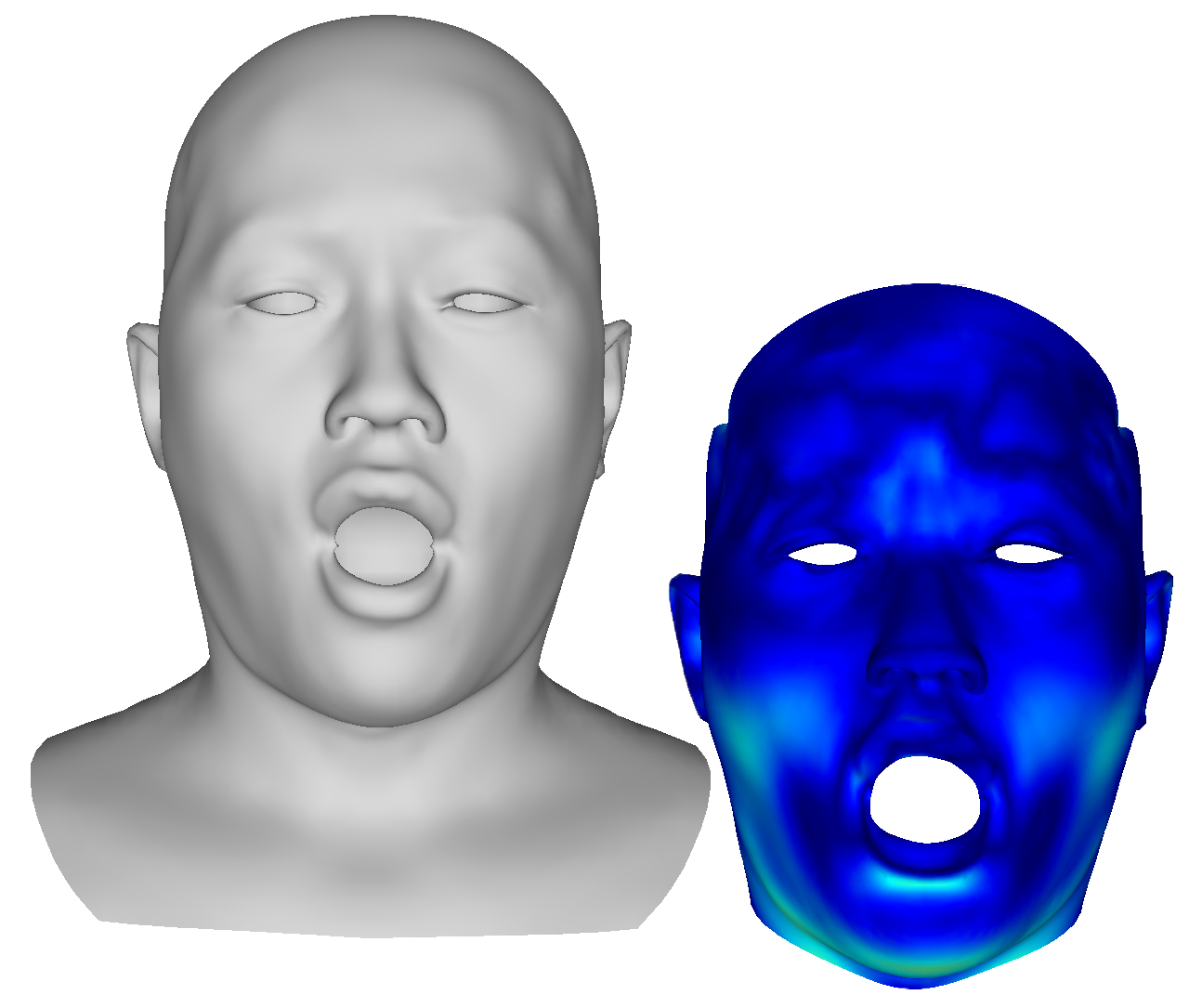} &
  \includegraphics[height=0.14\linewidth]{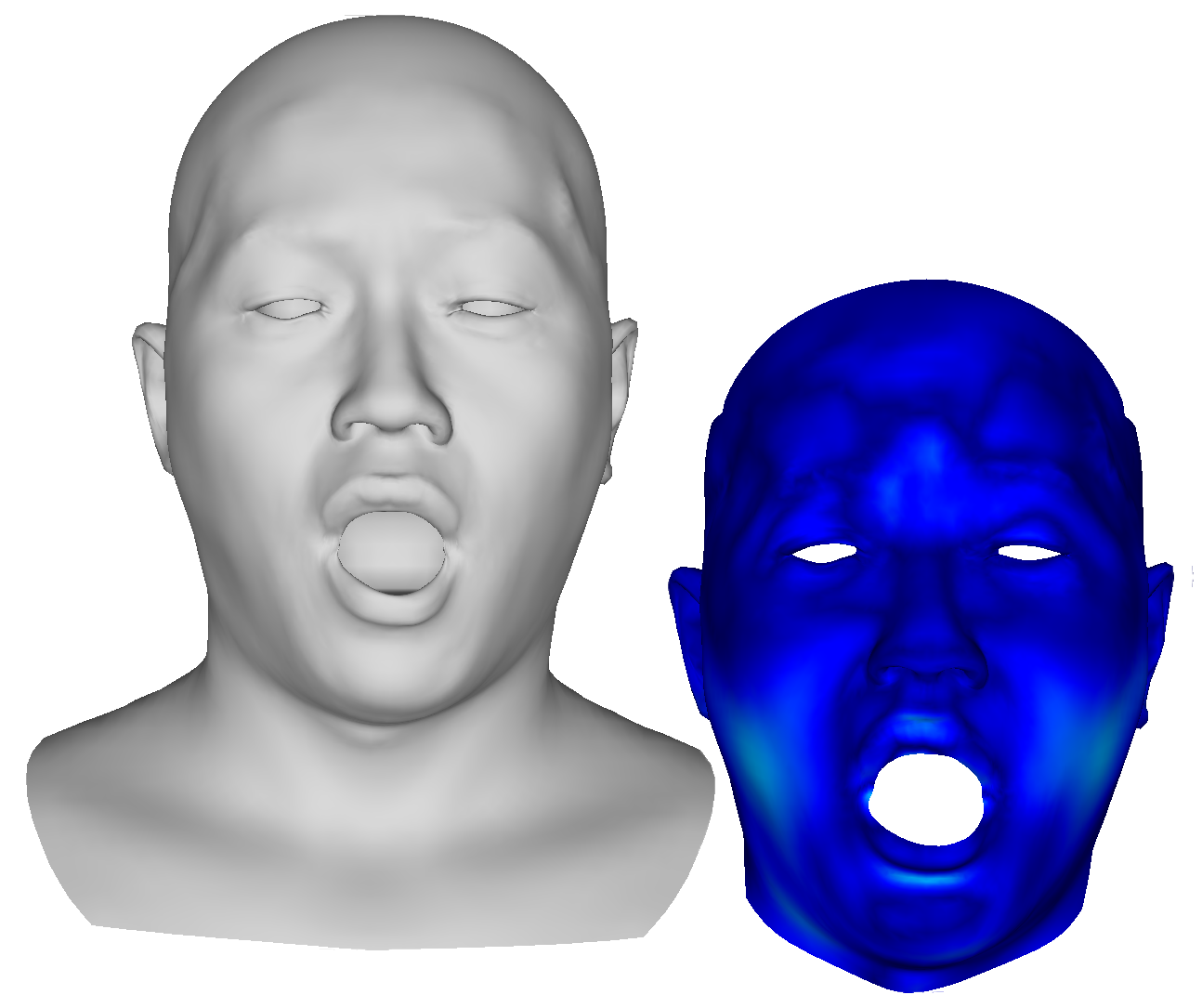} &
  \includegraphics[height=0.14\linewidth]{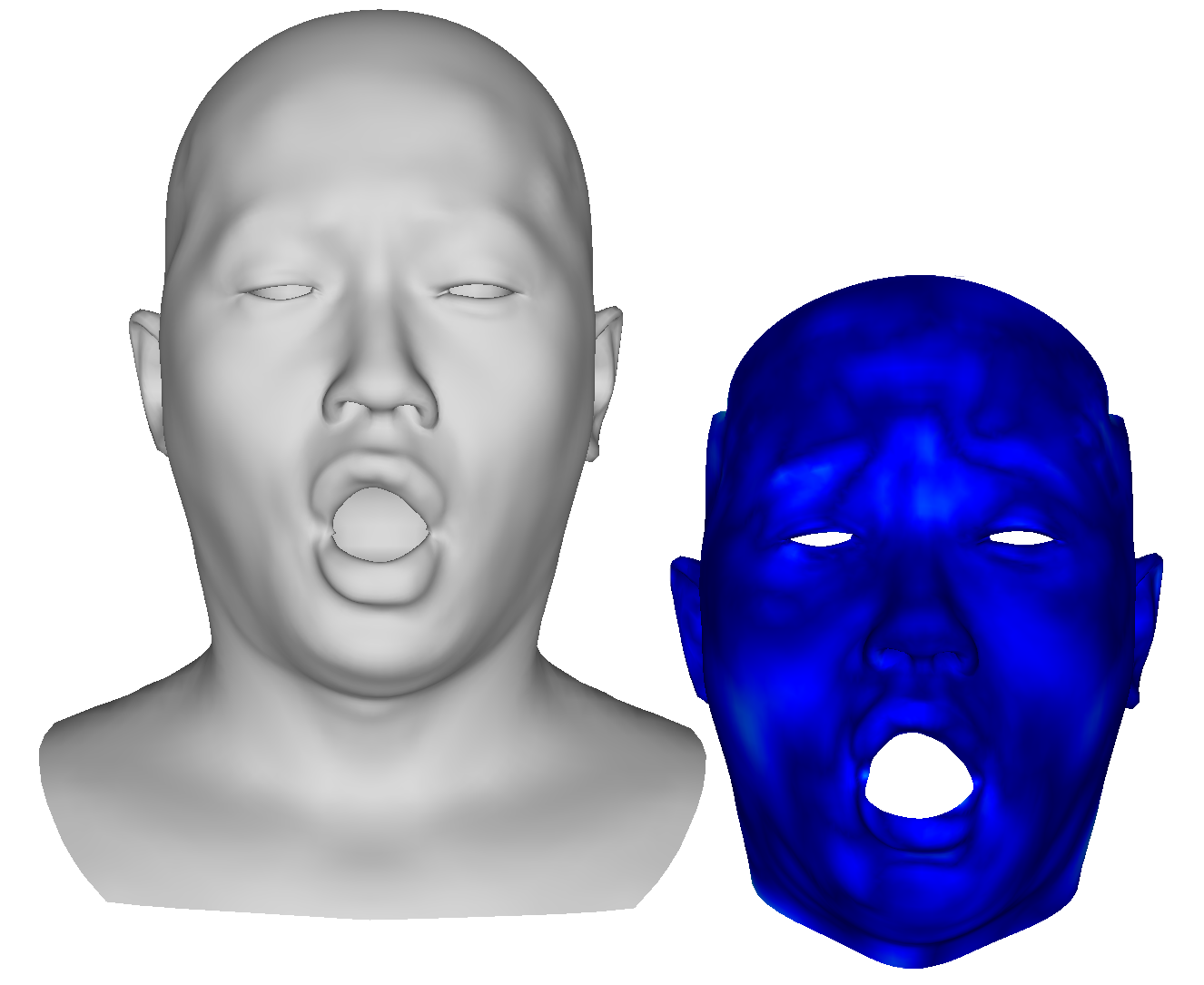} &
  \multirow{3}{*}{\includegraphics[height=0.25\linewidth]{figs/4_results/heatmap.png}} \\
  
  \includegraphics[height=0.14\linewidth]{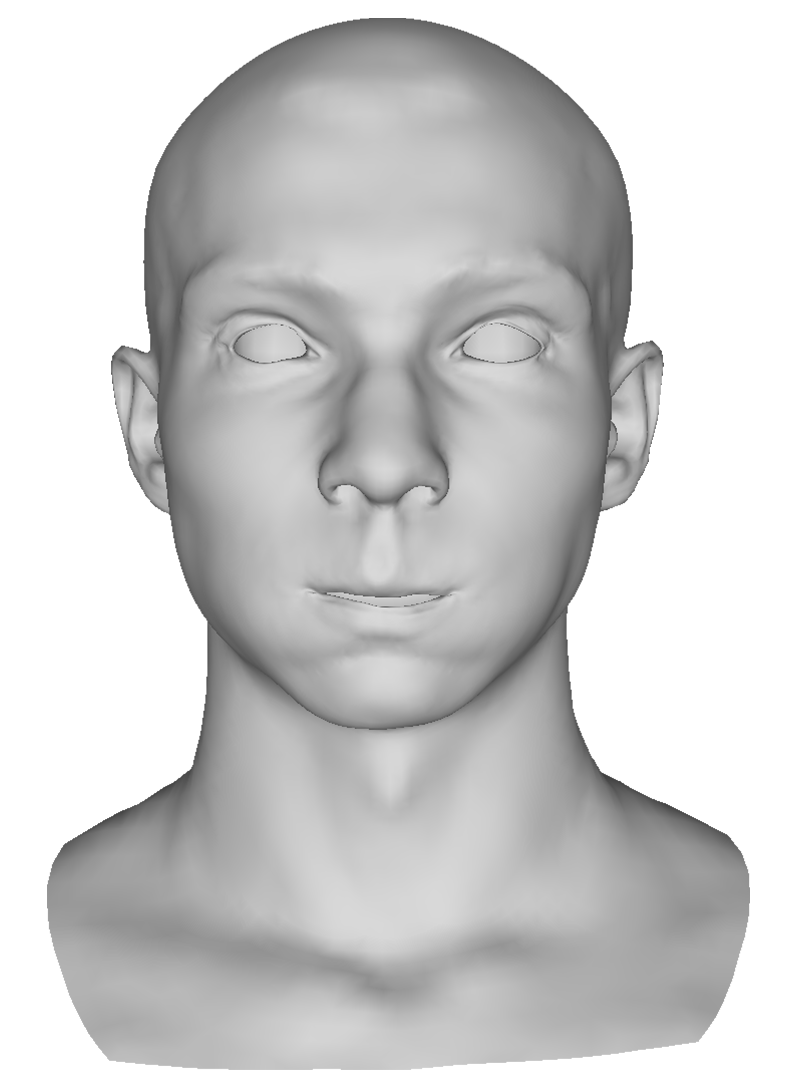} &
  \includegraphics[height=0.14\linewidth]{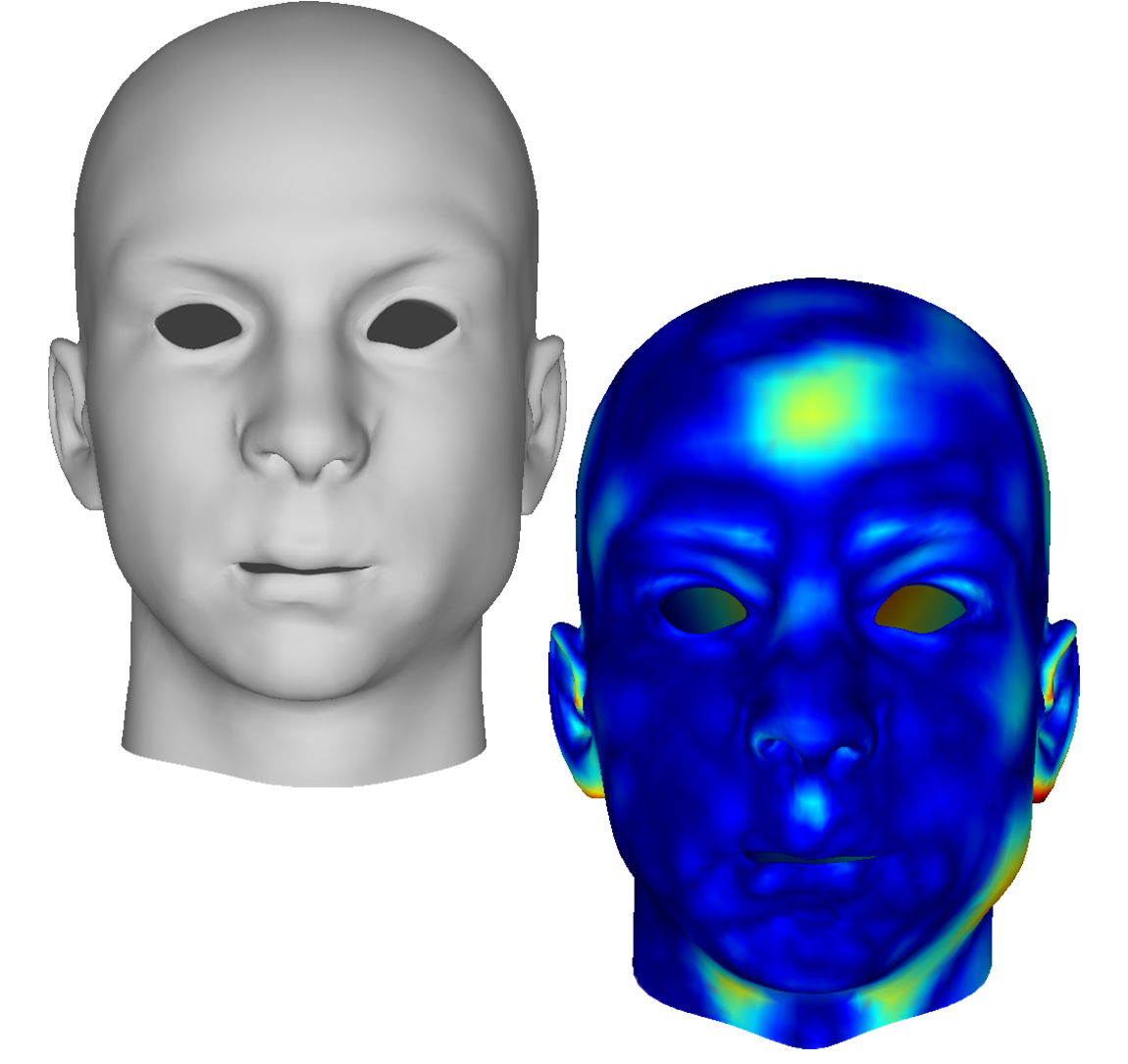} &
  \includegraphics[height=0.14\linewidth]{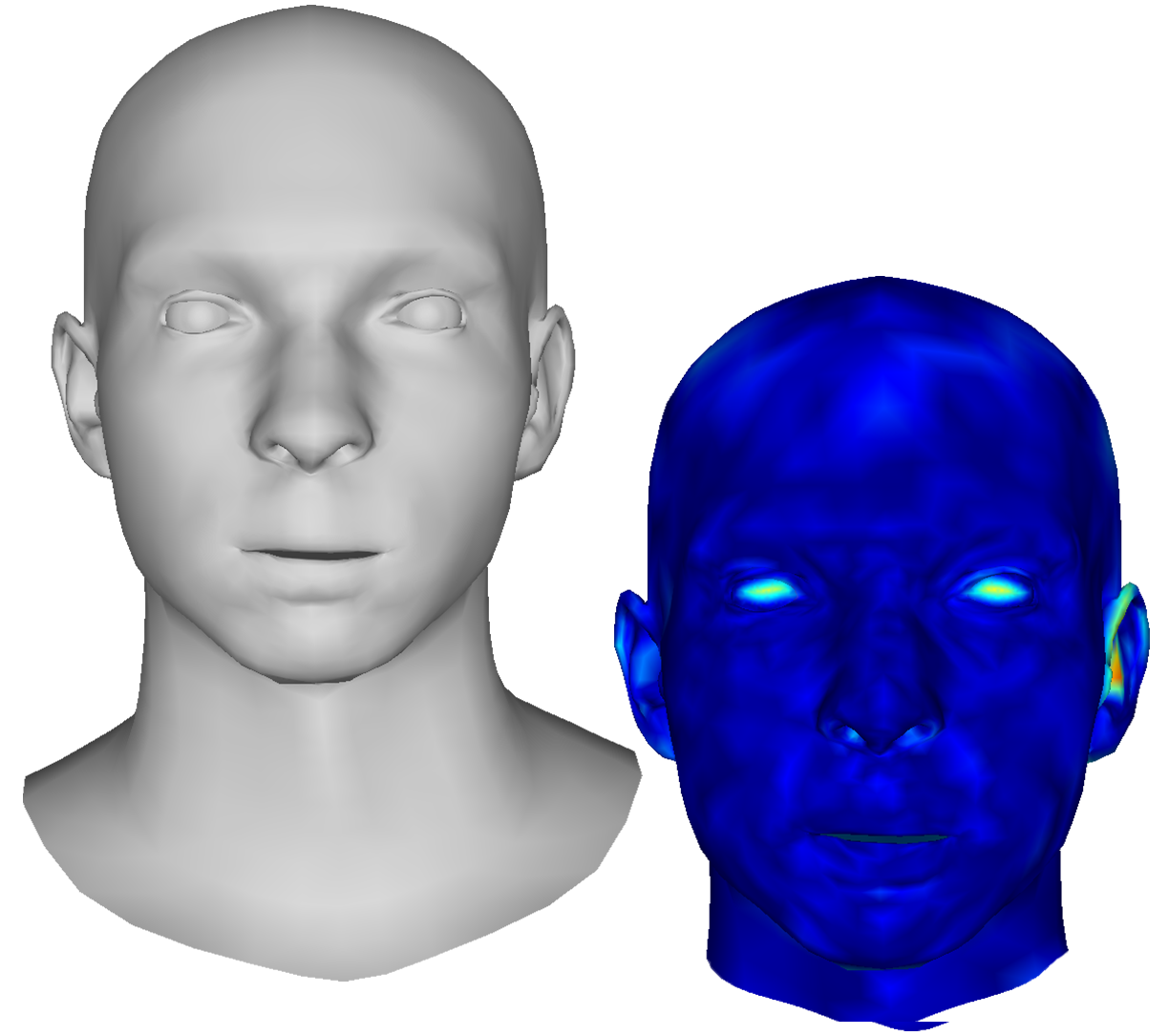} & 
  \includegraphics[height=0.14\linewidth]{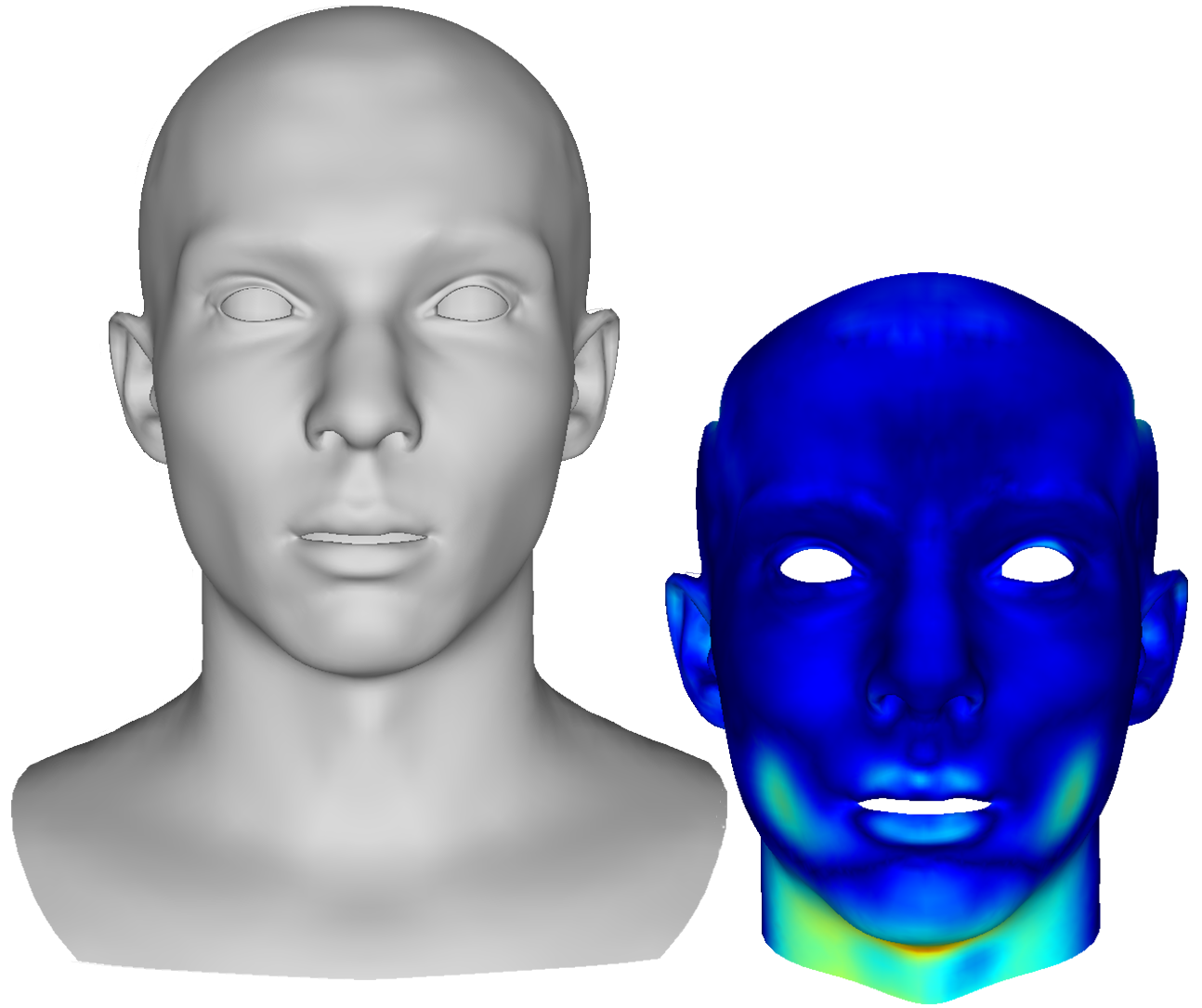} &
  \includegraphics[height=0.14\linewidth]{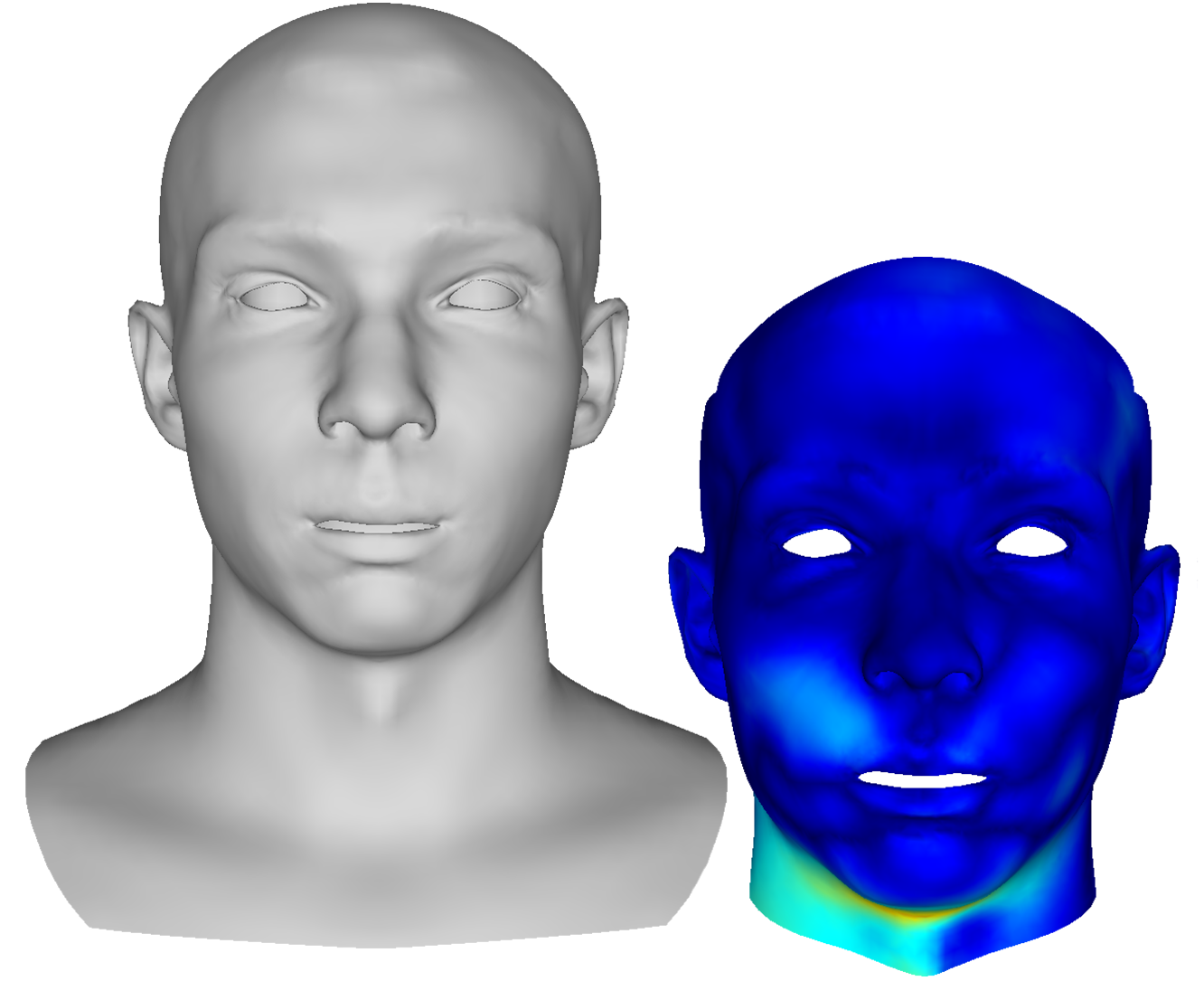} &
  \includegraphics[height=0.14\linewidth]{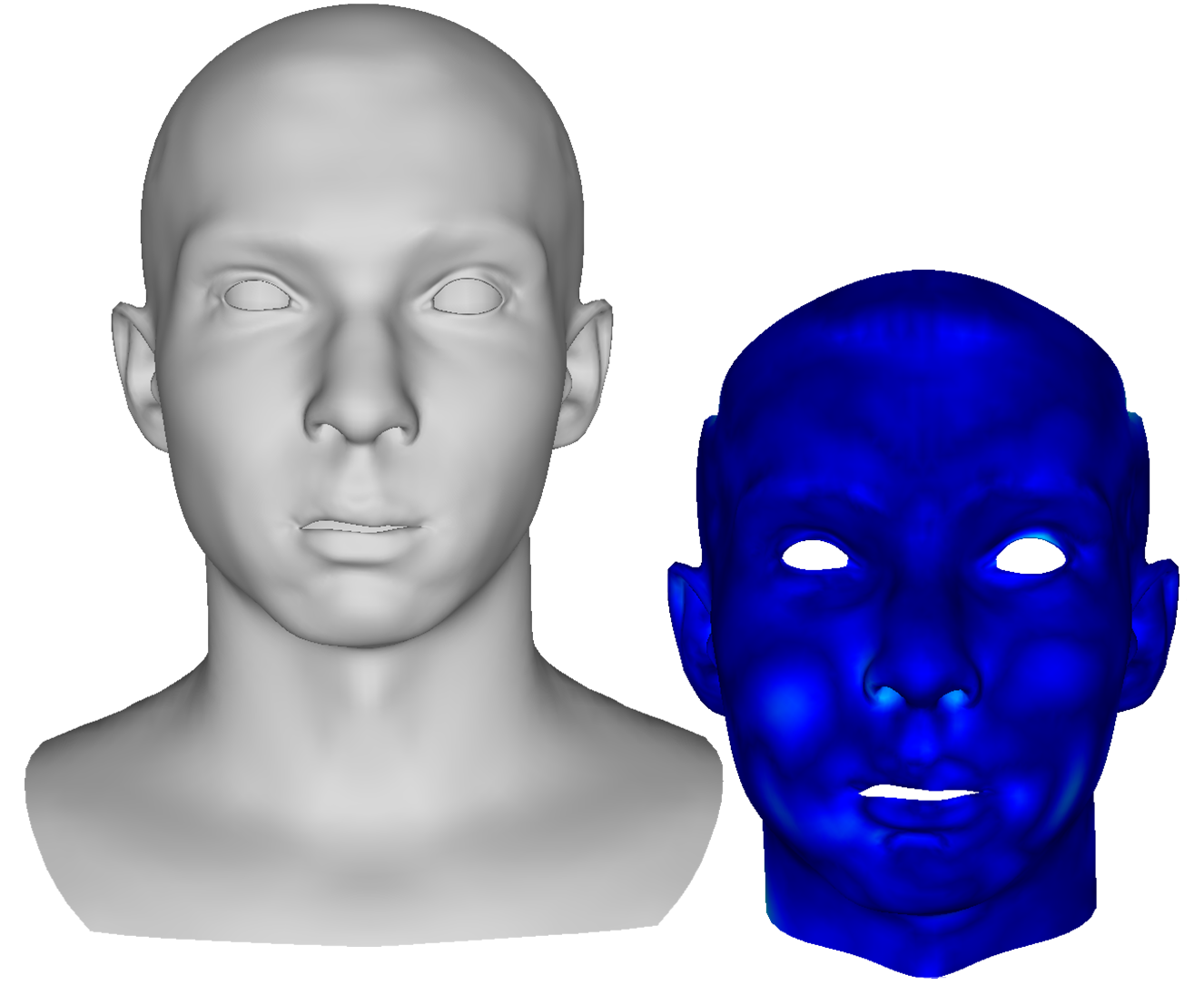} & 
  \\
  
  \includegraphics[height=0.14\linewidth]{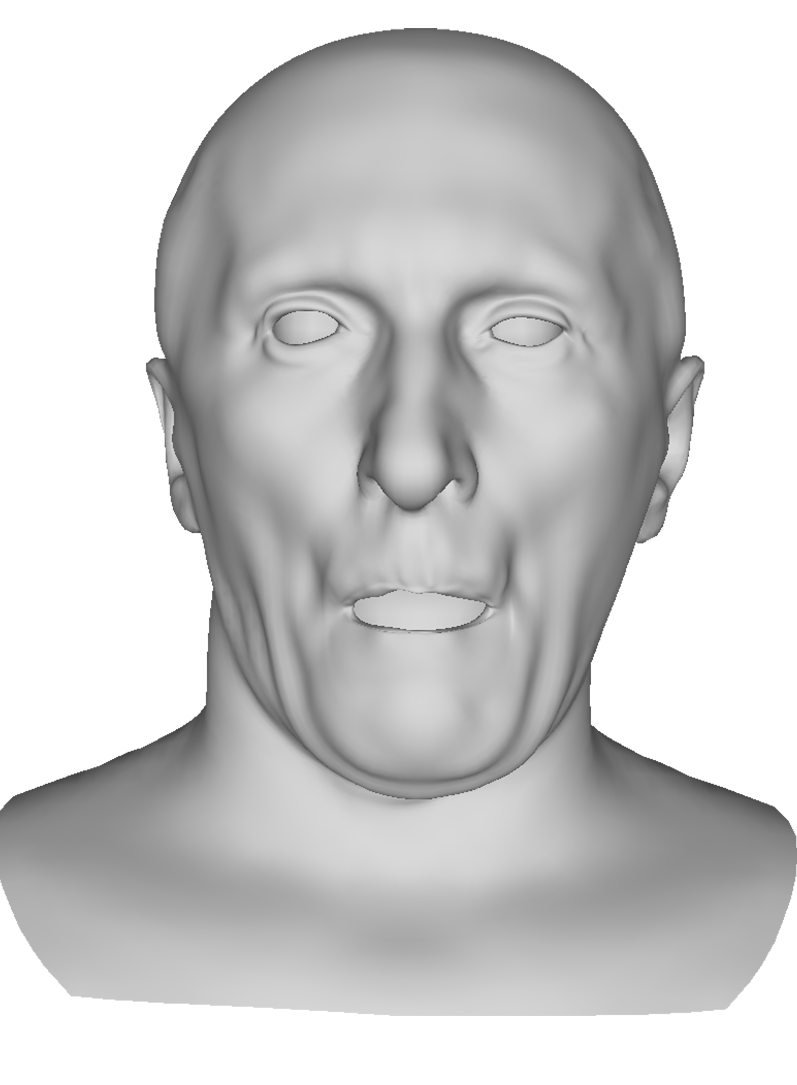} &
  \includegraphics[height=0.14\linewidth]{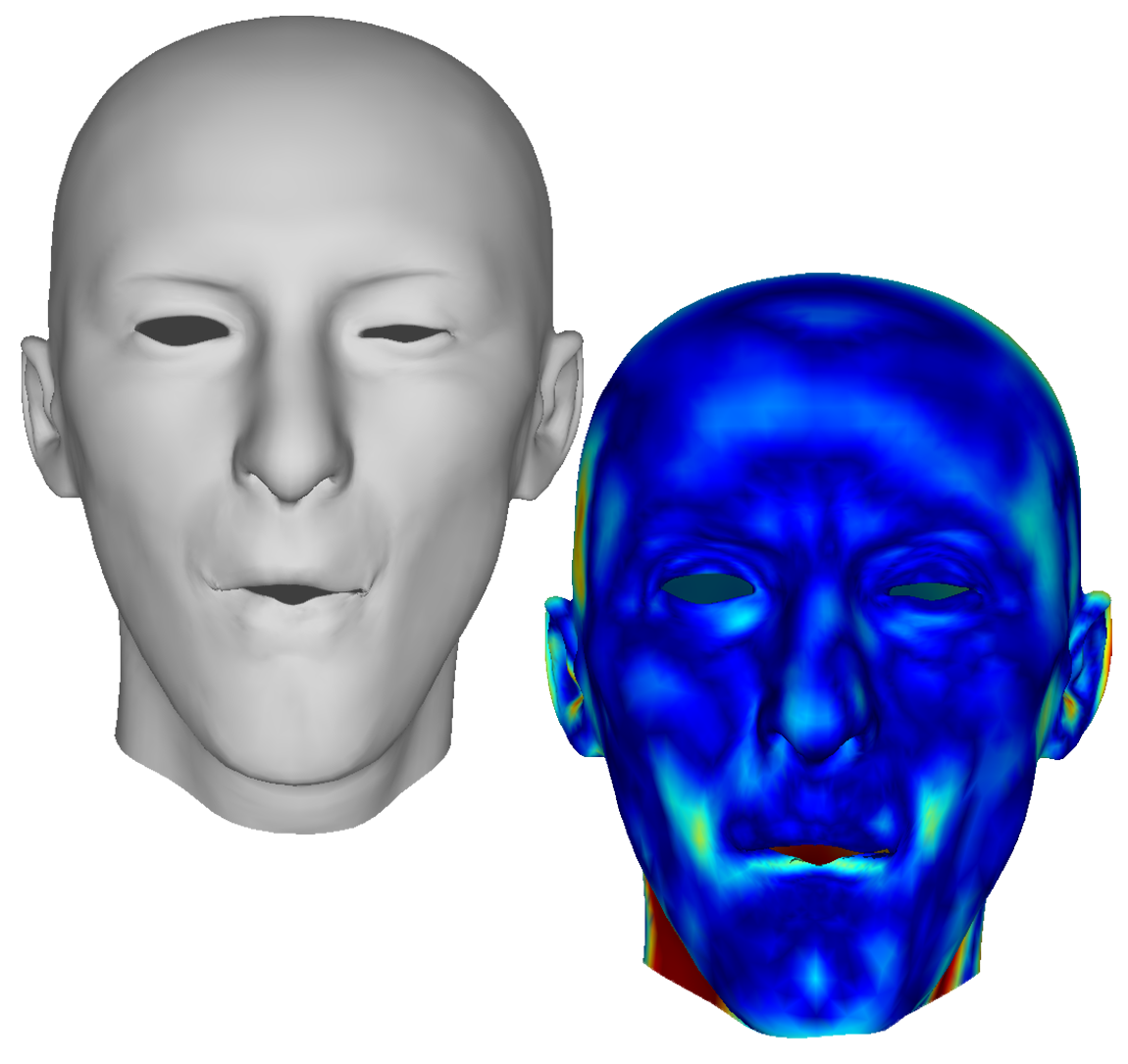} &
  \includegraphics[height=0.14\linewidth]{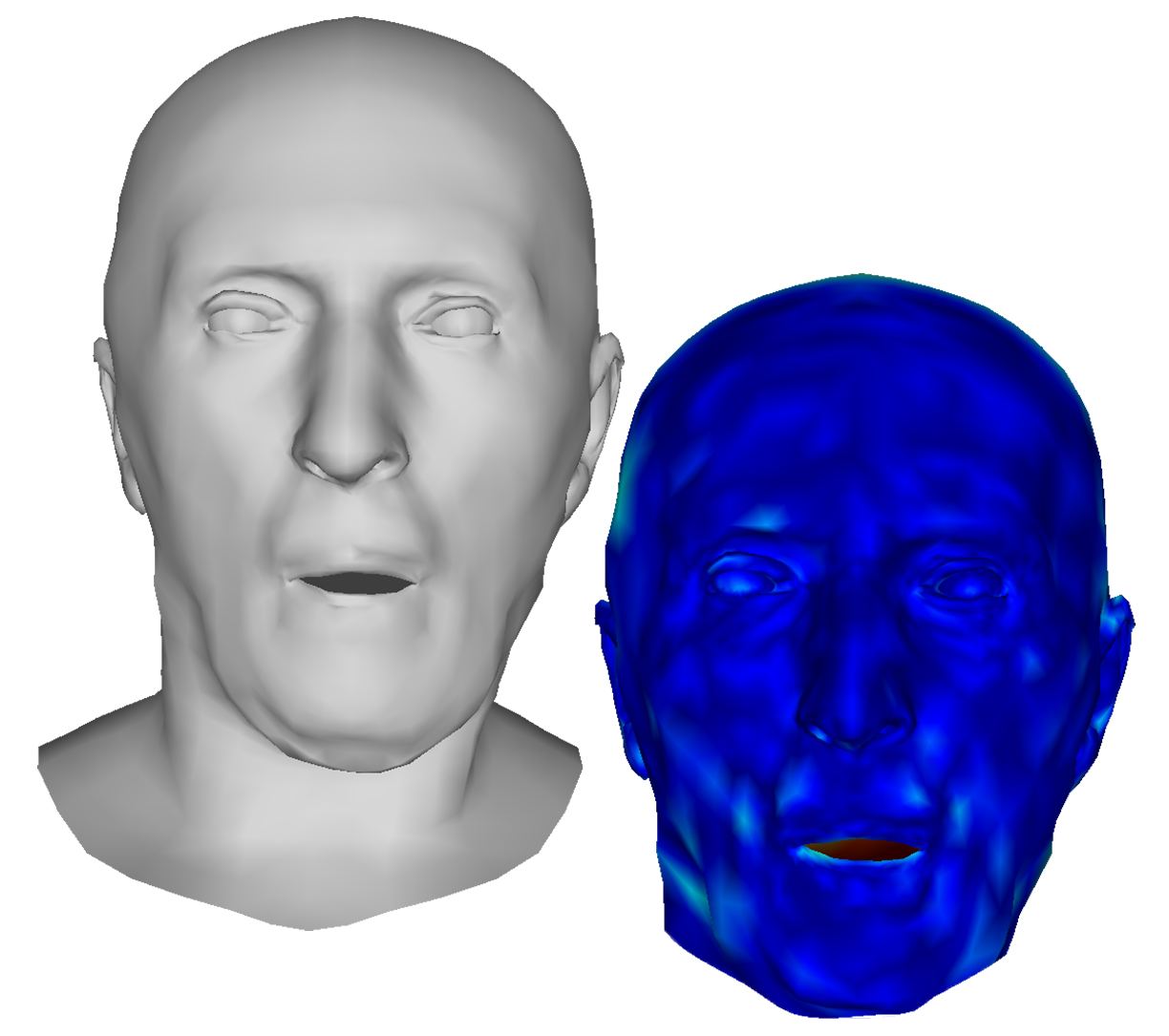} & 
  \includegraphics[height=0.14\linewidth]{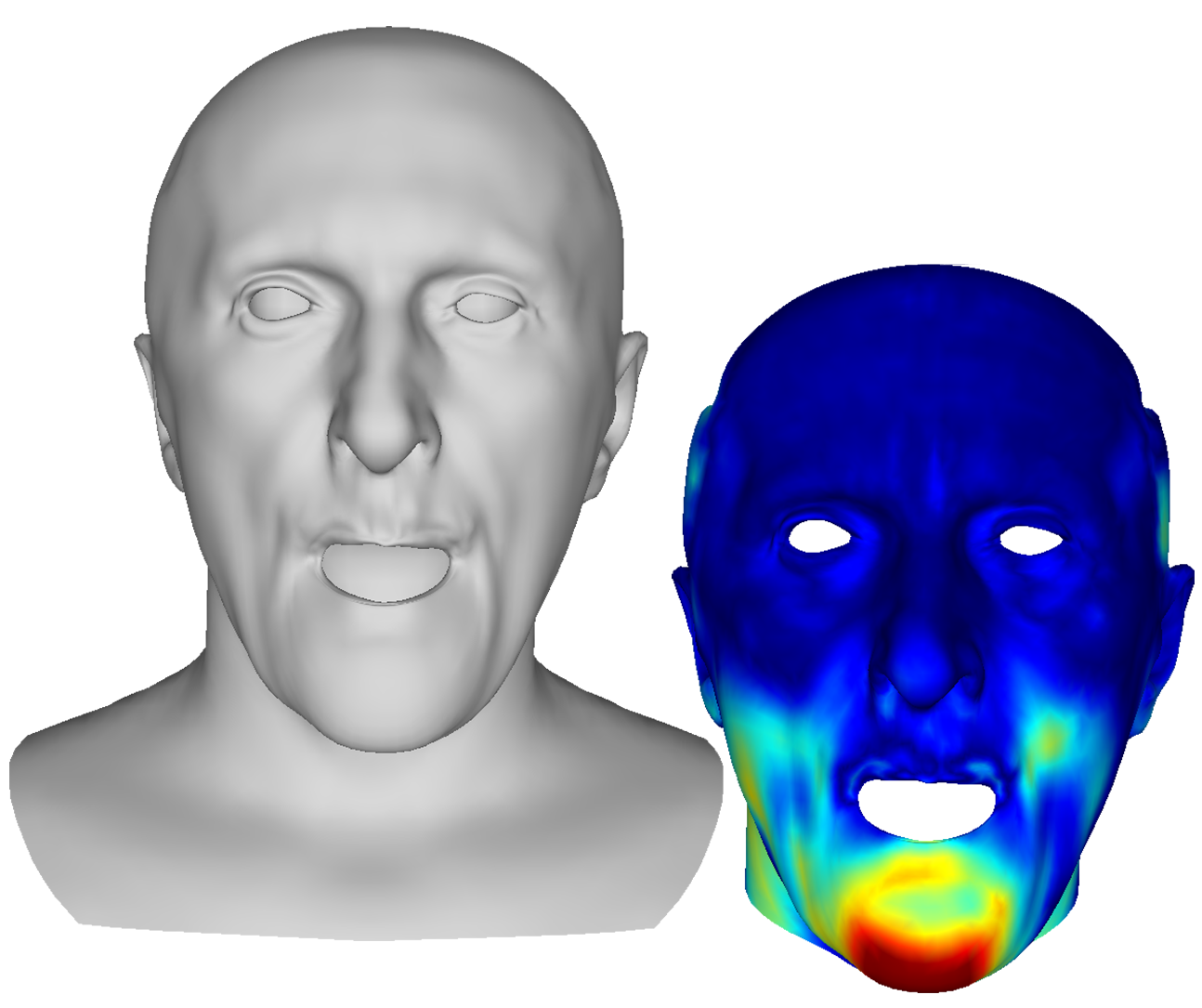} &
  \includegraphics[height=0.14\linewidth]{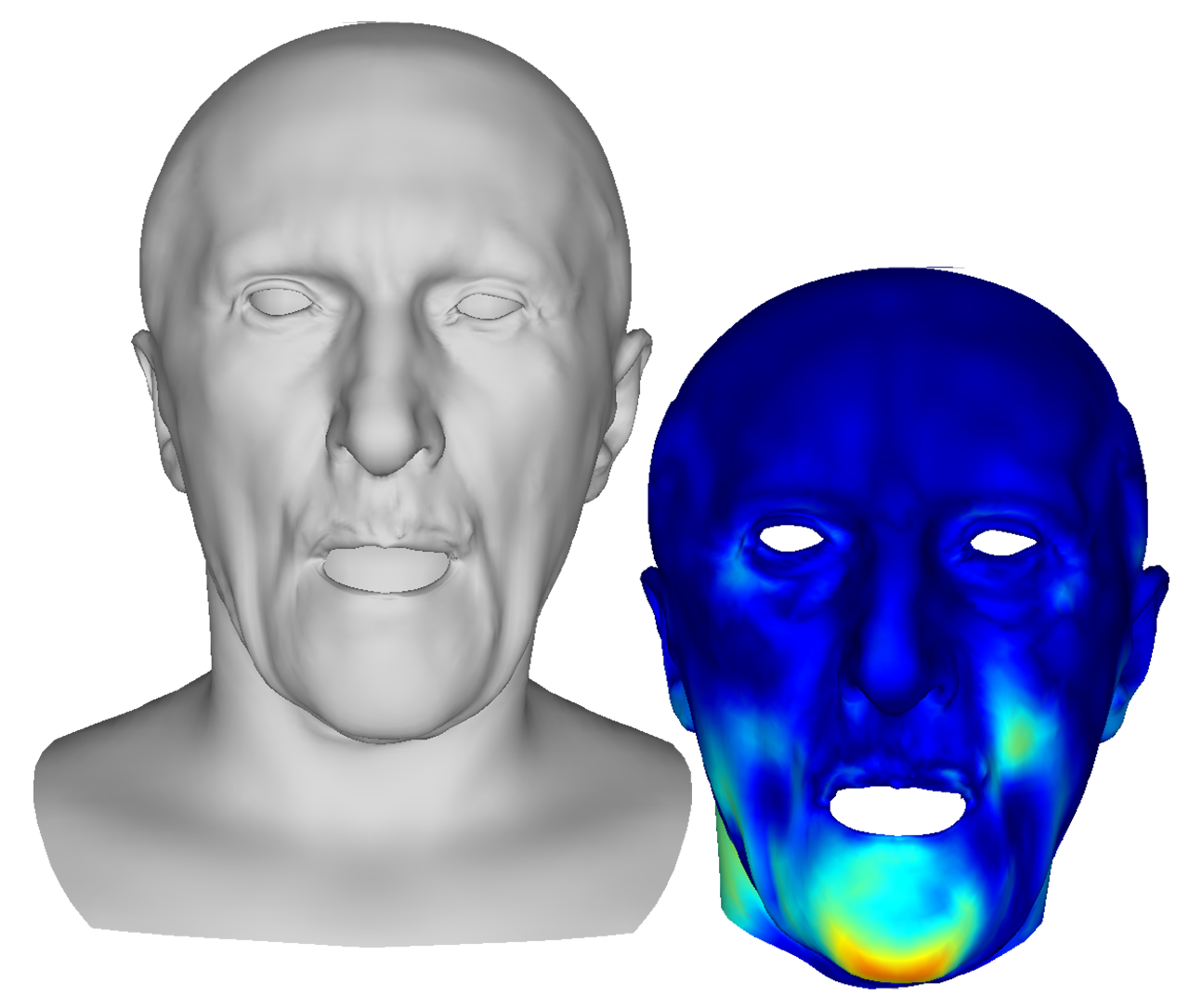} &
  \includegraphics[height=0.14\linewidth]{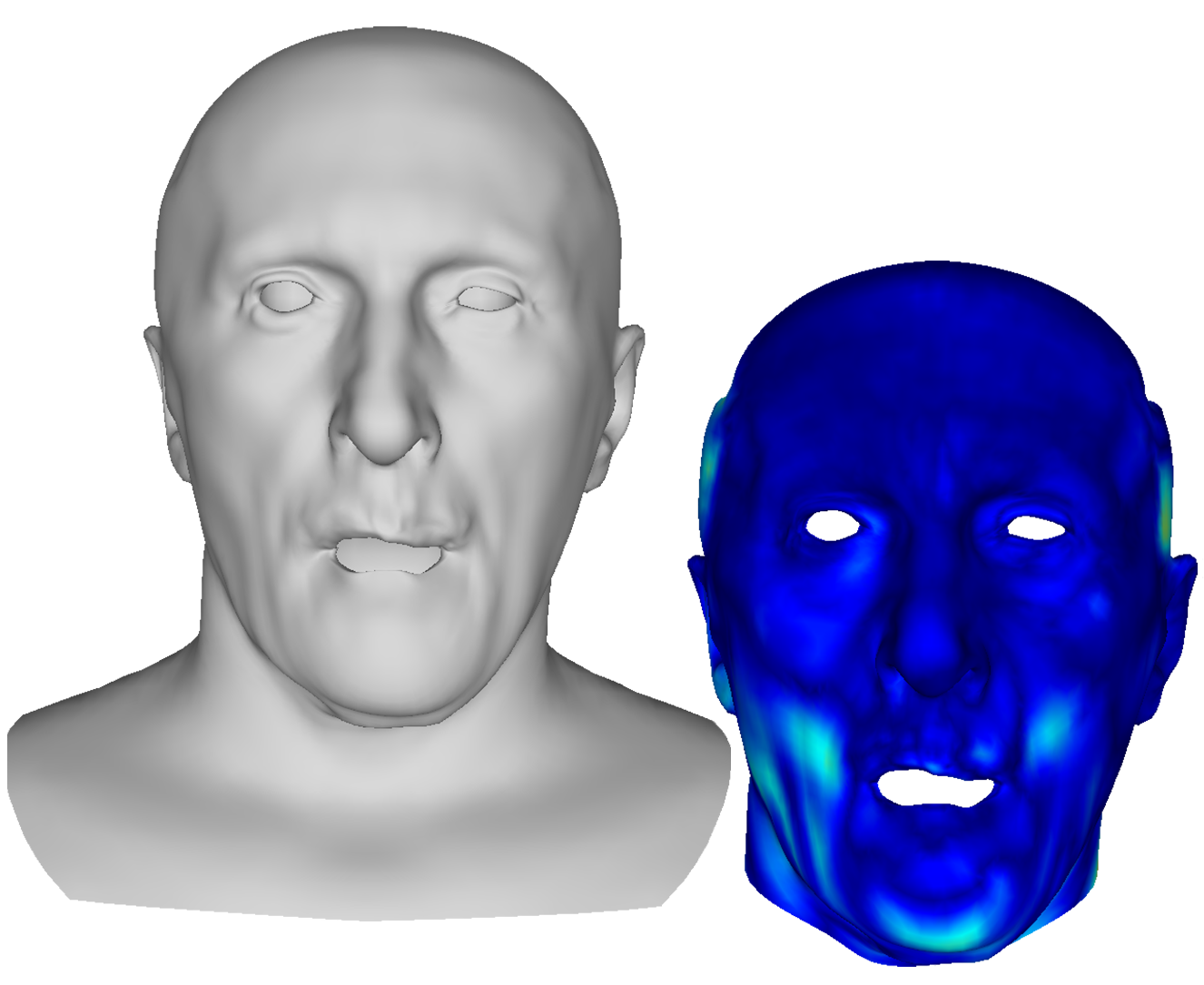} & \\
  
  GT expression & FaceWarehouse & FLAME & Template & \citet{Li_2010_SIGGRAPH} & Ours \\
  & \cite{Cao_2014_TVCG} & \cite{Li_2017_SIGGRAPH} & & & \\
  \end{tabular}}
 \caption{Comparison on the task of face fitting using different methods.}{}
 \label{fig:error_map}
\end{figure*}

\begin{figure*}
 \centering
 \includegraphics[width=0.8\linewidth]{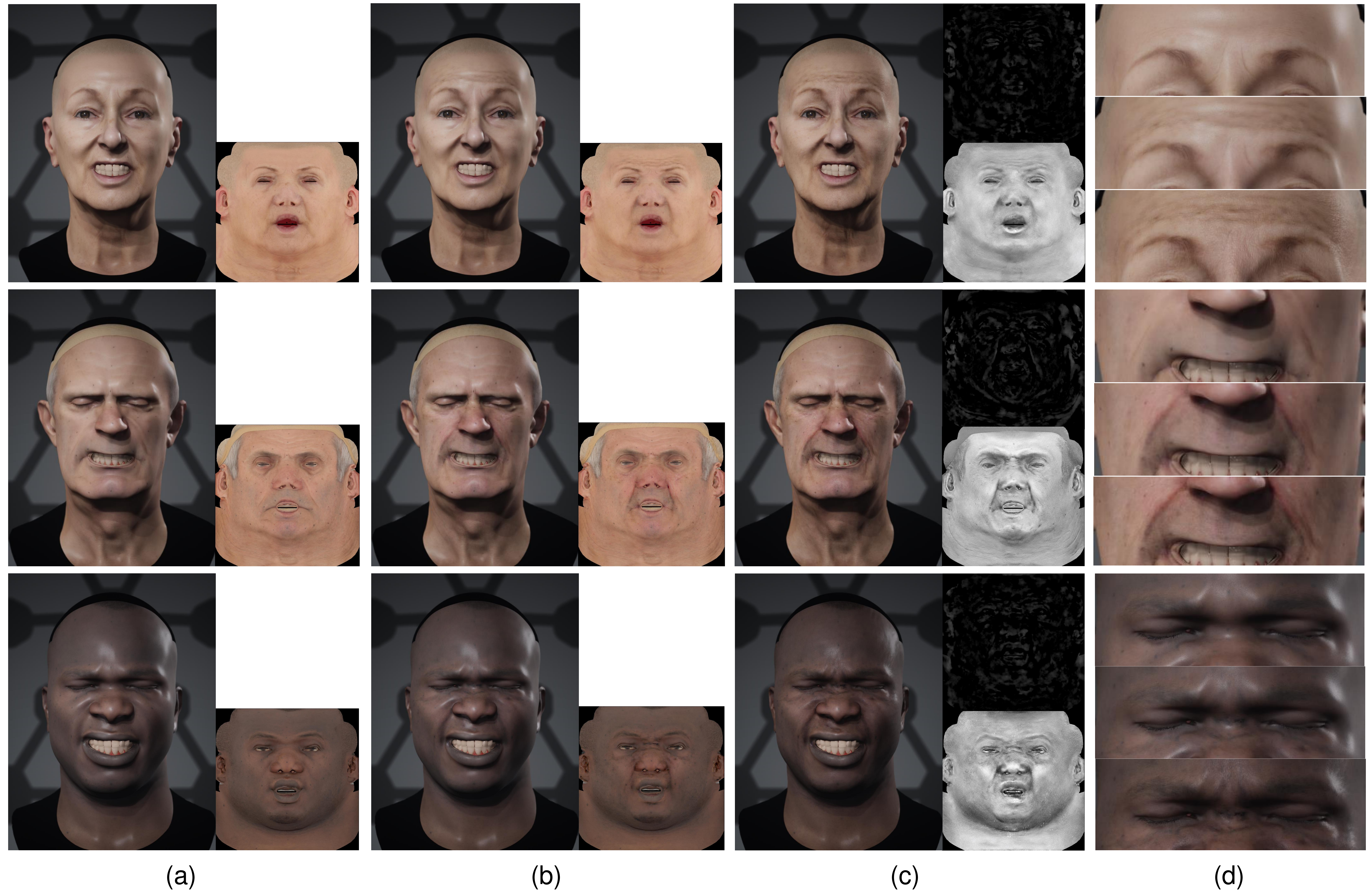}\\
 \caption{Results and comparison on Dynamic Textures. (a) Input static albedo and expression renders. (b) Our generated dynamic albedo for the specific expression and renders. (c) Our generated dynamic specular and displacement maps and renders using full set of generated assets (dynamic albedo, specular intensity and displacement). (d) From top to bottom: close-up of skin details of (a), (b) and (c).
 }{}
 \label{fig:dynamic_texture_cmp}
\end{figure*}

\begin{figure*}
  \centering
  \includegraphics[width=0.8\linewidth]{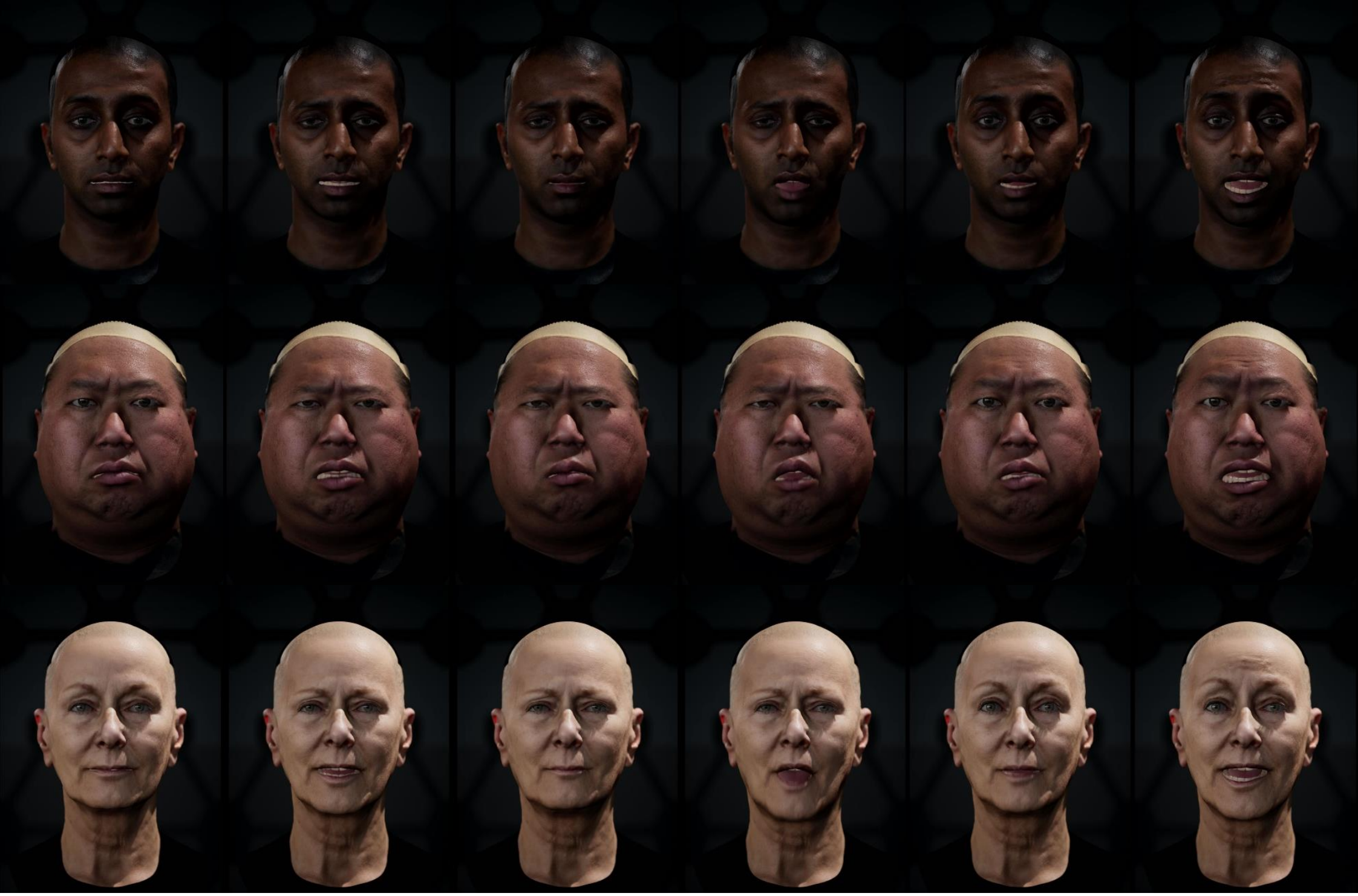}
  \caption{Animation sequences using the the full set of our generated assets.~Row 1: source sequences.~Row 2 to Row 3: target sequences driven by source sequences.}{}
  \label{fig:animation_res}
 \end{figure*}
 
\begin{figure}
\begin{center}
\includegraphics[width=0.45\textwidth]{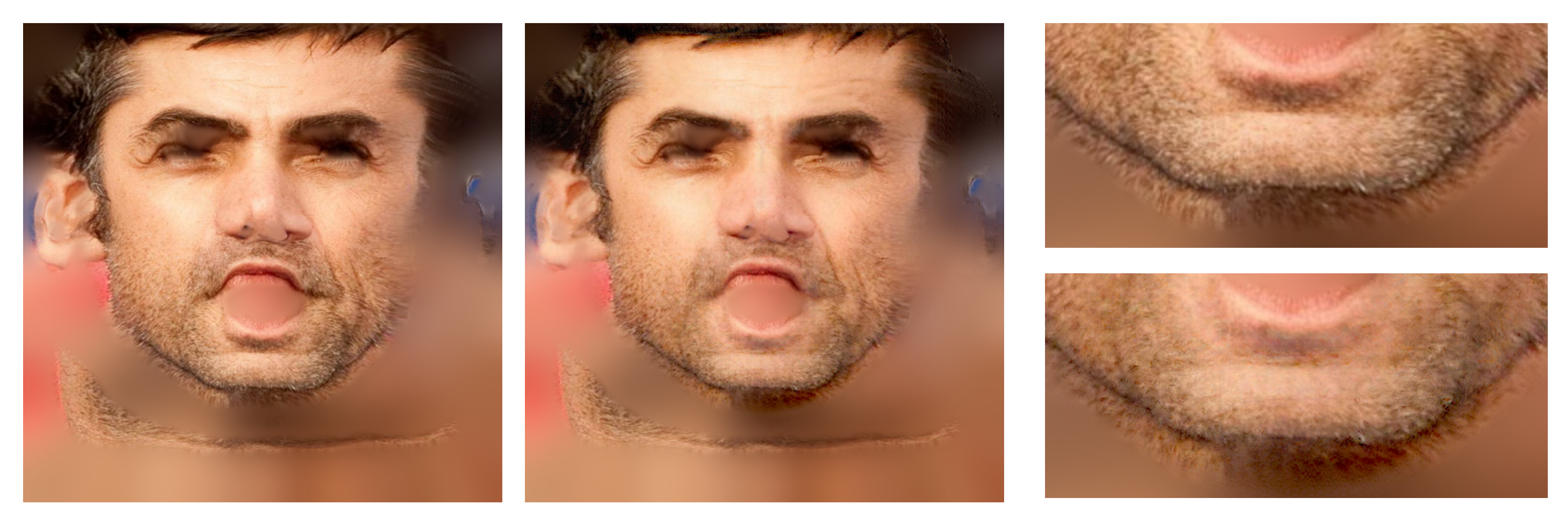}
\end{center}
  \vspace*{-8pt}
  \caption{A failure case of our texture generation model. In this case, we first extract albedo map from an image (taken from CelebA Dataset~\cite{liu2015faceattributes}), then feed this map to our texture generation network. From left to right columns: input neutral static albedo map; generated dynamic albedo of one expression by our network; close-up details of static (Top) and dynamic (Down) albedo. Note that our result has slight distortion and miscoloration in some area. This is mainly due to the limited quality of the input image, baked-in lighting and the individual's beard, which our network has not learned how to handle during the training.}
    \vspace{-10pt}
\label{fig:failure_case}
\end{figure}

\subsection{Face Model Registration.}
\paragraph{Neutral Scans Registration.} First, a linear 3D morphable face (PCA) model is used to fit the neutral face of a scan subject, reconstructed using multi-view stereo \cite{Hsieh2015UnconstrainedRF}. Secondly, the fitted model is further deformed using the non-rigid iterative closest point method \cite{Li:2008:GCO} constrained by facial landmarks \cite{sagonas2016300}. Additional Laplacian mesh surface warping is applied for surface detail reconstruction \cite{li2009robust}.
\paragraph{Expression Scans Registration.} We first estimate the blendshape expressions from our template set using the same algorithm, but varying blendshape weights in composite to identity PCA weights and followed by landmarks refinement step. We further introduce a Laplacian deformation step with dense constraints based on multi-view 2D optical flow between the current expression and neutral expression to densely correspond expressions belonging to the same subject \cite{fyffe2017multi}, see Fig. \ref{fig:textureDrifting}. 
\subsection{Texture Data Generation}
By leveraging \textit{polarized spherical gradient illumination} \cite{Ma_2007_EG,ghosh2011multiview} we can compute skin micro-structure, and material intrinsics such as diffuse albedo and specularity as we have seen inferred by our pixel translation networks. Specifically, these maps are computed on a fixed aligned topology provided by the before mentioned morphable face model.
\subsection{Template Blendshape Model}
Our blendshape model is based on the naming convention of Apple's ArKit with additional modifications enabling asymmetries for eyebrow shapes. The shapes were computed by fitting a set of around 50 scanned face neutrals along with their performed FACS shapes. By computing averages over all subjects, keeping each expression fixed, we could find reasonable averages of each shape which could be artistically isolated to keep linear independence and semantic meaning; and to avoid self-intersection. 

%% file: Results.tex
\section{Results}
\subsection{Implementation Details}
We split our data into two subsets: training set (137 subjects) and testing set (41 subjects). Each of the subsets covers a wide span of age, gender, and race. We learn our Blendshape generation networks using the RMSProp optimizer with a fixed learning rate of 0.0001 and a batch size of 4. For the texture generation network, it is optimized by the Adam optimizer with a fixed learning rate of 0.0002, batch size of 1. We train Estimation Stage and Tuning Stage for about 50,000 and 60,000 iterations respectively on an NVIDIA GeForce RTX 2080 GPU. And we train texture generation model on NVIDIA Tesla V100.
\subsection{Experiments}
\label{sec:results}
\paragraph{Run Time.} We record the run time of each component for an end-to-end system test (Table~\ref{tab:infer_time}). Testing of our blendshape generation model was performed on an NVIDIA GeForce RTX 2080 GPU while texture generation was performed on an NVIDIA Tesla V100. 

Compared to the standard high resolution avatar generation pipeline, that requires intensive manual work of weeks or months of time along with many reference expressions to be captured, our proposed approach is fast, low-cost, and robust (high-resolution training data ensures the output avatar quality).

\begin{table}
\begin{center}
\caption{Run time for each component in our framework.}
 \vspace*{-.05in}
 \begin{tabular}{l|c}
 \toprule
 Component & Time (\textit{ms})\\
 \hline
 Estimation Stage (Single Branch) & 2.386 \\
 Tuning Stage & 2.200 \\
 Texture Generation - Albedo map & 130.9 \\
 Texture Generation - Displacement \& Specular & 398.1 \\
 Texture Generation - Up-scaling & 3801\\
  \bottomrule
 \end{tabular} 
 \label{tab:infer_time}
\end{center}
\end{table}
 
\paragraph{Results}

In Fig.~\ref{fig:render_res}, we show selected expressions of novel subjects rendered using all the assets automatically generated by our framework from different sources of input data. Results show that our generated dynamic textures capture the middle-frequency details such as wrinkles and folds. In particular, the generated blendshapes of different individuals show that our approach captures the user-specific motion properties (\textit{e.g.} \textit{Mouth Right} in row two, four, six) with the semantics preserved. Note that all the generated subjects are unseen by the networks. Input test data from \citet{3DScanstore} and low-quality data captured by a mobile device are from a different domain and have never been observed by our networks. Hence, these results indicate the robustness of our framework.

\paragraph{Comparison and Evaluation.} In Fig.~\ref{fig:diff_bs_combo}, by combining the same neutral with the corresponding personalized Blendshapes units (\textit{Jaw Open} and \textit{Mouth Right}) belonging to different individuals, we showcase that our network is successful in imposing user-specific motion features to the template blendshapes. 

In Fig.~\ref{fig:extreme_exp_fit}, we show an extreme expression's fitting results with template blendshapes and our generated personalized blendshapes separately. Results indicate that our generated personalized blendshapes perform better in the non-rigid deformation (\textit{e.g.} double-chin when open mouth).

In Fig.~\ref{fig:diff_sub_bs_cmp}, we demonstrate the influence of personalized blendshapes on reconstruction/tracking accuracy by swapping blendshapes of two subjects during expression tracking. Results show that personalized blendshapes will be more expressive to the input identity regarding tracking accuracy, especially in the facial part with more non-linear and large motion (\textit{e.g.} Mouth). This result also demonstrates the effectiveness of our network: One of our network objective is to achieve better reconstruction of scanned expression.   

In Fig.~\ref{fig:bs_cmp}, we further compare our generated blendshapes with template blendshapes and  the method of \citet{Li_2010_SIGGRAPH}. Results show that our approach is comparable to \citet{Li_2010_SIGGRAPH} in the task of imposing personality to template blendshapes. Note that in \citet{Li_2010_SIGGRAPH}, 26 references scanned expression are used for optimization purposes. On the other side, our results are obtained based on a single neutral scan. Another observation is that our deep learning-based method shows more robust results with fewer artifacts (\textit{e.g.} the left mouth corner on the blendshape \textit{Mouth Left}).

In Fig.~\ref{fig:displacement}, we show dynamic displacement generated by our framework on novel subjects. Results show the effectiveness of our displacement network, which infers middle frequency details (\textit{e.g.} wrinkles) as well as high-frequency mesoscopic details.

In Fig.~\ref{fig:dynamic_texture_cmp}, we show the results and comparison of our generated dynamic textures on different subjects. Compared to static albedo from input neutral, our generated dynamic albedo predicts wrinkles, and folds caused by local self-occlusion of middle-frequency geometry change during deformation. The results also show that our predicted dynamic specular and displacement maps add mesoscopic details on top of diffuse albedo. It greatly improves the visual realism of rendering, which is important for high-end applications.

In Fig.~\ref{fig:avatar_cmp}, we compare our generated full set of face rig assets with the  state-of-the-art paGAN~\cite{Nagano_2018_SIGGRAPH}. Note the the base geometry used by paGAN~\cite{Nagano_2018_SIGGRAPH} are reconstructed from a single frontal image while ours are based on a high-quality scan. Compared to paGAN, our avatar shows better quality and much more details, which indicates that a good quality neutral scan serves better in the task of high-end avatar generation. The results also shows the unique physically-based skin assets will greatly improve the avatar rendering quality. The displacement map in our assets captures the middle frequency and pore-level details.

\subsection{Applications}
\paragraph{Expression Reconstruction/ Face Tracking.} In Fig.~\ref{fig:error_map}, we compare our generated personalized blendshapes on fitting of performance capture sequences with other methods. As shown in Fig.~\ref{fig:diff_sub_bs_cmp}, smaller fitting errors indicates better personality on blendshapes. Results show that our generated personalized blendshapes outperform baseline methods (Template and optimization-based method in \citet{Li_2010_SIGGRAPH} on accuracy of the face tracking task using the same solver. To provide better quantitative evidence, we evaluate face reconstruction on 2,548 expressions in training dataset and 626 expressions in testing datasets. The results are listed in Table~\ref{tab:our_data_cmp}. Blendshapes optimized by~\citet{Li_2010_SIGGRAPH} and ours show smaller reconstruction errors in both training and testing data. 

\begin{table}
\caption{Reconstruction errors between the ground truth expressions and the reconstructed expressions using blendshapes by different methods on training and testing datasets.}
 \vspace*{-.05in}
 \begin{tabular}{l|cc}
  \toprule
  Method & Training $\downarrow$ & Testing $\downarrow$\\
  \hline
  Template blendshapes & 1.661 & 1.638 \\
  Optimization method~\cite{Li_2010_SIGGRAPH} & 1.389 &  1.483 \\
  Ours & \textbf{1.341} & \textbf{1.372} \\
  \bottomrule
 \end{tabular} 
 \label{tab:our_data_cmp}
 \end{table}

\paragraph{Animation}

In Fig.~\ref{fig:animation_res}, we show that our generated face rig assets can be used directly for animation. \textit{Please refer to accompanying video material for more results}.

%% file: Conclusion.tex
\section{Conclusion}
\label{sec:Con}
We have demonstrated an end-to-end framework for high-quality personalized face rig and asset generation from a single scan. Our face rig assets include a set of personalized blendshapes, physically based dynamic textures and secondary facial components (including teeth, eyeballs, and eyelashes). Compared to previous automatic avatar and facial rig generation approaches, which either require a considerable number of person-specific scans or can only produce a relatively low-fidelity avatar, our framework only requires a single neutral scan as input and can produce plausible identity attributes including physically-based dynamic textures of facial skins. This characteristic is key to creating compelling animation-ready avatars at scale.

We achieve the above objective by modeling the correlation between identity and personalized blendshapes using an extensive dataset of high-resolution facial scans. In particular, our generated dynamic textures add details from mid-frequencies (wrinkles) to mesoscopic ones (pore level). Our automatically generated face rig assets are valuable for real-world production pipelines, as these high-fidelity initial models can be provided to artists for fine-tuning or simply used as secondary characters for crowds. Our proposed method is fast, robust, and lightweight, allowing production studios to simply scan a neutral face of a person and immediately obtain a high-quality facial rig. An interesting insight from our experiments is that the identity seems to be enough for a plausible inference of personalized facial appearance and dynamic expressions. In addition to our framework, we have also introduced a novel self-supervised deep neural network training approach to deal with the case when no ground truth data is available, which in our case are the personalized blendshapes.

\paragraph{Limitations and Future Work}
As a deep learning approach, the effectiveness of our algorithm relies on the variety and volume of training data of our database. In particular, facial expressions that are specific to young subjects could be improved, due to the lack of young subjects in our current database. For the same reason, our framework also does not perform well on subjects with facial hair or beard as shown in Fig.~\ref{fig:failure_case}. We plan to augment our database to cover more diversity and appearance variations.

Our template model consists of 55 blendshape vectors, which can recover most of the expressions in daily life and is commonly used in lightweight applications. However, certain extreme expressions still cannot be represented by our model. Our proposed network architecture can be adapted for arbitrary template blendshapes. Thus, we are interested in exploring more sophisticated blendshape rigs that consist of hundreds to thousands of expressions, such as the ones used in film production.
We use generic eyes and teeth models for all the generated avatars. An interesting direction would be to explore how to generate personalized eyes \cite{Berard_2016_SIGGRAPH, Berard_2019_EG} and teeth \cite{Wu_2016_SIGGRAPH, Velinov_2018_SIGGRAPH} automatically as well. 

\begin{acks}
We thank Liwen Hu from Pinscreen for the fruitful discussions and helping with this paper. This research is funded by in part by the ONR YIP grant N00014-17-S-FO14, the CONIX Research Center, one of six centers in JUMP, a Semiconductor Research Corporation (SRC) program sponsored by DARPA, the Andrew and Erna Viterbi Early Career Chair, the U.S. Army Research Laboratory (ARL) under contract number W911NF-14-D-0005, Adobe, and Sony. The content of the information does not necessarily reflect the position or the policy of the Government, and no official endorsement should be inferred.
\end{acks}